\documentclass[
    10pt,
    prx,
    superscriptaddress,
    a4paper,
    twocolumn,
    floatfix,
    longbibliography,
    showkeys
]{revtex4-2}

\usepackage[american]{babel}
\usepackage[utf8x]{inputenc}

\usepackage{grffile} % allow for period characters in filename

\usepackage{newtxtext,newtxmath}
\usepackage{microtype}

\usepackage{amsmath}
\usepackage{bbm}% bold math
\usepackage{bm}
\usepackage{mathtools}
\usepackage{physics}
\usepackage{dsfont}% einheitsoperator
\usepackage{braket}
\usepackage{cancel}
\usepackage{slashed}
\usepackage{xparse}
\usepackage{stackengine}

\usepackage{graphicx}% I Fig.~files
\usepackage[svgnames]{xcolor}

\makeatletter
  \g@addto@macro \normalsize{%
    \setlength\abovedisplayskip{3pt plus 0pt minus 0pt}%
    \setlength\belowdisplayskip{3pt plus 0pt minus 0pt}}%
\makeatother

\definecolor{cset-aps-blueberry}{RGB}{28,128,158}
\definecolor{cset-aps-blue}{RGB}{46,44,184}
\definecolor{cset-aps-turquoise}{RGB}{0,67,88}
\definecolor{cset-aps-limegreen}{RGB}{190,219,67}
\definecolor{cset-aps-green}{RGB}{31,138,112}
\definecolor{cset-aps-yellow}{RGB}{255,225,25}
\definecolor{cset-aps-orange}{RGB}{253,116,0}
\definecolor{cset-aps-red}{RGB}{219,0,43}

\definecolor{cset-aps-kobalt-medium}{RGB}{62,54,222}
\definecolor{cset-aps-kobalt-dark}{RGB}{28,24,150}

\definecolor{cset-aps-my-label-red}{RGB}{202,0,17}
\definecolor{cset-aps-my-label-blue}{RGB}{53,71,140}
\definecolor{cset-aps-my-label-gray}{RGB}{145,145,145}

\usepackage{circledsteps}

\usepackage{siunitx} %Fuer einheiten und nicefrac

\usepackage{ragged2e}
\usepackage{array}
\usepackage{tabularx}
\usepackage{booktabs}
\usepackage{multirow}
\usepackage[autostyle]{csquotes}

\usepackage[cal=cm,scr=dutchcal]{mathalpha}

\usepackage{hyperref}
\hypersetup{%
    colorlinks=true,
    linkcolor={cset-aps-red},
    linkbordercolor={cset-aps-red},
    filecolor={cset-aps-orange},
    filebordercolor={cset-aps-orange},
    citecolor={cset-aps-kobalt-medium},
    citebordercolor={cset-aps-kobalt-medium},
    urlcolor={cset-aps-kobalt-dark},
    urlbordercolor={cset-aps-kobalt-dark},
    menucolor={cset-aps-limegreen},
    menubordercolor={cset-aps-limegreen},
    breaklinks=true,
    pdfborderstyle={/S/U/W 2},
    pdfpagemode=UseOutlines,
    pdfstartpage={1},
}

\usepackage[capitalise]{cleveref}

\crefname{secinapp}{appendix}{appendices}
\Crefname{secinapp}{Appendix}{Appendices}

\crefname{equation}{}{}
\Crefname{equation}{}{}

\newcommand{\I}{\mathrm{i}}
\newcommand{\E}{\mathrm{e}}

\DeclareFontFamily{OMS}{oasy}{\skewchar\font48 }
\DeclareFontShape{OMS}{oasy}{m}{n}{%
         <-5.5> oasy5     <5.5-6.5> oasy6
      <6.5-7.5> oasy7     <7.5-8.5> oasy8
      <8.5-9.5> oasy9     <9.5->  oasy10
      }{}
\DeclareFontShape{OMS}{oasy}{b}{n}{%
       <-6> oabsy5
      <6-8> oabsy7
      <8->  oabsy10
      }{}
\DeclareSymbolFont{oasy}{OMS}{oasy}{m}{n}
\SetSymbolFont{oasy}{bold}{OMS}{oasy}{b}{n}

\DeclareMathSymbol{\smallleftarrow}     {\mathrel}{oasy}{"20}
\DeclareMathSymbol{\smallrightarrow}    {\mathrel}{oasy}{"21}
\DeclareMathSymbol{\smallleftrightarrow}{\mathrel}{oasy}{"24}

\newcommand\barbelow[1]{\stackunder[0.8pt]{$#1$}{\rule{.5em}{.075ex}}}
\newcommand{\TensorMatrixUnderlined}[1]{\barbelow{\barbelow{#1}}}
\newcommand{\SecondOrderTensor}[1]{\TensorMatrixUnderlined{#1}}

\DeclarePairedDelimiter\Abs{\lvert}{\rvert}

\DeclareMathOperator{\VLaplace}{\boldsymbol{\Delta}}

\newcommand{\DeltaTransversal}{{\SecondOrderTensor{\Vect{\delta}}^{\perp}}}
\newcommand{\DeltaLongitudinal}{{\SecondOrderTensor{\Vect{\delta}}}^{\parallel}}

\newcommand{\D}{\mathrm{d}}
\newcommand{\FuncD}{\mathscr{D}}

\NewDocumentCommand{\Ds}{sm}{%
  \IfBooleanTF{#1}
   {% * variant, we are at the beginning
    \mathrm{d}#2\,%
   }
   {% normal variant
    \mathop{}\!\mathrm{d}#2%
   }%
}

\NewDocumentCommand{\FuncDs}{sm}{%
  \IfBooleanTF{#1}
   {% * variant, we are at the beginning
    \mathscr{D}#2\,%
   }
   {% normal variant
    \mathop{}\!\mathscr{d}#2%
   }%
}

\newcommand{\Vect}[1]{\boldsymbol{#1}}
\NewDocumentCommand{\intRlap}{e{_^}}{%
  \DOTSI
  \int_{\IfValueT{#1}{\mathrlap{#1}}}^{\IfValueT{#2}{\mathrlap{#2}}}%
  \mspace{-4mu}%
}

\newcommand{\QmOp}[1]{\hat{#1}}

\newcommand{\PhaseSpaceX}{\chi}
\newcommand{\PhaseSpaceP}{\wp}
\newcommand{\PhaseSpaceXVec}{\Vect{\chi}}
\newcommand{\PhaseSpacePVec}{\Vect{\wp}}
\newcommand{\PhaseSpaceAction}{\mathcal{A}}
\NewDocumentCommand{\PhaseSpaceArgs}{ O{} O{}}{\PhaseSpaceX^{#1}_{#2},\PhaseSpaceP^{#1}_{#2},\PhaseSpaceAction^{#1}_{#2}}
\NewDocumentCommand{\PhaseSpaceArgsVec}{ O{} O{}}{\PhaseSpaceXVec^{#1}_{#2},\PhaseSpacePVec^{#1}_{#2},\PhaseSpaceAction^{#1}_{#2}}

\makeatletter
\newcommand{\xsize}[1]{\bBigg@{#1}}
\newcommand{\vast}{\bBigg@{4}}
\newcommand{\Vast}{\bBigg@{5}}
\makeatother

\makeatletter
\newcommand{\Eqlfill@}{\arrowfill@\Relbar\Relbar\Relbar}
\newcommand{\eql}[2][]{\ext@arrow 0066\Eqlfill@{#1}{#2}}
\makeatother

\makeatletter
\def\leftharpoonfill@{\scriptsize\bf\bf\arrowfill@\leftharpoonup\relbar\relbar}
\def\rightharpoonfill@{\scriptsize\bf\bf\arrowfill@\relbar\relbar\rightharpoonup}
\def\overarrow@#1#2#3{\vbox{\ialign{##\crcr#1#2\crcr
 \noalign{\nointerlineskip}$\m@th\hfil#2#3\hfil$\crcr}}}
\newcommand{\overrightharpoon}{%
  \mathpalette{\overarrow@\rightharpoonfill@}}
\newcommand{\overleftharpoon}{%
  \mathpalette{\overarrow@\leftharpoonfill@}}
\makeatother

\newcommand{\orcid}[1]{\href{https://orcid.org/#1}{\includegraphics[width=7pt]{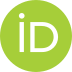}}}

\makeatletter
\renewcommand*{\thesection}{\arabic{section}}
\renewcommand*{\thesubsection}{\thesection.\arabic{subsection}}
\renewcommand*{\p@subsection}{}

\renewcommand*{\p@subsubsection}{}
\makeatother

\newcommand{\OrangelineChange}[1]{#1}

\newcommand{\FlabelNoIdx}{F}

\newcommand{\affULM}{\address{Institut f{\"u}r Quantenphysik and Center for Integrated Quantum Science and Technology (\OrangelineChange{IQST}), Universit\OrangelineChange{{\"a}}t Ulm, Albert-Einstein-Allee 11, D-89069 Ulm, Germany}}
\newcommand{\affHagler}{\address{Hagler Institute for Advanced Study and Department of Physics and Astronomy, Institute for Quantum Science and Engineering (IQSE), Texas A{\&}M University, College Station, Texas 77843-4242, USA}}
\newcommand{\affTexas}{\address{Texas A{\&}M AgriLife Research, Texas A{\&}M University, College Station, Texas 77843-4242, USA}}

\begin{document}
\title{The Wave Functional of the Vacuum in a Resonator
\\[1ex]
\small\normalfont{Published in \href{https://doi.org/10.12693/APhysPolA.143.S52}{Acta Phys. Pol. A 143, 52 (2023)}}
}

\keywords{\OrangelineChange{Wave functional, Vacuum, Wightman tensor, Cavity Quantum Electrodynamics}}

\author{Alexander Friedrich~\orcid{0000-0003-0588-1989}}
\affULM
\email{alexander.friedrich@uni-ulm.de}

\author{Daniela Moll}
\affULM

\author{Matthias Freyberger}
\affULM

\author{Lev Plimak~\orcid{0000-0003-4533-9254}}
\affULM

\author{Wolfgang P. Schleich~\orcid{0000-0002-9693-8882}}
\affULM
\affTexas
\affHagler

\begin{abstract}
\noindent We show that despite the fundamentally different situations, the wave functional of the vacuum in a resonator is identical to that of free space. The infinite product of Gaussian ground state wave functions defining the wave functional of the vacuum translates into an exponential of a sum rather than an integral over the squares of mode amplitudes weighted by the mode volume and a power of the mode wave number.~\OrangelineChange{We express this sum by an integral of a bilinear form of the field containing a kernel given by a function of the square root of the negative Laplacian acting on a transverse delta function.} For transverse fields it suffices to employ the familiar delta function which allows us to obtain explicit expressions for the kernels of the vector potential, the electric field and the magnetic induction. We show~\OrangelineChange{for} the example of the vector potential that different mode expansions lead to different kernels. Lastly, we show that the kernels have a close relationship with the Wightman correlation functions of the fields.
\end{abstract}

\maketitle

\section{Introduction}
The standard approach~\cite{Fermi1932,Lamb1995} towards the quantization of the electromagnetic field is straightforward: decomposition of the field into modes and quantization of the resulting harmonic oscillator amplitudes by the canonical commutation relations. The wave functional of the vacuum proposed by John Archibald Wheeler~\cite{Wheeler1957,Wheeler1962,Misner2017} and extended~\cite{Bialynicka-Birula1987,Bialynicki-Birula2000,Bialynicki-Birula1996,Bialynicki-Birula2003,Bialynicki-Birula2023} and refined by Iwo Bialynicki-Birula does not rely on a mode expansion but involves the complete electromagnetic field. The essence of the wave functional is best summarized by the following quote from Bialynicki-Birula's article~\cite{Bialynicki-Birula2000} employing the wave functional to obtain the Wigner phase space distribution of the whole electromagnetic field:
\blockquote{%
    \emph{''The whole electromagnetic field is treated as one huge, infinitely dimensional harmonic oscillator. The wave function and the corresponding Wigner function become then functionals of the field variables.''}%
}
The recent impressive progress in cavity and circuit quantum electrodynamics invites us to reconsider the wave functional of the vacuum in case of a resonator. Indeed, so far, the investigations have concentrated exclusively on free space. In the present article, we show that the expressions for the wave functional of the vacuum in the two situations are identical. 

\subsection{The cradle of the quantum theory of fields}
The year 1925 marks not only the birth of modern quantum mechanics, but is also arguably the beginning of quantum electrodynamics (QED). Indeed,~\OrangelineChange{the} ''Drei-M\"{a}nner-Arbeit''~\cite{Born1926} not only provided the foundations of matrix mechanics, but also presented for the first time the quantization of the \emph{free} electromagnetic field. \OrangelineChange{This was extended only two years later to include the interaction with quantized matter~\cite{Dirac1927}.}

The discovery of the Lamb shift~\cite{Lamb1947} and the anomalous magnetic moment~\cite{Foley1948} in 1947 demonstrated that the theory, so far plagued by infinities, contained some truth. The renormalization theory~\cite{Schwinger1958,Schweber1994}
developed shortly after, removed these infinities, and gave rise to the field of QED, a theory~\cite{Bialynicki-Birula1975} with unprecedented agreement with experiment. 

Almost 40 years later, new experimental manifestations of QED emerged from the use of high-Q microwave cavities~\cite{Haroche2013,Walther2006} and the interaction of individual atoms with single modes of the radiation field. Whereas in the first era of cavity QED, the experiments were only in the microwave domain, the optical domain \OrangelineChange{soon} followed.
The last 20 years have seen the development of a new rapidly moving branch of quantum optics summarized by circuit QED~\cite{Blais2021} and\OrangelineChange{,} recently\OrangelineChange{,} waveguide QED~\cite{Sheremet2023}.

Ever since the proposal of quantized electrodynamics\OrangelineChange{,} there \OrangelineChange{has been} a constant drive \OrangelineChange{toward} a deeper understanding of the associated vacuum fluctuations and the measurability of the field components. For example, Lev Davidovich Landau and Rudolf Peierls \cite{Landau1931} applied the uncertainty principle to relativistic quantum theory and concluded:
\blockquote{%
    \emph{''The assumptions of wave mechanics which have been shown to be necessary in section 2 are therefore not fulfilled in the relativistic range and the application of wave mechanics methods to this range goes beyond their scope. It is therefore not surprising that the formalism leads to various infinities; it would be surprising if the formalism bore any resemblance to reality.''}%
}
Needless to say, this grim outlook was not shared by Niels Bohr\OrangelineChange{,} who\OrangelineChange{,} together with L{\'e}on Rosenfeld\OrangelineChange{,} immediately started to correct this article. However, it took them two years to achieve this goal for the case of free fields~\cite{Bohr1933}, and they stated:
\blockquote{%
    \emph{''Not only is it an essential complication of the problem of field measurements that, when comparing field averages over different space-time regions, we cannot in an unambiguous way speak about a temporal sequence of the measurement process;''}%
}
After the discovery of renormalization Bohr and Rosenfeld returned~\cite{Bohr1950} to this problem and included charges. For an interesting commentary by Rosenfeld providing the historical context of both articles we refer to~\cite{Wheeler2016}.

The analogous question of the measurability of the gravitational field, pioneered by Helmut Salecker and Eugene Paul Wigner~\cite{Salecker1958}\OrangelineChange{,} led to Wheeler's Geometrodynamics~\cite{Wheeler1957} and the quantum fluctuations of gravity and the quantum foam. It was in this context that he proposed \OrangelineChange{to} consider the wave functional~\cite{Wheeler1962,Misner2017} of electromagnetism \OrangelineChange{as a} guide for linearized gravity. Armed with the insights from electromagnetism\OrangelineChange{,} he was able to derive an estimate for the fluctuations of the space-time geometry at distances of the Planck length. For a detailed discussion of the wave functional of linearized gravity, we refer to the classic paper by \OrangelineChange{Karel Kucha\u{r}}~\cite{Kuchar1970}. 

Similarly, but on more general grounds, Julian Schwinger investigated the effect of so-called \emph{fluctuating sources}~\OrangelineChange{(i.e., transient fields)} in quantum field theories~\cite{Schwinger1967}. Some of these ideas\OrangelineChange{~\cite{Weinberg1995,Padmanabhan2016}} eventually found their way into the framework which later became effective (quantum) field theory. 

\OrangelineChange{Recent} years have seen a renaissance of the wave functional of the vacuum. It now appears not only in the Schr\"{o}dinger representation of quantum field theory~\cite{Jackiw1987,Hatfield2018} but also in possible realizations~\cite{Bose2017,Marletto2017} of the \emph{Gedanken Experiment} of Richard P.~Feynman~\cite{Rickles2011} addressing the question of measurability~\cite{Chen2023} of entanglement between two quantum systems due to gravity which has recently attracted significant attention. This field has become~\OrangelineChange{quite an} active area of research, due to the emerging technical possibility of preparing almost macroscopic systems in motional quantum states, and~\OrangelineChange{also} because direct tests of the quantum nature of gravity via the detection of gravitons \OrangelineChange{seem} highly unlikely\OrangelineChange{,} as suggested in Ref.~\cite{Dyson2013} by yet another founding father of QED\OrangelineChange{,} Freeman Dyson.

For this reason we find it appropriate to revisit the wave functional of the vacuum and analyze it for the case of a resonator. This situation is not only timely, but the set of discrete modes makes the derivation much cleaner. On the other hand the discreteness adds a different complication arising from the sum over the modal indices confirming the well-known adage: "There ain't no such thing as a free lunch".

\subsection{Road to the wave functional}
\begin{figure*}
    \centering
    \includegraphics[width=0.91009434\textwidth]{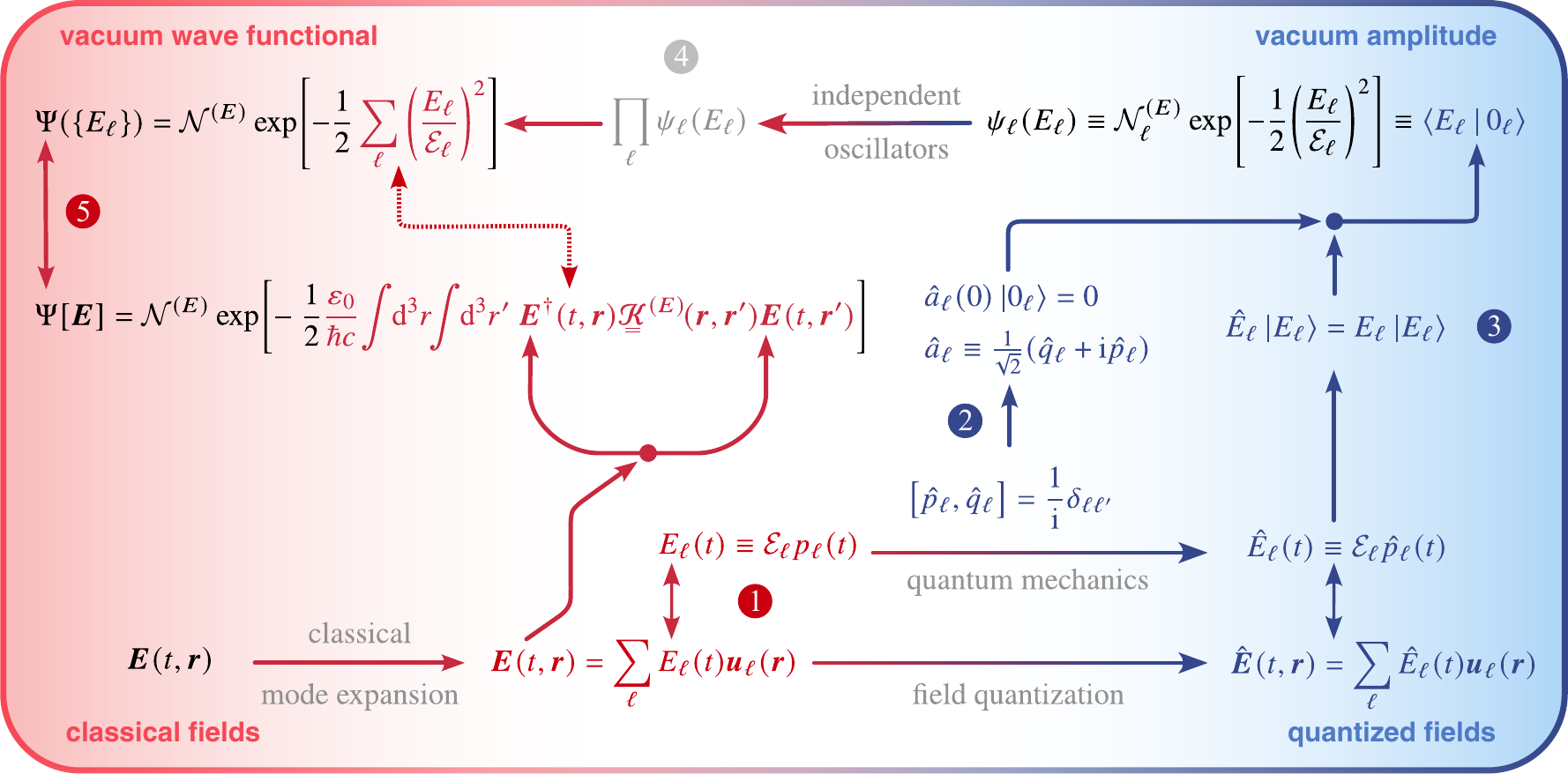}
    \caption{%
    {\bfseries Road to the wave functional $\Psi[\Vect{E}]$ of the vacuum in a resonator.} \normalfont
    We identify five different ingredients marked by numbers: \Circled[fill color=cset-aps-my-label-red,inner color=white, outer color=white]{\footnotesize{1}} mode expansion of the electric field $\Vect{E}=\Vect{E}(t,\Vect{r})$, \Circled[fill color=cset-aps-my-label-blue,inner color=white, outer color=white]{\footnotesize{2}} quantization of the field $E_\ell$ in each mode $\Vect{u}_\ell$ according to the canonical commutation relations, \Circled[fill color=cset-aps-my-label-blue,inner color=white, outer color=white]{\footnotesize{3}} definition of the electric field eigenstates $\Ket{E_\ell}$ and the ground state $\Ket{0_\ell}$ of the $\ell$-th mode, \Circled[fill color=cset-aps-my-label-gray,inner color=white, outer color=white]{\footnotesize{4}} infinite product of all ground state wave functions  \OrangelineChange{$\psi_\ell(E_\ell)$}, and \Circled[fill color=cset-aps-my-label-red,inner color=white, outer color=white]{\footnotesize{5}} wave functional $\Psi[\Vect{E}]$ of the vacuum after the \OrangelineChange{elimination} of the mode decomposition. The last ingredient, that is the connection between the discrete sum over the modes and the space integrals of the bilinear form of the electric field $\Vect{E}$ and a kernel  \OrangelineChange{$\SecondOrderTensor{\mathbfscr{K}}^{(E)}$}, indicated by a dashed line, constitutes the topic of our article.
    }
    \label{fig:figure1-how-everything-works-together}
\end{figure*}

We now summarize our path to the wave functional of the vacuum in a resonator using the example of the electric field representation.~\OrangelineChange{In} Figure~\ref{fig:figure1-how-everything-works-together} we start from the decomposition of the electric field $\Vect{E}\equiv \Vect{E}(t,\Vect{r})$ (left lower corner) into a discrete set of modes $\Vect{u}_\ell$. \OrangelineChange{Here,}~the subscript $\ell$ combines the polarization index as well as the indices characterizing the wave vector $\Vect{k}_\ell$ enforced by the boundary conditions on the Helmholtz equation by the shape of the resonator.

The subsequent quantization of the corresponding electric field amplitudes $E_\ell\equiv \mathcal{E}_\ell p_\ell$ using the canonical commutation relations leads us to the eigenvalue equation of the electric field operator $\QmOp{E}_\ell$ in mode $\Vect{u}_\ell$. Together with the definition of the ground state $\ket{0_\ell}$ of the $\ell$-th mode in terms of the annihilation operator $\QmOp{a}_\ell$, we find the Gaussian wave function $\psi_\ell(E_\ell) \equiv \Braket{E_\ell|0_\ell}$ in the electric field representation.

In the absence of matter and interactions, the modes are independent of each other and correspond to a product state with all modes in the ground state. Hence, we arrive at an \emph{infinite product} of Gaussian wave functions. Due to the functional equation of the exponential function, this product reduces to a \emph{single} exponential of an \emph{infinite sum} over the squares of the scaled fields $E_\ell/\mathcal{E}_\ell$ in the modes.

The mode expansion we started \OrangelineChange{with} shows that this sum is identical to an appropriate space integral consisting of a bilinear form of the electric field and a kernel. In this way, we have eliminated the mode decomposition and have arrived at an expression containing the quantum mechanics of the vacuum as well as the complete electric field distribution $\Vect{E}=\Vect{E}(t,\Vect{r})$ without resorting to modes.

We conclude this section by briefly addressing the differences and detours enforced by free space due to the \emph{continuous} superposition of plane wave modes.

In the case of free space\OrangelineChange{,} the continuous superposition of the plane waves\OrangelineChange{,} rather than the discrete set of modes\OrangelineChange{,} involves an \emph{integration} over the wave vector rather than a \emph{summation} over mode indices $\ell$. The quantization of the field is identical to that in a resonator\OrangelineChange{,} with the exception of the commutation relation where the Kronecker delta in $\ell$ and $\ell^\prime$ is replaced by the Dirac delta function in the difference of the wave vectors $\Vect{k}$ and $\Vect{k}^\prime$.

However, the \emph{infinite continuous} product of the ground state wave functions now requires either a discretization of the continuum of the wave vectors, or a more sophisticated technique. Once the functional equation of the exponential function has transformed the \emph{infinite product} into \OrangelineChange{an} \emph{infinite sum}\OrangelineChange{,} we can continue with the integral\OrangelineChange{, which} is a continuous superposition. In free space as well as in the resonator\OrangelineChange{,} we arrive at the same expression for the wave functional $\Psi[\Vect{E}]$ of the vacuum.

\subsection{In a nutshell}
Before we dive into the mathematics, we motivate our results without detailed derivations and summarize them in \cref{tab:table1,tab:table2}. We start our discussion by recalling in~\cref{tab:table1} the essential ingredients of the expansion of a vector field into modes.

Throughout the article, we focus on an expansion of the vector potential $\Vect{A}$, the electric field $\Vect{E}$, and the magnetic induction $\Vect{B}$ into a set of discrete mode functions.
%%%%%%%%%%%%%%%%%%%%%%%%%%%
% TABLE 
%%%%%%%%%%%%%%%%%%%%%%%%%%%
\begin{table*}%[b]
\caption{%
Elements of the expansion of a vector field $\Vect{F}$ such as the vector potential $\Vect{A}$, the electric field $\Vect{E}$, and the magnetic induction $\Vect{B}$ into a discrete set of mode functions $\Vect{u}_\ell$ of the vector potential. Here $\ell$ denotes the mode index consisting of the polarization and three integers
characterizing the wave vector $\Vect{k}_\ell$ determined by the boundary conditions of the Helmholtz equation imposed by the shape of the resonator. The field strengths $A_\ell$, $E_\ell$ and $B_\ell$ in the $\ell$-th mode are given by the products $\mathcal{A}_\ell q_\ell$,
\OrangelineChange{$\mathcal{E}_\ell p_\ell$} and $\mathcal{B}_\ell q_\ell$ of the vacuum field strengths $\mathcal{A}_\ell$, $\mathcal{E}_\ell$ and $\mathcal{B}_\ell$ as well as the dimensionless \OrangelineChange{quadrature variables} $q_\ell$ and $p_\ell$ of a harmonic mode oscillator of frequency $\omega_\ell$.
Here we have also included the mode expansions in terms of the eigenmodes of the individual fields defined by the solution of the Helmholtz equation for each field. These are the eigenmodes $\Vect{u}_\ell$ of the vector potential $\Vect{A}$, the eigenmodes $\Vect{v}_\ell$ of the electric field  and the eigenmodes $\Vect{w}_\ell$ of the magnetic induction.
}
\label{tab:table1}
\begin{ruledtabular}
\begin{tabular}{llllll}
    \textbf{Field}  $\Vect{F}$ & \textbf{Eigenmodes} $\qty{\Vect{f}_\ell}$ & \textbf{Eigenmode expansion} & $\qty{\Vect{u}_\ell}$-\textbf{Mode expansion} & \textbf{Mode field strength} & \textbf{Vacuum field} \\
    \midrule
     $\Vect{A}$ & 
     $\qty{\Vect{u}_\ell}$ &
     $\displaystyle{\sum_\ell A_\ell \Vect{u}_\ell}$ & 
     $\displaystyle{\sum_\ell A_\ell \Vect{u}_\ell}$ & 
     $A_\ell = \mathcal{A}_\ell q_\ell$ & 
     $\mathcal{A}_\ell \equiv \qty(\dfrac{\hbar}{\varepsilon_0 \omega_\ell \mathcal{V}_\ell})^{1/2}$ \\[3ex]
     $\Vect{E}$ & 
     $\qty{\Vect{v}_\ell}$ &
     $\displaystyle{\sum_\ell E_\ell \Vect{v}_\ell}$ & 
     $\displaystyle{\sum_\ell E_\ell \Vect{u}_\ell}$ & 
     $E_\ell = \mathcal{E}_\ell p_\ell$ & 
     $\mathcal{E}_\ell = \mathcal{A}_\ell \omega_\ell $ \\[3ex]
     $\Vect{B}$ & 
     $\qty{\Vect{w}_\ell}$ &
     $\displaystyle{\sum_\ell B_\ell \Vect{w}_\ell}$ &
     $\displaystyle{\sum_\ell B_\ell k_\ell^{-1} \big(\nabla \times \Vect{u}_\ell\big)}$ & 
     $B_\ell = \mathcal{B}_\ell q_\ell$ & $\mathcal{B}_\ell = \dfrac{1}{c} \mathcal{E}_\ell $
\end{tabular}
\end{ruledtabular}
\end{table*}
%%%%%%%%%%%%%%%%%%%%%%%%%%%
% TABLE 
%%%%%%%%%%%%%%%%%%%%%%%%%%%

Whereas the decomposition of $\Vect{A}$ and $\Vect{E}$ involve\OrangelineChange{s} the mode functions $\Vect{u}_\ell$, the one \OrangelineChange{of} $\Vect{B}$ brings in the curl of $\Vect{u}_\ell$ due to the fact that there are no magnetic monopoles. In order to make the curl of $\Vect{u}_\ell$ dimensionless, we have introduced the inverse of the wave number $k_\ell$.

The field strengths $A_\ell$,~$E_\ell$, and $B_\ell$ of $\Vect{A}$, $\Vect{E}$, and $\Vect{B}$ in the mode $\ell$ are determined by the products of the corresponding vacuum fields $\mathcal{A}_\ell$,$\mathcal{E}_\ell$ and $\mathcal{B}_\ell$, and a dimensionless amplitude. In the case of $A_\ell$ and $B_\ell$, this amplitude is given by $q_\ell$, whereas for $E_\ell$ it is $p_\ell$. They are \OrangelineChange{analogs} of the familiar coordinate and momentum variables of a harmonic oscillator. 

We recall from the Maxwell equations that in \OrangelineChange{the} Coulomb gauge\OrangelineChange{,} the electric field is determined by the time derivative of the vector potential. As a result, the vacuum electric field $\mathcal{E}_\ell$ differs from that  of the vector potential $\mathcal{A}_\ell$ by the frequency $\omega_\ell$ of the mode. 

In general, the ratio of the magnetic induction to the electric field is governed by the speed of light $c$. This property also holds true for the corresponding vacuum fields. Thus, the ratio between the magnetic induction and the vector potential is given by the wave number $k_\ell$ due to the dispersion relation $k_\ell\equiv \omega_\ell/c$ of light. 

This difference in the wave number dependence of the vacuum fields has important implications when we now make the transition to quantum mechanics and motivate the wave functional of the vacuum in a resonator. We summarize our path to this expression in \cref{tab:table2}.
%%%%%%%%%%%%%%%%%%%%%%%%%%%
% TABLE 
%%%%%%%%%%%%%%%%%%%%%%%%%%%
\begin{table*}
\caption{%
Building blocks of the wave functional $\Psi[\Vect{F}]\equiv \mathcal{N}^{(F)}\exp(-\frac{1}{2}\beta^{(F)} \mathcal{I}^{(F)}[\OrangelineChange{\Vect{G}}])$ of the vacuum in a resonator for a free field $\Vect{F}=\Vect{F}(t,\Vect{r})$, given either by the electric field $\Vect{E}$, the magnetic induction $\Vect{B}$ or the vector potential $\Vect{A}$ emerging from the infinite product of ground state wave functions $\psi_\ell(F_\ell)\equiv \OrangelineChange{\mathcal{N}_\ell^{(F)}} \exp[-(F_\ell/\mathcal{F}_\ell)^2/2]$ of the $\ell$-th mode. For the example of $\Vect{A}$ we obtain two different kernels and two different fields in the double integral. For the modes $\Vect{u}_\ell$, we find a kernel $\sim 1/r^4$ with $\Vect{A}$ in the integral whereas for $\Vect{w}_\ell$, we arrive at the same kernel as in $\Vect{E}$ and $\Vect{B}$, but now $\nabla \times \Vect{A}$ appears. Here $\mathcal{N}^{(F)}\equiv \prod_\ell \mathcal{N}_\ell^{(F)}$ denotes a normalization constant, and the bilinear form $\mathcal{I}^{(F)} \equiv \int \D^3 r \int \D^3 r^\prime \OrangelineChange{\Vect{G}^\dagger(\Vect{r})\SecondOrderTensor{\mathbfscr{K}}^{(F)}(\Vect{r},\Vect{r}^\prime)\Vect{G}(\Vect{r}^\prime)}$ associated with $\Vect{F}$ can be reduced to a scalar kernel \OrangelineChange{$\mathbfscr{K}^{(F)}\equiv\mathbfscr{K}^{(F)}(\Vect{r})$} in the mode basis $\qty{\Vect{f}_\ell}$ and is given by the Fourier integral
$\OrangelineChange{\mathbfscr{F}\qty{\FlabelNoIdx(k)}\equiv (2\pi)^{-3} \int \D^3 k~ \FlabelNoIdx(k)~\E^{\I\Vect{k}\Vect{r}}}$ extending over all space.}
% COMMENT: Find out how to do formulas in caption
% COMMENT: Fractions
\label{tab:table2}
\begin{ruledtabular}
%%begin novalidate
\OrangelineChange{%
\begin{tabular}{llllllcc}
        \textbf{Field} $\Vect{F}$ & 
        \textbf{Mode basis} $\qty{\Vect{f}_\ell}$ &
        $F_\ell/\mathcal{F}_\ell$ &
        $\qty(F_\ell/\mathcal{F}_\ell)^2$  &
        $\beta^{(F)}$ &
        $\FlabelNoIdx(k_\ell)$ & 
        $\Vect{G}$ &
        Scalar kernel~$\mathbfscr{K}^{(F)}$ \\
        \midrule
        %%%%%%%%%%%%%
        $\Vect{A}$ & 
        $\qty{\Vect{u}_\ell}$ &
        $\dfrac{A_\ell}{\mathcal{A}_\ell}$ & 
        $\beta^{(A)} A_\ell^2 k_\ell\mathcal{V}_\ell $ & 
        $\dfrac{\varepsilon_0 c}{\hbar}$ & 
        $k_\ell$ &
        $\Vect{A}$ &
        $\mathbfscr{F}\qty{k}\sim -1/r^4$ \\[2ex]
        \midrule
        %%%%%%%%%%%%%
        $\Vect{E}$ &
        $\qty{\Vect{v}_\ell}=\qty{\Vect{u}_\ell}$ &
        $\dfrac{E_\ell}{\mathcal{E}_\ell}=\dfrac{E_\ell}{\mathcal{A}_\ell \omega_\ell}$ & 
        $\beta^{(E)} E_\ell^2  k_\ell^{-1} \mathcal{V}_\ell$ &
        $\dfrac{\varepsilon_0}{\hbar c}$ &
        $ 1 /k_\ell $ &
        $\Vect{E}$ &
        $\mathbfscr{F}\qty{1/k} \sim 1/r^2 $  \\[3ex]
        %%%%%%%%%%%%%
        $\Vect{B}$ &
        $\qty{\Vect{w}_\ell}=\qty{k_\ell^{-1}\nabla \times\Vect{u}_\ell}$ &
        $\dfrac{B_\ell}{\mathcal{B}_\ell}=\dfrac{c B_\ell}{\mathcal{E}_\ell}$ & 
        $\beta^{(B)} B_\ell^2  k_\ell^{-1} \mathcal{V}_\ell$ & 
        $\dfrac{\varepsilon_0 c}{\hbar}$ &
        $1/k_\ell$ &
        $\Vect{B}$ &
        $\mathbfscr{F}\qty{1/k}\sim 1/r^2$ \\[3ex]
       $\Vect{A}$ & 
        $\qty{\Vect{w}_\ell}=\qty{k_\ell^{-1}\nabla \times\Vect{u}_\ell}$ &
        $\dfrac{A_\ell^{(w)}}{\mathcal{A}_\ell}$ & 
        $\beta^{(A)} {A_\ell^{(w)}}^2 k_\ell^{-1} \mathcal{V}_\ell $ &
        $\dfrac{\varepsilon_0 c}{\hbar}$ & 
        $1/k_\ell$ &
        $\nabla \times \Vect{A}$ &
        $\mathbfscr{F} \qty{1/k} \sim 1/r^2$ \\[3ex]
        %%%%%%%%%%%%%
\end{tabular}
}
%%end novalidate
\end{ruledtabular}
\end{table*}
%%%%%%%%%%%%%%%%%%%%%%%%%%%
% TABLE 
%%%%%%%%%%%%%%%%%%%%%%%%%%%

\OrangelineChange{We} start by recalling that the ground state wave function $\psi_\ell$ of a single mode is determined by a Gaussian. Since its argument $\mathscr{f}_\ell$ has to be dimensionless, it must involve the ratio of the field strength $F_\ell$ divided by the associated vacuum field $\mathcal{F}_\ell$. 

The wave function of the complete electromagnetic field describing a quantum state with every mode in the ground state is defined by the infinite product of the corresponding single mode wave functions. Due to the functional equation of the exponential function, this product of exponentials reduces to a single exponential whose argument is determined by the sum of the arguments of the individual exponentials. Hence, we arrive at a sum of the squares of the dimensionless variables $\mathscr{f}_\ell$ over all modes.

When we recall from ~\cref{tab:table1},~\OrangelineChange{the} definitions of these vacuum fields, we~\OrangelineChange{obtain} for $\mathscr{f}_\ell^2$ \OrangelineChange{the} the product of the parameter $\beta^{(F)}$, determined by fundamental constants such as the dielectric constant $\varepsilon_0$,~\OrangelineChange{reduced} Planck's constant $\hbar$, speed of light $c$, and resonator specific parameters such as the square of the field strengths $F_\ell^2$, the mode volume $\mathcal{V}_\ell$ and the wave number $k_\ell$, or its inverse. 

Since $F_\ell^2$ emerges in this sum, it is tempting to replace it \OrangelineChange{with} an integral of a bilinear form of the complete field. Indeed, this sum over modes is reminiscent of the energy of the electromagnetic field in a resonator. However, in contrast to the present discussion, where the sums \OrangelineChange{involve either} the mode wave number or its inverse, the expression for the energy contains the square of it.

It is at this point that the difference in the descriptions of the electromagnetic field in terms of a continuous or a discrete superposition of modes enters the stage. This subtle point originates from the definition of the frequency of the mode. 

Indeed, when we use a continuous superposition of plane waves, the wave number given by the absolute value of the wave vector is directly related to the integration variable representing the superposition. In contrast, for a discrete superposition of mode functions\OrangelineChange{,} the summation is over the mode indices defining the frequency\OrangelineChange{,} which is determined by the boundary conditions for the Helmholtz equation. 

It is this distinct feature \OrangelineChange{that} forces us to take advantage of the concept of a fractional root of the negative Laplacian. This tool allows us to represent the kernel as an operator acting on the completeness relation\OrangelineChange{,} which is ultimately a Dirac delta function.

Hence, the difference between the kernels of the vector potential and the electric \OrangelineChange{field or} the magnetic induction manifests itself in an additional factor to the Fourier representation of what would normally be the Dirac delta function by the same power of the wave number as in the mode sum. This feature stands out most clearly in~\cref{tab:table2}.

\subsection{Overview}
Our article is organized as follows: In~\cref{MAIN:sec:ground-state-wavefunctions}, we derive an expression for the wave function of the vacuum in a resonator in terms of a sum over modes. For this purpose, we start from the corresponding probability amplitudes of every mode being in the ground state for $\Vect{A}$,~$\Vect{E}$ and ~$\Vect{B}$. Since these expressions are identical in their form for the three fields of interest, we confine ourselves to a general field $\Vect{F}$.

The wave function of the complete field in the vacuum is then the infinite product of all Gaussian wave functions\OrangelineChange{,} which translates into an exponential whose argument is a sum of all field strengths weighted by a function \OrangelineChange{$\FlabelNoIdx(k_\ell)$} whose form depends on the field $\Vect{F}$ that we consider. 

We devote~\cref{MAIN:sec:bilinear-forms-and-kernels} to the elimination of the modes in the infinite product of the ground state wave functions by \OrangelineChange{expressing} the sum over modes \OrangelineChange{by} a double integral over space containing a bilinear form of the fields and a kernel. For this purpose\OrangelineChange{,} we replace the expansion coefficient $F_\ell$ \OrangelineChange{by} the integral over the product of the field and mode functions $\Vect{f}_\ell$ and arrive\OrangelineChange{,} due to the appearance of the square of $F_\ell$ in the mode sum, at a double integral of a bilinear form of $\Vect{F}$ and a kernel. The kernel is then determined by the function $F$ of the square root of the negative Laplacian acting on the completeness relation of the modes given by the transverse delta function. Since the field $\Vect{F}$ is already transverse, it \OrangelineChange{suffices} to work with the familiar delta function which allows us to derive an explicit expression for the kernel, and thus for the wave functional of the vacuum in a resonator.

This analysis demonstrates that the kernels of $\Vect{E}$ and $\Vect{B}$ are identical, but different from the one \OrangelineChange{of} $\Vect{A}$.In ~\cref{MAIN:sec:vector-potential-once-more} we show that when we use the eigenmodes of $\Vect{B}$ to expand $\Vect{A}$\OrangelineChange{,} we find the same kernel as for $\Vect{E}$ and $\Vect{B}$.

We dedicate~\cref{MAIN:sec:discussion-of-wave-functionals} to a comparison of the resulting expressions for \OrangelineChange{the} wave functional of the vacuum in the different representations. Moreover, we connect our results to the literature.

In~\cref{MAIN:sec:functional-stuff}\OrangelineChange{,} we calculate the Wightman tensor of the vacuum fields and show how it is related to the kernels of $\Vect{E}$, $\Vect{B}$ and $\Vect{A}$. Furthermore, we sketch how vacuum expectation values can be expressed in terms of the wave functional.

We conclude in~\cref{MAIN:sec:conclusion} by summarizing our results \OrangelineChange{and providing} an outlook.

In order to keep our article \OrangelineChange{self-contained}, we have included additional material that is helpful \OrangelineChange{in understanding the} main sections \OrangelineChange{and keeping} track of factors of 2. For example, in~\cref{APP:sec:mode-decomposition}, we summarize the essential building blocks of the free electromagnetic field. Here we concentrate on the expansions of $\Vect{A}$,~$\Vect{E}$ and $\Vect{B}$ into a complete set of discrete modes. Moreover, we define the corresponding vacuum electric fields by equating the energy in a single mode of a given frequency to that of a quantized harmonic oscillator of the same frequency. 

In~\cref{APP:sec:field-energy-em-field}, we re-derive the energy of an electromagnetic field in a resonator. This calculation also most \OrangelineChange{clearly brings} out the difference in the powers of the mode frequency in the energy and the infinite product of the ground state wave functions. Moreover, we verify that the mode volumes of the $\Vect{u}_\ell$-modes and the $\Vect{w}_\ell$-modes are identical.

We devote~\cref{APP:sec:wavefunctional-groundstate} to \OrangelineChange{the} derivation of the ground state wave function in three different representations, that is, in the variables of $\Vect{E}$,~$\Vect{B}$ and $\Vect{A}$. In each case, we find a Gaussian whose dimensionless argument is determined by the ratio of the variable and the vacuum field strength.

In~\cref{APP:sec:reductionscheme} we present an alternative derivation of the double integral containing the bilinear form of the field and the kernel by \emph{reverse engineering}. In contrast to the derivation of ~\cref{MAIN:sec:bilinear-forms-and-kernels}, we start by already assuming that the kernel is a scalar function and given by \OrangelineChange{the} Fourier integral of the function $F$.
We then reduce this double integral to a single one of the square of the \OrangelineChange{fourth} root of the negative Laplacian acting on $\Vect{F}$. The mode expansion of $\Vect{F}$ then leads us straight to the mode sum of the wave function of the vacuum. We also point out a curious analogy to the $P$- and $R$-distributions~\cite{Glauber1963} of quantum optics.

\OrangelineChange{In ~\cref{APP:sec:explicit-j-kernels} we provide an explicit expression for the kernel by performing the relevant integrations with the help of a convergence factor.}

\OrangelineChange{Finally, in ~\cref{APP:sec:mathematical-complement} we derive an identity for the scalar product of two curls, evaluated at different positional arguments, needed in the evaluation of the Hamiltonian density of the electromagnetic field in~\cref{APP:sec:field-energy-em-field}.}

%%%%%%%%%%%%%%%%%%%%%%%%%%%%%%%%%%%%%%%%%%%%%%%%%%%%%%%%%%%%%%%%%%%%%%%%%%%%%
%MAIN PAPER STARTS HERE
%%%%%%%%%%%%%%%%%%%%%%%%%%%%%%%%%%%%%%%%%%%%%%%%%%%%%%%%%%%%%%%%%%%%%%%%%%%%%

\section{Infinite product of ground state wave functions}\label{MAIN:sec:ground-state-wavefunctions}
In this section, we derive the wave function of the electromagnetic vacuum in a resonator in terms of an infinite product of the ground state wave functions. Throughout the section\OrangelineChange{,} we use the field $\Vect{F}$\OrangelineChange{,} which represents either the electric field $\Vect{E}$, the magnetic induction $\Vect{B}$, or the vector potential $\Vect{A}$, and rely on the expansion 
\begin{align}\label{MAIN:eq:f-mode-decomposition}
    \Vect{F}(t,\Vect{r}) \equiv \sum_\ell F_\ell(t) \Vect{f}_\ell(\Vect{r})
\end{align}
of these fields into their natural modes $\Vect{f}_\ell$ determined by the Helmholtz equation subjected to the boundary conditions of the resonator as outlined in~\cref{APP:sec:mode-decomposition}. For the sake of simplicity in notation\OrangelineChange{,} we have not attached a superscript $F$ on the modes $\Vect{f}_\ell$ but emphasize that they depend on the choice of $\Vect{F}$.

The expansion coefficient $F_\ell$ denotes the field strength in the mode $\Vect{f}_\ell$. Hence, $F_\ell$ depends on the choice of the modes. Obviously, in a different mode expansion\OrangelineChange{,} the field strength would be different. Again\OrangelineChange{,} for the sake of simplicity in notation\OrangelineChange{,} we suppress this dependence in $F_\ell$ but keep it in mind.

\subsection{Wave function of the ground state}
In ~\cref{APP:sec:wavefunctional-groundstate} we have recalled the expressions for the ground state wave functions $\psi_\ell$ in the representations of the electric field $E_\ell$, the magnetic induction $B_\ell$, or the vector potential $A_\ell$ in the $\ell$-th natural mode given by $\Vect{f}_\ell=\Vect{f}_\ell(\Vect{r})$. Since the~\OrangelineChange{not yet normalized} ground state is completely symmetric in phase space\OrangelineChange{,} it takes the same form in each of these representations and reads
\begin{align}\label{MAIN:eq:dimensionless-wave-function}
    \psi_\ell(\mathscr{f}_\ell) \equiv \OrangelineChange{\frac{1}{\sqrt[4]{\pi}}} \exp(-\frac{1}{2}\mathscr{f}_\ell^2)
\end{align}
where the dimensionless variable
\begin{align}\label{MAIN:eq:dimensionless_variable_fl}
    \mathscr{f}_\ell \equiv \frac{F_\ell}{\mathcal{F}_\ell}
\end{align}
involves the field $F_\ell$ in the $\ell$-th mode $\Vect{f}_\ell$, and $\mathcal{F}_\ell$ is the corresponding field strength of the vacuum. Here $F_\ell$ is either $E_\ell$, $B_\ell$ or $A_\ell$.

The quantities $\mathcal{F}_\ell$ are different for the three fields. Indeed, the strength
\begin{align}\label{MAIN:eq:vacuum-amplitude-vector-potential}
    \mathcal{A}_\ell \equiv \sqrt{\frac{\hbar}{\epsilon_0\omega_\ell \mathcal{V}_\ell}}
\end{align}
of the vector potential\OrangelineChange{,} which involves the mode volume $\mathcal{V}_\ell$, is defined by postulating the electromagnetic energy of the ground state of the mode to be identical to $\hbar \omega_\ell/2$ where $\omega_\ell$\OrangelineChange{,} denotes the frequency of the $\ell$-th mode $\Vect{f}_\ell$. 

We emphasize that also the mode volume $\mathcal{V}_\ell$ depends on the choice of modes. For this reason\OrangelineChange{,} it should also carry a superscript indicating the type of eigenmodes used, such as $\Vect{u}_\ell$ for the eigenmodes of $\Vect{A}$, $\Vect{v}_\ell$ for the eigenmodes of $\Vect{E}$ or $\Vect{w}_\ell$ for the eigenmodes of $\Vect{B}$. However, for the sake of simplicity in notation\OrangelineChange{,} we suppress it.

The strength 
\begin{align}\label{MAIN:eq:scriptE-link-scriptA}
    \mathcal{E}_\ell
    \equiv
    \mathcal{A}_\ell \omega_\ell
    =
    \sqrt{\frac{\hbar \omega_\ell}{\varepsilon_0 \mathcal{V}_\ell}}
\end{align}
follows from the Maxwell equations, that is from the fact that in \OrangelineChange{the} Coulomb gauge without currents and charges, $\Vect{E}$ is the time derivative of $\Vect{A}$.

Moreover, for $\Vect{B}$ we obtain in ~\cref{APP:sec:mode-decomposition} the expression
\begin{align}\label{MAIN:eq:link-between-script-fields}
    \mathcal{B}_\ell
    \equiv 
    \mathcal{A}_\ell \frac{\omega_\ell}{c}
    =
    \frac{\mathcal{E}_\ell}{c}
\end{align}
for the field strength $B_\ell$ of $\Vect{B}$. Hence, apart from  a factor of $c$\OrangelineChange{,} the field strengths $\mathcal{B}_\ell$ and $\mathcal{E}_\ell$ are identical.

When we substitute the dimensionless variable $\mathscr{f}_\ell$ given by~\cref{MAIN:eq:dimensionless_variable_fl} into~\cref{MAIN:eq:dimensionless-wave-function}\OrangelineChange{,} the probability amplitude $\psi_\ell=\psi_\ell(F_\ell)$ of finding the field $F_\ell$ of the mode $\ell$ in the ground state of this mode reads
\begin{align}\label{MAIN:eq:abstract-wave-functional-ground-state-single-amplitude}
    \psi_\ell(F_\ell)
    \equiv
    \mathcal{N}_\ell^{(F)}
    \exp\qty[-\frac{1}{2}\qty(\frac{F_\ell}{\mathcal{F}_\ell})^{2}],
\end{align}
where the normalization constant $\mathcal{N}_\ell^{(F)}$ takes the form
\begin{align}\label{MAIN:eq:abstract-f-wavefunction-normalization}
    \OrangelineChange{
    \mathcal{N}_\ell^{(F)}
    \equiv
    \frac{1}{\sqrt[4]{\pi}\sqrt{\mathcal{F}_\ell}}.
    }
\end{align}
Due to the presence of $\mathcal{F}_\ell$\OrangelineChange{,} the normalization constant $\pi^{-1/4}$ \OrangelineChange{of} the Gaussian in~\cref{MAIN:eq:dimensionless-wave-function} is \OrangelineChange{modified to} achieve the condition
\begin{align}
    \OrangelineChange{%
    \int_{\mathrlap{-\infty}}^{\mathrlap{\infty}} 
    }
    \D F_\ell \Abs{\psi_\ell(F_\ell)}^2 = 1,
\end{align}
dictated by the Born interpretation.

\subsection{Sum over modes}
Hence, the corresponding probability amplitude $\Psi(\qty{F_\ell})$ for finding the field $F_{\ell_1}$ in the mode $\ell_1$, $F_{\ell_2}$ in the mode $\ell_2$, \OrangelineChange{etc.}, in the ground state is the infinite product
\begin{align}\label{MAIN:eq:abstract-wave-functional-product}
    \Psi(\qty{F_\ell}) \equiv \prod_\ell \psi_\ell (F_\ell),
\end{align}
of the ground state wave functions $\psi_\ell$ of all modes.

With the expression~\cref{MAIN:eq:abstract-wave-functional-ground-state-single-amplitude} for $\psi_\ell$ and the elementary property $\E^a \E^b=\E^{a+b}$ of the exponential function\OrangelineChange{,} we arrive at the formula
\begin{align}\label{MAIN:eq:wavefunction-frepresentation-ground-state}
    \Psi(\qty{F_\ell})
    =
    \mathcal{N}^{(F)}
    \exp\qty[
        -\frac{1}{2}\beta^{(F)} \Sigma^{(F)}
    ]
\end{align}
where we have defined
\begin{align}
    \mathcal{N}^{(F)}
    \equiv
    \prod_\ell
    \mathcal{N}_\ell^{(F)}
\end{align}
representing the infinite product of all normalization factors $\mathcal{N}^{(F)}_\ell$, and introduced the abbreviation
\begin{align}\label{MAIN:eq:sigmaF-exponential-sum}
    \Sigma^{(F)}
    \equiv
    \sum_\ell
    F_\ell^2 \FlabelNoIdx(k_\ell) \mathcal{V}_\ell
\end{align}
for the sum over all modes. Here we have taken into account that $\mathcal{F}_\ell$ is slightly different for the three fields. For this reason\OrangelineChange{,} the factor $\beta^{(F)}$ containing constants of nature such as $\hbar$,~$\varepsilon_0$ and $c$\OrangelineChange{,} and listed in \cref{tab:table2}, depends on the choice of \OrangelineChange{$\FlabelNoIdx$}.

Moreover, since $\mathcal{A}_\ell$ and $\mathcal{E}_\ell$ depend differently on $\omega_\ell$,\OrangelineChange{,} as shown by~\cref{MAIN:eq:vacuum-amplitude-vector-potential} and ~\cref{MAIN:eq:scriptE-link-scriptA}, we have a different dependence of $\Sigma^{(F)}$ on the wave number $k_\ell\equiv \Abs{\Vect{k}_\ell} \equiv \omega_\ell/c$ of the mode indicated in~\cref{MAIN:eq:sigmaF-exponential-sum} by the contribution $\FlabelNoIdx(k_\ell)$. Indeed, for $\Vect{E}$ and $\Vect{B}$ we find
\begin{align}\label{MAIN:eq:FEB-definition}
    \FlabelNoIdx^{(E/B)}(k_\ell) = k_\ell^{-1},
\end{align}
while for $\Vect{A}$ we obtain
\begin{align}\label{MAIN:eq:FA-definition}
    \FlabelNoIdx^{(A)}(k_\ell) =k_\ell.
\end{align}
It is this difference in $F$ that leads to different expressions for the wave functional of the vacuum in a resonator, as we shall show in the next section.

\subsection{Connection to free space}
We conclude this analysis of the product of all ground state wave functions \OrangelineChange{with} a side, but not snide\OrangelineChange{,} remark about the corresponding calculation in free space. Since in this case we have a \emph{continuous} superposition of modes\OrangelineChange{,} we have to deal with a \emph{continuous product} of ground state wave functions. One possibility to describe this unusual quantity which is fundamentally different from the discrete product arising in the case of a resonator, is to employ the Volterra-Schlesinger product integral ~\cite{Rasch1934} used to define in QED the quantum state after a time-dependent interaction~\cite{Salecker1950}.

However, a much more elementary approach to overcome this complication of a continuous product is to first discretize the modes, perform the discrete product and then replace the sum over modes again \OrangelineChange{with} the appropriate integral. Hence, in free space, we retreat from the continuous superposition of modes to a discrete set and then return again to the continuous one.

In contrast, \OrangelineChange{in the case of} a resonator\OrangelineChange{,} we always deal with a discrete set, and the complication of the infinite product never occurs. We note that it would be interesting to perform the calculation in free space evaluating the continuous product, for example, with the help of the Volterra-Schlesinger product integral.

\section{Bilinear forms and kernels}\label{MAIN:sec:bilinear-forms-and-kernels}
The goal of the present section is to \emph{construct} from the mode expansion and the mode sum $\Sigma^{(F)}$, given by ~\cref{MAIN:eq:sigmaF-exponential-sum} an equivalent expression in terms of the complete field $\Vect{F}$ rather than the field amplitudes $F_\ell$. For this purpose\OrangelineChange{,} we note that the terms in $\Sigma^{(F)}$, are quadratic in the fields $F_\ell$. Therefore, $\Sigma^{(F)}$ might be represented by a quadratic form of the total field $\Vect{F}$. Since $\Sigma^{(F)}$ is independent of the coordinate, there must be an integration over space involved. 

However, this integral cannot just contain $\Vect{F}^2\equiv\Vect{F}^\dagger \Vect{F}= \Vect{F}\cdot \Vect{F}$, since that would lead to a quantity proportional to the energy in the resonator. Indeed, as shown in ~\cref{APP:sec:field-energy-em-field}, the contribution of the electric field or the magnetic induction to the energy scale \OrangelineChange{with} $\omega_\ell^2$ in the field oscillator frequency. Hence, a bilinear form of $\Vect{F}$ and a position-dependent kernel are necessary to obtain the scaling in $k_\ell$ required by the function $\FlabelNoIdx(k_\ell)$ given by~\cref{MAIN:eq:FEB-definition} and~\cref{MAIN:eq:FA-definition}.

In the present section we pursue this approach in four steps: (\textit{i}) We first obtain an explicit expression for the expansion coefficients $F_\ell$ of $\Vect{F}$ into the natural modes $\Vect{f}_\ell$ and establish the completeness relation of $\Vect{f}_\ell$\OrangelineChange{.} (\textit{ii}) Then we cast the mode sum $\Sigma^{(F)}$ into a double integral of the fields $\Vect{F}$ and $\Vect{F}^\prime$ together with a matrix kernel. (\textit{iii}) Since $\Vect{F}$ is transverse\OrangelineChange{,} this kernel reduces to a scalar, and (\textit{iv}) we finally evaluate this kernel.

\subsection{Completeness relation of transverse modes}
Central to the representation of $\Sigma^{(F)}$ by a double integral of a bilinear form of $\Vect{F}$ and a kernel\OrangelineChange{, is the expansion}
\begin{align}\label{MAIN:eq:mode-expansion-fmodes}
    \Vect{F} 
    =
    \sum_\ell F_\ell \Vect{f}_\ell
\end{align}
of the free field $\Vect{F}$ into the modes $\Vect{f}_\ell$ discussed in ~\cref{APP:sec:mode-decomposition}.

Indeed, the strength $F_\ell$ of $\Vect{F}$ in the mode $\Vect{f}_\ell$, which appears quadratically in $\Sigma^{(F)}$, follows from~\cref{MAIN:eq:mode-expansion-fmodes} by multiplication of $\Vect{f}_m$, integration over space\OrangelineChange{,} and using the orthonormality relation
\begin{align}\label{MAIN:eq:mode-expansion-fmodes-orthonormality}
     \frac{1}{\mathcal{V}_{\ell}}
     \int \D^3 r 
     ~\Vect{f}_{\ell}^{\dagger} (\Vect{r})
      \Vect{f}_{m} (\Vect{r}) 
     = 
     \delta_{\ell m} 
\end{align}
of the modes. Moreover, the integration extends over the resonator volume, unless specified otherwise. 

Indeed,~\OrangelineChange{the modes $\Vect{f}_\ell$} form a complete and orthonormal basis of \emph{transverse mode-space} since they are eigenfunctions of the self-adjoint Helmholtz operator applied to the field $\Vect{F}$ as discussed in \cref{APP:sec:mode-decomposition}.

\OrangelineChange{We} arrive at the explicit form
\begin{align}
    F_\ell
    =
    \frac{1}{\mathcal{V}_\ell} 
    \int \D^3 r^\prime 
    \Vect{f}_\ell^\dagger (\Vect{r}^\prime) 
    \Vect{F}(t,\Vect{r}^\prime) 
\end{align}
or
\begin{align}\label{MAIN:eq:mode-coefficient-Fell-only}
    F_\ell = 
    \frac{1}{\mathcal{V}_\ell} 
    \int \D^3 r^\prime
    {\Vect{f}^\prime_\ell}^\dagger \Vect{F}^\prime.
\end{align}
Here we have attached a prime on $\Vect{F}$ and $\Vect{f}_\ell$ \OrangelineChange{to} emphasize the fact that both depend on the integration variable $\Vect{r}^\prime$ rather than $\Vect{r}$.

Since the field $\Vect{F}$ and the modes $\Vect{f}_\ell$ are \OrangelineChange{hermitian} fields, we have the identity $F_\ell^\dagger=F_\ell$\OrangelineChange{,} and thus
\begin{align}\label{MAIN:eq:mode-coefficient-Fell}
    \frac{1}{\mathcal{V}_\ell} 
    \int \D^3 r^\prime
    {\Vect{F}^\prime}^\dagger{\Vect{f}^\prime_\ell}
    =
    \frac{1}{\mathcal{V}_\ell} 
    \int \D^3 r
    {\Vect{f}_\ell}^\dagger \Vect{F}
\end{align}

When we substitute,~\cref{MAIN:eq:mode-coefficient-Fell-only}, into the expansion~\cref{MAIN:eq:mode-expansion-fmodes}\OrangelineChange{,} we find
\begin{align}\label{MAIN:eq:mode-coefficients-Fell-abstract}
    \Vect{F} 
    =
    \sum_{\ell} 
    \frac{1}{\mathcal{V}_\ell} 
    \int \D^3 r^\prime
    ~\qty({\Vect{f}_\ell^\prime}^\dagger \Vect{F}^\prime ) \Vect{f}_\ell 
\end{align}
which when we interchange the sum and the integral reduces to
\begin{align}\label{MAIN:eq:orthonormality-condition-is-delta-kernel}
    \Vect{F} 
    &=
    \int \D^3 r^\prime \qty[ \sum_\ell \frac{1}{\mathcal{V}_\ell} \Vect{f}(\Vect{r})\Vect{f}^\dagger(\Vect{r}^\prime)] \Vect{F}^\prime
\end{align}
or
\begin{align}\label{MAIN:eq:kernel-idenity-equation}
    \Vect{F}&\equiv 
    \int \D^3 r^\prime 
    \SecondOrderTensor{\mathbfscr{D}}(\Vect{r},\Vect{r}^\prime)\Vect{F}^\prime,
\end{align}
where we have introduced the term
\begin{align}\label{MAIN:eq:kernel-of-projector-in-modes}
    \SecondOrderTensor{\mathbfscr{D}}(\Vect{r},\Vect{r}^\prime)
    \equiv
    \sum_\ell 
    \frac{1}{\mathcal{V}_\ell} 
    \Vect{f}_\ell(\Vect{r})
    \Vect{f}_\ell^\dagger (\Vect{r}^\prime).
\end{align}

In order to maintain the identity $\Vect{F}=\Vect{F}$ in~\cref{MAIN:eq:kernel-idenity-equation} the kernel $\SecondOrderTensor{\mathbfscr{D}}$ has to act as a delta-function-like object with respect to the spatial coordinates. However, since our modes are in Coulomb gauge and are thus transverse, $\SecondOrderTensor{\mathbfscr{D}}$ cannot be an ordinary delta function, but must be a transverse delta function $\DeltaTransversal$. Thus~\cref{MAIN:eq:kernel-of-projector-in-modes} takes the form
\begin{align}\label{MAIN:eq:orthogonality-kernel-is-transverse-delta}
    \SecondOrderTensor{\mathbfscr{D}}(\Vect{r},\Vect{r}^\prime) 
    \equiv
    \DeltaTransversal(\Vect{r}-\Vect{r}^\prime)
\end{align}
with the expansion~\cref{MAIN:eq:kernel-of-projector-in-modes} in terms of the modes $\Vect{f}_\ell$. 

Hence, the matrix $\SecondOrderTensor{\mathbfscr{D}}$ defines a completeness relation and represents the kernel of a projection operator $\SecondOrderTensor{\mathcal{P}}^\perp$ onto the (function) space spanned by the transverse (generalized Fourier) modes, \OrangelineChange{i.e.,} 
\begin{align}\label{MAIN:eq:completeness-relation-kernel-delta-relation}
    \SecondOrderTensor{\mathcal{P}}^\perp(\bullet)
    = \int \D^3 r^\prime\SecondOrderTensor{\mathbfscr{D}}(\Vect{r},\Vect{r}^\prime)~\bullet
\end{align}
where $(\bullet)$ \OrangelineChange{acts} as a placeholder for an arbitrary vector field to be projected onto that space.

\subsection{Mode sum as double integral}

We are now in the position to cast the sum $\Sigma^{(F)}$ over modes defined by~\cref{MAIN:eq:sigmaF-exponential-sum} into a double integral containing a bilinear form of $\Vect{F}$ and a kernel $\SecondOrderTensor{\mathbfscr{K}}$. In particular, we can obtain an exact expression for $\SecondOrderTensor{\mathbfscr{K}}$. 

For this purpose\OrangelineChange{,} we substitute the expression~\cref{MAIN:eq:mode-coefficient-Fell-only} for $F_\ell$ combined with the symmetry relation~\cref{MAIN:eq:mode-coefficient-Fell} of $F_\ell$ into $\Sigma^{(F)}$ and find the identity
\begin{align}\label{MAIN:eq:double-integral-FKF}
    \Sigma^{(F)}
    \equiv
    \Sigma^{(F)}[\Vect{F}]
    =
    \int \D^3 r
    \int \D^3 r^\prime~
    \Vect{F}^\dagger
    \SecondOrderTensor{\mathbfscr{K}}(\Vect{r},\Vect{r}^\prime)
    \Vect{F}^\prime
\end{align}
with the kernel
\begin{align}\label{MAIN:eq:defining-eq-kernel}
    \SecondOrderTensor{\mathbfscr{K}}(\Vect{r},\Vect{r}^\prime)
    \equiv
    \sum_\ell
    \frac{1}{\mathcal{V}_\ell} 
    \FlabelNoIdx(k_\ell)
    \Vect{f}_\ell(\Vect{r})
    \Vect{f}_\ell^\dagger(\Vect{r}^\prime).
\end{align}

When we compare $\SecondOrderTensor{\mathbfscr{K}}$ to the completeness relation~\cref{MAIN:eq:kernel-of-projector-in-modes}\OrangelineChange{,} we find that, apart from the appearance of $\FlabelNoIdx(k_\ell)$ from~\cref{tab:table2}, which is due to the different powers of $k_\ell$ in the vacuum field strength $\mathcal{F}_\ell$, they are identical. Therefore, we want to eliminate $\FlabelNoIdx(k_\ell)$ from the sum over modes in~\cref{MAIN:eq:defining-eq-kernel} by recalling the Helmholtz equation in the form
\begin{align}\label{MAIN:eq:Laplacian-on-mode-function}
    (-\VLaplace)\Vect{f}_\ell = k_\ell^2 \Vect{f}_\ell,
\end{align}
which shows that $\Vect{f}_\ell$ is the eigenvector of the negative Laplacian associated with the eigenvalue $k_\ell^2$.

As a result, we find the identity
\begin{align}\label{APP:eq:scalarkernel-defining-equation-laplacian}
    \FlabelNoIdx(k_\ell) \Vect{f}_\ell = \FlabelNoIdx \qty(\OrangelineChange{\sqrt{-\VLaplace}}) \Vect{f}_\ell,
\end{align}
and the kernel $\SecondOrderTensor{\mathbfscr{K}}$ given by ~\cref{MAIN:eq:defining-eq-kernel} reduces to
\begin{align}\label{MAIN:eq:kernel-as-deltatransverse-derivative}
    \SecondOrderTensor{\mathbfscr{K}}(\Vect{r},\Vect{r}^\prime)
    =
     \FlabelNoIdx\qty(\OrangelineChange{\sqrt{-\VLaplace}})
     ~\DeltaTransversal(\Vect{r}-\Vect{r}^\prime)\OrangelineChange{,}
\end{align}
or equivalently
\begin{align}\label{MAIN:eq:kernel-as-deltatransverse-derivative-components}
    \SecondOrderTensor{\mathbfscr{K}}_{mn}(\Vect{r},\Vect{r}^\prime)
    =
    \FlabelNoIdx\qty(\OrangelineChange{\sqrt{-\VLaplace}})
    ~\DeltaTransversal_{\!\!mn} (\Vect{r}-\Vect{r}^\prime)
\end{align}
in component notation. Here we have recalled ~\cref{MAIN:eq:orthogonality-kernel-is-transverse-delta}.

We emphasize that in~\cref{MAIN:eq:kernel-as-deltatransverse-derivative,MAIN:eq:kernel-as-deltatransverse-derivative-components} the differentiation in the Laplacian could be with respect to $\Vect{r}$ or $\Vect{r}^\prime$. This fact follows directly from the definition,~\cref{MAIN:eq:defining-eq-kernel}, of the kernel or from the argument of the transverse delta function. For this reason\OrangelineChange{,} we have not attached a subscript $\Vect{r}$ to the Laplacian.

\subsection{Simplification of the kernel}
Next\OrangelineChange{,} we recall that the tensorial version $\delta(\Vect{r}) \SecondOrderTensor{\mathbbm{1}}_3$ of the familiar Dirac delta function $\delta(\Vect{r})$ contains not only the transverse part $\DeltaTransversal(\Vect{r})$, but also the longitudinal part $\DeltaLongitudinal(\Vect{r})$,
and reads in components
\begin{align}
    \delta(\Vect{r})\delta_{mn} 
    = 
    {\DeltaTransversal}_{\!\!mn} (\Vect{r}) +
    {\DeltaLongitudinal}_{\!\!mn} (\Vect{r}),
\end{align}
or
\begin{align}
    {\DeltaTransversal}_{\!\!mn} (\Vect{r})
    = 
    \delta(\Vect{r})\delta_{mn} -
    {\DeltaLongitudinal}_{\!\!mn} (\Vect{r}).
\end{align}

The operator $\FlabelNoIdx(\OrangelineChange{\sqrt{-\VLaplace}})$ acting on $\DeltaLongitudinal$ does not change the directionality of the longitudinal part. This property stands out most clearly in its Fourier representation
\begin{align}
    {\DeltaLongitudinal}_{\!\!mn}(\Vect{r})
    \equiv
    \frac{1}{(2\pi)^3}
    \int \D^3 k~
    \E^{\I\Vect{k}\Vect{r}}
    \frac{k_m k_n}{k^2}.
\end{align}
Indeed, we find
\begin{align}\label{MAIN:eq:FLaplace-applied-to-delta-parallel}
    \FlabelNoIdx\qty(\OrangelineChange{\sqrt{-\VLaplace}})~{\DeltaLongitudinal}_{\!\!mn}
    =
    \frac{1}{(2\pi)^3}
    \int \D^3 k~
    \E^{\I\Vect{k}\Vect{r}}
    \FlabelNoIdx(k)
    \frac{k_m k_n}{k^2},
\end{align}
where we have used the fact that a plane wave is also an eigenfunction of the negative Laplacian in free space corresponding to the eigenvalue $k^2$, \OrangelineChange{i.e.,}
\begin{align}\label{MAIN:eq:vlaplace-of-plane-wave}
    (-\VLaplace)\E^{\I\Vect{k}\Vect{r}}=k^2 \E^{\I\Vect{k}\Vect{r}}.
\end{align}

As a result, the kernel $\SecondOrderTensor{\mathbfscr{K}}_{mn}$ given by ~\cref{MAIN:eq:kernel-as-deltatransverse-derivative-components} reads
\begin{align}
    \SecondOrderTensor{\mathbfscr{K}}_{mn} 
    = 
    \delta_{mn} \mathbfscr{K}^{(F)} +
    \SecondOrderTensor{\mathbfscr{K}}^{\parallel}_{mn}
\end{align}
where the part
\begin{align}\label{MAIN:eq:scalar-kernel-as-deltatransverse-derivative}
    \mathbfscr{K}^{(F)}(\Vect{r}-\Vect{r}^\prime)
    \equiv
    \FlabelNoIdx\qty(\OrangelineChange{\sqrt{-\VLaplace}})~\delta(\Vect{r}-\Vect{r}^\prime)
\end{align}
of the kernel, which is diagonal, arises from the operator $\FlabelNoIdx(\OrangelineChange{\sqrt{-\VLaplace}})$ acting on the familiar Dirac delta function. 

On the other hand, according to ~\cref{MAIN:eq:FLaplace-applied-to-delta-parallel}, the expression
\begin{align}
    \mathbfscr{K}^{\parallel}_{mn}
    =
    \FlabelNoIdx\qty(\OrangelineChange{\sqrt{-\VLaplace}})~\DeltaLongitudinal_{mn}(\Vect{r}-\Vect{r}^\prime)
\end{align}
is still longitudinal.

Next\OrangelineChange{,} we recall that in the double integral,~\cref{MAIN:eq:double-integral-FKF}, the fields $\Vect{F}$ and $\Vect{F}^\prime$\OrangelineChange{,} on which the kernel acts, are already transverse, since they are expanded into the transverse modes $\Vect{f}_\ell$. Hence $\SecondOrderTensor{\mathbfscr{K}}^\parallel$ does not contribute, and we arrive at the expression
\begin{align}\label{MAIN:eq:mode-sum-kernelF-link}
    \Sigma^{(F)}
    =
    \int \D^3{r}
    \int \D^3{r}^\prime
    \mathbfscr{K}^{(F)} \Vect{F}^\dagger \Vect{F}^\prime, 
\end{align}
where we have made use of the fact that $\mathbfscr{K}^{(F)}$ is a scalar\OrangelineChange{,} which can now be moved out of the matrix products.

\subsection{Evaluation of the kernel}
Finally, we evaluate the scalar kernel $\mathbfscr{K}^{(F)}$ given by ~\cref{MAIN:eq:scalar-kernel-as-deltatransverse-derivative}. Here\OrangelineChange{,} two possibilities offer themselves:
(i) We recall the Green's function relation
\begin{align}
    (-\VLaplace)\frac{1}{r} = 4\pi \delta(r)
\end{align}
which leads us to the expression
\begin{align}
    \mathbfscr{K}^{(F)}(\Vect{r})
    =
    \frac{1}{4\pi}
    \FlabelNoIdx\qty(\OrangelineChange{\sqrt{-\VLaplace}}) (-\VLaplace) \frac{1}{\Abs{\Vect{r}}}
\end{align}
for $\mathbfscr{K}^{(F)}$, or (ii) we employ the Fourier representation
\begin{align}\label{APP:eq:dirac-delta-definition}
    \delta(\Vect{r})
    \equiv
    \frac{1}{(2\pi)^3}
    \int \D^3 k ~
    \E^{\I \Vect{k}\Vect{r}}
\end{align}
of the Dirac delta function to evaluate $\mathbfscr{K}^{(F)}$. 

In this article\OrangelineChange{,} we pursue the second approach since it is straightforward. Indeed, from ~\cref{MAIN:eq:scalar-kernel-as-deltatransverse-derivative} we immediately find with ~\cref{MAIN:eq:vlaplace-of-plane-wave} the representation
\begin{align}
    \mathbfscr{K}^{(F)}(\Vect{r}) = \frac{1}{(2\pi)^3} \int \D^3 k \FlabelNoIdx(k) \E^{\I \Vect{k}\Vect{r}}.
\end{align}

In ~\cref{APP:sec:explicit-j-kernels} we evaluate this integral for the two cases $\FlabelNoIdx(k)=k^{-1}$ or $\FlabelNoIdx(k)=k$ corresponding to the fields $\Vect{E}$ and $\Vect{B}$ or $\Vect{A}$, and \OrangelineChange{we} find
\begin{align}\label{MAIN:eq:kernels-efield-bfield}
    \mathbfscr{K}^{(E)}(\Vect{r}) =  \mathbfscr{K}^{(B)}(\Vect{r}) = \frac{1}{2\pi^2}\frac{1}{\Abs{\Vect{r}}^2}
\end{align}
or
\begin{align}\label{MAIN:eq:kernels-afield}
    \mathbfscr{K}^{(A)}(\Vect{r}) = -\frac{1}{\pi^2}\frac{1}{\Abs{\Vect{r}}^4}.
\end{align}

We note that apart from slightly different prefactors\OrangelineChange{,} the power laws of the two kernels in~\cref{MAIN:eq:kernels-efield-bfield} and~\cref{MAIN:eq:kernels-afield} are different. \OrangelineChange{While} $\mathbfscr{K}^{(E)}=\mathbfscr{K}^{(B)}$ decays as $\mathbfscr{K}^{(E/B)}\sim 1/r^2$, the one for $\Vect{A}$,~\OrangelineChange{i.e.,} $\mathbfscr{K}^{(A)}$\OrangelineChange{,} decays as $\mathbfscr{K}^{(A)} \sim 1/r^4$. Moreover, they also differ in sign. While $\mathbfscr{K}^{(E/B)}$ is positive, $\mathbfscr{K}^{(A)}$ is negative.

\OrangelineChange{At} first sight this sign change might cause a problem in the exponential. However, when we recall that the double integral with the bilinear form of $\Vect{A}$ and $\mathbfscr{K}^{(A)}$ is identical to the mode sum $\Sigma^{(A)}$ in which each term is positive, we recognize that here is really no problem \OrangelineChange{here}.

\subsection{Wave functional}
We conclude by combining our results to obtain the wave functional $\Psi[\Vect{F}]$ of the vacuum in a resonator expressed by the field $\Vect{F}$. Indeed, when we use the connection
~\cref{MAIN:eq:mode-sum-kernelF-link} between the mode sum $\Sigma^{(F)}$ and the double integral, we find the expression
\begin{align}\label{MAIN:eq:wave-functional-F-representation}
\Psi[\Vect{F}] 
=
\mathcal{N}^{(F)}
\exp(-\frac{1}{2}\beta^{(F)} \!\!
\int \D^3 r 
\int\D^3 r^\prime ~
\mathbfscr{K}^{(F)} \Vect{F} \cdot \Vect{F}^\prime
)
\end{align}
where the kernel $\mathbfscr{K}^{(F)}$ involves the difference $\Vect{r}-\Vect{r}^\prime$ of the two integration \OrangelineChange{variables only}.

We emphasize that in \OrangelineChange{contrast} to the infinite product $\Psi(\qty{F_\ell})$\OrangelineChange{,} which is in terms of the set $\qty{F_\ell}$ of field strengths in all modes and given by ~\cref{MAIN:eq:wavefunction-frepresentation-ground-state}, we now have the complete field $\Vect{F}$. Hence, \OrangelineChange{the quantity} $\Psi$ defined by ~\cref{MAIN:eq:wave-functional-F-representation} represents a \emph{functional} of $\Vect{F}$ as indicated by the square brackets in $\Psi[\Vect{F}]$.

\section{Vector potential once more}\label{MAIN:sec:vector-potential-once-more}
In the preceding section\OrangelineChange{,} we have derived the wave functional $\Psi[\Vect{A}]$ in terms of the vector potential $\Vect{A}$ and have found a kernel, ~\cref{MAIN:eq:kernels-afield} which is different from the ones of $\Vect{E}$ and $\Vect{B}$, given by~\cref{MAIN:eq:kernels-efield-bfield}.
However, it has been argued~\cite{Wheeler1962,Bialynicki-Birula1998,Bialynicki-Birula2000} that an expression for a wave functional solely in terms of $\Vect{A}$ is problematic since the vacuum, and hence the wave functional should be gauge invariant, and the full vector potential is not gauge invariant. This line of reasoning was first used by Wheeler~\cite{Wheeler1962} in his original article on the wave functional where we find the quote:
\blockquote{%
\emph{''Often the dynamics of the electromagnetic field is discussed in terms of the vector potential $\Vect{A}$, connected with $\Vect{H}$, by the equation}
\begin{align*}
    \Vect{H}=\text{curl}~\Vect{A}.
\end{align*}
\emph{Then the probability amplitude is \OrangelineChange{evaluated} in the first instance as a functional of} $\Vect{A}$\emph{. Only later is it discovered as a consequence of gauge invariance, that} $\Vect{A}$ \emph{comes into evidence in the state functional only in the form of }$\Vect{H}=\text{curl}~\Vect{A}$\emph{.''}
}

Other authors~\cite{Bialynicki-Birula1998,Bialynicki-Birula2000} have argued in the same vein and thus concentrated their effort on expressions for the wave functional of the vacuum in terms of $\nabla \times \Vect{A}\equiv \Vect{B}$ instead of $\Vect{A}$. However, we have found in~\cref{MAIN:eq:wave-functional-F-representation} exactly such a wave functional $\Psi[\Vect{A}]$ and a corresponding kernel,~\cref{MAIN:eq:kernels-afield}. Hence, we are lead to the question \OrangelineChange{of} how to reconcile these opposing points of view.

Our answer to this question rests on the fact that the appearance of $\nabla \times \Vect{A}$ is not a consequence of gauge invariance but a specific choice of the mode expansion. Indeed, we first argue that due to the expansion in transverse modes\OrangelineChange{,} our expression is already gauge invariant. We then obtain an expression for the wave functional $\Psi[\Vect{A}]$ in terms of $\nabla\times\Vect{A}$ by use of the eigenmodes $\qty{\Vect{w}_\ell}$ of the magnetic induction $\Vect{B}$ without appealing to gauge invariance.

\subsection{Field functionals, quantization and gauge invariance}
While the gauge invariance argument seems superficially sound, it contains a very subtle flaw and is thus not applicable. Indeed, we start by \OrangelineChange{noticing that} electromagnetism is a gauge field theory~\cite{Weinberg1995}\OrangelineChange{,} and it is thus essential to remove redundant gauge degrees of freedom during the quantization procedure. It is then, and only then, that we can identify the actual physical degrees of freedom of the theory. Any observable, such as correlation functions or the wave functional\OrangelineChange{,} are \OrangelineChange{afterwards} expressed solely in terms of the quantized physical degrees of freedom.

In contrast to the earlier works of Wheeler~\cite{Wheeler1962} and Bialynicki-Birula~\cite{Bialynicki-Birula1998,Bialynicki-Birula2000}, we state and rely on a specific gauge choice from the start. Accidentally, the~\emph{gauge-fixing} of Coulomb gauge directly isolates \OrangelineChange{easy-to-interpret} physical degrees of freedom in non-relativistic situations for the electromagnetic field. However, this comes at the cost of sacrificing manifest Lorentz invariance of the theory. This procedure partitions the electromagnetic degrees of freedom into quantized (transverse) and non-quantized (longitudinal) degrees of freedom by enforcing the conditions $A_0 \equiv 0$ and $\nabla \cdot \Vect{A}\equiv0$ for the vector potential.

These quantized physical degrees of freedom are exactly our transverse fields~$\Vect{A}$,~$\Vect{E}$ and $\Vect{B}$. \OrangelineChange{Only} these fields\OrangelineChange{,} are associated with quantum states\OrangelineChange{,} that is \OrangelineChange{the} wave functions \OrangelineChange{of our theory.} 

In ~\cref{APP:sec:wavefunctional-groundstate}, we determine these wave functions for the ground state of the respective fields. 

Since these wave functions form the starting point of our derivation, any expression we obtain from them is naturally expressed in terms of \emph{gauge invariant quantities}, even if the \emph{transverse part} of the vector potential, namely $\Vect{A}$\OrangelineChange{,} appears in it. \OrangelineChange{Consequently, our expression for the wave functional $\Psi[\Vect{A}]$ of the vector potential~\cref{MAIN:eq:wave-functional-F-representation}, together with the associated kernel~\cref{MAIN:eq:kernels-afield}, is perfectly valid.}

We conclude, by returning to the subtle flaw in the argument of gauge invariance we have alluded to. Ultimately, a wave functional can only be defined \emph{after quantization} of a gauge theory \OrangelineChange{such as} electromagnetism has already been achieved, as it is a fundamentally quantum object. More specifically, the fields $\Vect{F}$ appearing in it are \emph{not classical fields} and\OrangelineChange{,} in general\OrangelineChange{, do not even} obey the \emph{classical field equations}\OrangelineChange{,} but are mere c-number fields \OrangelineChange{that} parameterize all \emph{quantum mechanically valid} field configurations \emph{interfering} in an appropriate functional integral.

Simultaneously, at this point in the development of the theory\OrangelineChange{,} the gauge-freedom is already incorporated in the choice of the quantized degrees of freedom, since all physically relevant quantities that appear, are by construction expressed without the gauge-redundant degrees of freedom. As a consequence, we cannot argue about \OrangelineChange{the} gauge-invariance of a quantity like \OrangelineChange{a} wave functional anymore when it is expressed in these quantities. Thus ultimately, it is the simple oversight that not the vector potential but only its transverse part can appear in a field functionals\OrangelineChange{,} which leads to the demise of any post-quantization argument relying on gauge transformations/invariance. 

Finally, although we worked \OrangelineChange{in} Coulomb gauge throughout this article, our reasoning applies to any gauge-fixing \OrangelineChange{chosen} during quantization.
Moreover, it translates to the wave-functional of other theories featuring gauge-invariances~\cite{Anastopoulos2022}\OrangelineChange{, e.g.,} the quantization of weak field gravity~\cite{Chen2023}. However, we note that when one is studying such cases, starting from a more modern path-integral formulation seems preferable~\cite{Weinberg1996} since gauge-fixings are implemented more easily via functional $\delta$-functions inside the path integral.

With these ideas in mind, we briefly comment on possible generalizations of our calculation to relativistic situations using the standard QED approach. While we have sacrificed manifest Lorentz covariance by our choice of Coulomb gauge, this was simply due to our interest in the cavity QED situation of the quantization in a resonator. If wanted, retaining Lorentz covariance and determining relativistically invariant analogs of the expressions,~\cref{MAIN:eq:wave-functional-F-representation}, for the wave functionals is possible by resorting to the Gupta-Bleuler~\cite{Gupta1950,Bleuler1950} method or the more general approach of BRST quantization~\cite{Fuster2005,Falceto2022}. For a modern discussion contrasting these approaches as applied to electromagnetism in $\xi$-gauge, a generalization of Lorenz gauge, we refer to Ref.~\cite{Falceto2022}.

\subsection{Wave functional in eigenmodes of magnetic induction}\label{MAIN:sec:wave-functional-for-magnetic-induction-eigenmodes}
In order to reexpress the wave functional $\Psi[\Vect{A}]$, as suggested by Wheeler and Bialynicki-Birula, in terms of \OrangelineChange{$\nabla \times \Vect{A}$,} we expand $\Vect{A}$ into the eigenmodes
\begin{align}
    \Vect{w}_\ell
    \equiv
    k_\ell^{-1} \nabla \times \Vect{u}_\ell
\end{align}
of the wave equation for $\Vect{B}$, rather than the one for $\Vect{A}$, \OrangelineChange{i.e.,}
\begin{align}
    \Vect{A}
    \equiv
    \sum_\ell
    A_\ell^{(w)} \Vect{w}_\ell.
\end{align}
Here we have attached a superscript $w$ to the amplitude $A_\ell$ \OrangelineChange{to} reflect the fact that this expansion is in the set of modes $\qty{\Vect{w}_\ell}$.

When we now take the curl of this representation of $\Vect{A}$, recall the Coulomb gauge condition, as well as the Helmholtz equation for $\Vect{u}_\ell$, we find
\begin{align}
    \nabla \times \Vect{A}
    =
    \sum_\ell A_\ell^{(w)} k_\ell \Vect{u}_\ell.
\end{align}

Consequently, the expansion coefficient $A_\ell^{(w)}$ in the $w$-representation takes the form
\begin{align}\label{MAIN:eq:expansion-coefficient-A-w-modes}
    A_\ell^{(w)}
    = 
    \frac{1}{k_\ell}\frac{1}{\mathcal{V}_\ell}
    \int \D^3 r~
    \Vect{u}_\ell^\dagger(\nabla \times \Vect{A}).
\end{align}

When we compare this expression to the corresponding one for \OrangelineChange{$F_\ell$}, expressed in the natural modes $\Vect{f}_\ell$, \OrangelineChange{i.e.,} to ~\cref{MAIN:eq:mode-coefficient-Fell-only}, we note an additional factor $k_\ell^{-1}$\OrangelineChange{,} which allows us to regain the same kernel in the double integral as in $\Vect{E}$ and $\Vect{B}$.

Since the quantization of $\Vect{A}$ now takes place in the $\Vect{w}_\ell$-modes\OrangelineChange{,} the wave function of the vacuum in the resonator reads
\begin{align}
    \Psi[\Vect{A}]
    =
    \mathcal{N}^{(A)}
    \exp(-\frac{1}{2}\beta^{(A)} \Sigma^{({A^{(w)}})})
\end{align}
where now the sum
\begin{align}\label{MAIN:eq:modesum-aw-modes}
    \Sigma^{({A^{(w)}})}
    \equiv
    \sum_\ell
    \qty(A^{(w)}_\ell)^2 k_\ell^{-1} \mathcal{V}_\ell 
\end{align}
runs over the $\Vect{w}_\ell$-modes.

When we substitute the explicit form, ~\cref{MAIN:eq:expansion-coefficient-A-w-modes}, of the expansion coefficients $A^{(w)}_\ell$ into the mode sum, ~\cref{MAIN:eq:modesum-aw-modes} we arrive at
\begin{align}
    \Sigma^{({A^{(w)}})}
    =
    \int \D^3 r
    \int \D^3 r^\prime ~
    (\nabla\times\Vect{A})^\dagger
    \SecondOrderTensor{\mathbfscr{K}}(\Vect{r},\Vect{r}^\prime)
    (\nabla^\prime\times\Vect{A}^\prime)
\end{align}
where according to~\OrangelineChange{\cref{MAIN:eq:FEB-definition}}
the term $F(k_\ell)$ in the kernel $\SecondOrderTensor{\mathbfscr{K}}$ defined by~\cref{MAIN:eq:defining-eq-kernel}
takes the form
\begin{align}
    \FlabelNoIdx(k_\ell) = k_\ell^{-1},
\end{align}
and is thus identical to the one for $\Vect{E}$ and $\Vect{B}$ in their natural modes.

As a consequence, the kernel for the vector potential $\Vect{A}$ expanded into $\Vect{w}_\ell$- rather than $\Vect{u}_\ell$-modes is identical to that of $\Vect{E}$ and $\Vect{B}$. \OrangelineChange{However, now the} wave functional of the vacuum in the representation of $\Vect{A}$\OrangelineChange{, contains $\Vect{A}$} only in the form $\nabla\times \Vect{A}$. In this way\OrangelineChange{,} $\Psi[\Vect{A}]$ is expressed in terms \OrangelineChange{of the} magnetic induction, which is a gauge invariant quantity.

\section{Discussion of Wave functionals}\label{MAIN:sec:discussion-of-wave-functionals}
We are now in a position to present the explicit expressions for the wave functionals of the vacuum in a resonator\OrangelineChange{,} as summarized in~\cref{tab:table3}. Moreover, we compare and contrast the corresponding expressions \OrangelineChange{to the ones in the literature}.

\subsection{Dependence on mode expansion}
The central message of~\cref{tab:table3} is that the kernel of the wave functional depends on the mode expansion of the field. \OrangelineChange{At} first sight\OrangelineChange{,} this property is surprising since the creation of the bilinear form of the complete field removes the field expansion. However, the wave functional $\Psi[\Vect{A}]$ of the vacuum in the representation of the vector potential $\Vect{A}$, summarized in the \OrangelineChange{first and last} row of~\cref{tab:table3}\OrangelineChange{,} demonstrates this feature in a striking way.

Indeed, when we use the eigenmode expansion of $\Vect{A}$, given by $\qty{\Vect{u}_\ell}$, which is identical to the one of the electric field $\Vect{E}$, we find a kernel \OrangelineChange{that} is proportional to $1/r^4$ and negative. In this case\OrangelineChange{,} the bilinear form involves only $\Vect{A}$.

However, when we employ the eigenmode expansion of the magnetic induction $\Vect{B}$\OrangelineChange{, i.e.,} the modes $\qty{k_\ell^{-1}\nabla\times\Vect{u}_\ell}$, the kernel of $\Vect{E}$, which is identical to that of $\Vect{B}$ emerges and enjoys the decay $1/r^2$. In this case the kernel is positive. However, most importantly, the bilinear form does not involve $\Vect{A}$ but $\nabla \times \Vect{A}\equiv \Vect{B}$.

This dependence of the kernel on the mode representation, and the associated form of the bilinear form, is reminiscent of the different operator orderings in quantum mechanics and the associated quasi-probability distribution functions. \OrangelineChange{We recall~\cite{Schleich2001} that} a symmetric ordering requires the use of the Wigner function\OrangelineChange{,} whereas the anti-normal ordering leads us to the \OrangelineChange{Husimi} or $Q$-function. Normal ordering brings in the $P$-distribution.

Hence, the same quantum state can enjoy different phase space distribution functions depending on the choice of the operator ordering. Nevertheless, the quantum mechanical average of interest is always the same.

\OrangelineChange{%
This analogy draws attention to the quantity so far not addressed in our article, that is, the field operators. Indeed, we have concentrated excessively on the wave functional, which of course, could be employed to calculate expectation values of the field operators. In order to perform this evaluation in an effective way, it is necessary to have the operators to be averaged in the same modes as the wave functional. Indeed, an identical mode expansion in operators and wave functionals is necessary to express the operator in a c-number representation.
This requirement is analogous to the familiar technique of one-particle quantum mechanics to perform averages using wave functions in the eigenrepresentation of the operator. In this way, we can evaluate the expectation values by functional integration as discussed in the next section.
}

\subsection{Connection to free space}
We conclude by comparing and contrasting the form of the functionals in a resonator, to the ones in free space first suggested by Wheeler~\cite{Wheeler1962} and discussed and extended by Bialynicki-Birula. Here we confine ourselves to the one involving $\nabla\times\Vect{A}$\OrangelineChange{,} which according \OrangelineChange{to~\cite{Bialynicki-Birula1998,Bialynicki-Birula2000}} reads
\begin{widetext}
\begin{align}
    \Psi[\Vect{A}]
    =
    \mathcal{N}^{(A)}
    \exp(%
    -\frac{1}{4\pi^2\hbar}\sqrt{\frac{\varepsilon_0}{\mu_0}}
    \int \D^3 r
    \int \D^3 r^\prime
    \frac{(\nabla\times\Vect{A})\cdot(\nabla^\prime\times\Vect{A}^\prime)}{\Abs{\Vect{r}-\Vect{r}^\prime}^2}
    ).
\end{align}
\end{widetext}
The only difference to the expression in the \OrangelineChange{fourth row} of~\cref{tab:table3} is in the prefactor $\beta^{(A)}$ containing fundamental constants.
\OrangelineChange{Whereas we} always use $\varepsilon_0$ and $c$, Bialynicki-Birula's expression involves the ratio $\sqrt{\varepsilon_0/\mu_0}$. Here $\mu_0$ denotes the permeability of the vacuum.

However, the Kirchhoff identity
\begin{align}
    \frac{1}{\mu_0 \varepsilon_0} = c^2
\end{align}
immediately yields the connection \OrangelineChange{formula}
\begin{align}
\sqrt{\frac{\varepsilon_0}{\mu_0}}
=\varepsilon_0 c,
\end{align}
in complete agreement with our expression in~\cref{tab:table3}.
%%%%%%%%%%%%%%%%%%%%%%%%%%%
% TABLE 
%%%%%%%%%%%%%%%%%%%%%%%%%%%
\begin{table*}
\caption{%
Wave functional $\Psi[\Vect{F}]$ of the vacuum in a resonator for the three fields $\Vect{F}=\Vect{E},\Vect{B}$ or $\Vect{A}$ and their corresponding kernels \OrangelineChange{$\SecondOrderTensor{\mathbfscr{K}}\equiv\SecondOrderTensor{\mathbfscr{K}}(\Vect{r},\Vect{r}^\prime)$} when expressed in the mode basis $\qty{\Vect{f}_\ell}=\qty{\Vect{u}_\ell},\qty{\Vect{v}_\ell}$ or $\qty{\Vect{w}_\ell}$. Here the prime indicates the field at the integration variable $\Vect{r}^\prime$ rather than $\Vect{r}$.
%and the kernel $\mathbfscr{K}^{(1)}$ reads $\displaystyle{\mathbfscr{K}^{(1)}\equiv \frac{1}{(2\pi)^3}}\int \D^3 k k \exp(\I \Vect{k}\cdot \Vect{r})$. 
% COMMENT: Find out how to do formulas in caption
% COMMENT: Fractions
}
\label{tab:table3}
\OrangelineChange{%
\begin{ruledtabular}
\begin{tabular}{llll}
        %%%%%%%%%%%%
        \textbf{Field} $\Vect{F}$ & 
        \textbf{Mode basis} $\qty{\Vect{f}_\ell}$ &
        \textbf{Mode basis kernel $\SecondOrderTensor{\mathbfscr{K}}(\Vect{r},\Vect{r}^\prime)$}  &
        \textbf{Wave functional}~$\Psi[\Vect{F}]$~\textbf{in field basis}  \\
        \midrule
        $\Vect{A}$ &
        $\qty{\Vect{u}_\ell}$ &
        $\displaystyle{\sum_\ell \frac{k_\ell}{\mathcal{V}_\ell} \Vect{u}_\ell {\Vect{u}_\ell^\prime}^\dagger}$ &
        {%
        $\displaystyle{%
        \mathcal{N}^{(A)}\exp( \frac{1}{2\pi^2}\frac{\varepsilon_0 c}{\hbar}
       \int \D^3 r \int\D^3 r^\prime \frac{\Vect{A}\cdot\Vect{A}^\prime}{\Abs{\Vect{r}-\Vect{r}^\prime}^4}
        )
        }$%
        }\\[2ex] 
        \midrule        
        %%%%%%%%%%%%%
        $\Vect{E}$ &
        $\qty{\Vect{v}_\ell}=\qty{\Vect{u}_\ell}$ &
        $\displaystyle{\sum_\ell \frac{k_\ell^{-1}}{\mathcal{V}_\ell} \Vect{v}_\ell {\Vect{v}_\ell^\prime}^\dagger}$ &
        {%
        $\displaystyle{%
        \mathcal{N}^{(E)}\exp(-\frac{1}{4\pi^2}\frac{\varepsilon_0}{\hbar c}
         \int\D^3 r \int\D^3 r^\prime \frac{\Vect{E}\cdot\Vect{E}^\prime}{\Abs{\Vect{r}-\Vect{r}^\prime}^2})
        }$%
        }\\[3ex]
        %%%%%%%%%%%%%
        $\Vect{B}$ &
        $\qty{\Vect{w}_\ell}=\qty{k_\ell^{-1}\nabla \times\Vect{u}_\ell}$ &
        $\displaystyle{\sum_\ell \frac{k_\ell^{-1}}{\mathcal{V}_\ell} \Vect{w}_\ell {\Vect{w}_\ell^\prime}^\dagger}$ &
        {%
        $\displaystyle{%
        \mathcal{N}^{(B)}\exp(-\frac{1}{4\pi^2}\frac{\varepsilon_0 c}{\hbar}
         \int\D^3 r \int\D^3 r^\prime \frac{\Vect{B}\cdot\Vect{B}^\prime}{\Abs{\Vect{r}-\Vect{r}^\prime}^2})
        }$%
        }\\[3ex]
        %%%%%%%%%%%%%
        $\Vect{A}$ &
        $\qty{\Vect{w}_\ell}=\qty{k_\ell^{-1}\nabla \times\Vect{u}_\ell}$ &
        $\displaystyle{\sum_\ell \frac{k_\ell^{-1}}{\mathcal{V}_\ell} \Vect{w}_\ell {\Vect{w}_\ell^\prime}^\dagger}$ &
        {%
        $\displaystyle{%
        \mathcal{N}^{(A^{(w)})}\exp(-\frac{1}{4\pi^2} \frac{\varepsilon_0 c}{\hbar}
         \int\D^3 r \int\D^3 r^\prime \frac{\qty(\nabla \times \Vect{A})\cdot\qty(\nabla^\prime \times \Vect{A}^\prime)}{\Abs{\Vect{r}-\Vect{r}^\prime}^2}
        )
        }$%
        }
\end{tabular}
\end{ruledtabular}
}
\end{table*}
%%%%%%%%%%%%%%%%%%%%%%%%%%%
% TABLE 
%%%%%%%%%%%%%%%%%%%%%%%%%%%

\section{Wave functionals and expectation values}\label{MAIN:sec:functional-stuff}
In the preceeding sections we have made our way to explicit expressions for the wave functional of the electromagnetic vacuum, beginning with the quantization of the electromagnetic field in a resonator. Most of the expressions we have obtained coincide with the ones found previously by Wheeler~\cite{Wheeler1962} and Bialynicki-Birula~\cite{Bialynicki-Birula1998,Bialynicki-Birula2000} for free space, \OrangelineChange{although} were now obtained for the case of a resonator. However, one expression in terms of a bilinear functional of $\Vect{A}$ is new to \OrangelineChange{the best of} our knowledge.

While these functionals are certainly interesting from a fundamental point of view, ultimately we go through the trouble of setting up a field theory in order to calculate observables, that is scattering cross sections, correlation functions and their more complicated cousins. Naturally, we must thus face the question of how these calculations can be performed with the field wave functions and functionals. This \OrangelineChange{problem} constitutes the topic of this section and we shall show by the example of such a calculation for a specific correlation function how this can be done. 

We focus our effort on the Wightman tensor $\SecondOrderTensor{\mathbfscr{W}}^{(F)}_{\Vect{r}\Vect{r}^\prime}(t)$ for the field $\Vect{F}$, which contains all first order correlation functions of the vector field $\Vect{F}$ evaluated at two points\OrangelineChange{,} $\Vect{r}$ and $\Vect{r}^\prime$\OrangelineChange{,} in space. Furthermore, it is of specific interest because it can be used to easily determine the excitation probability~\cite{Lopp2021} for an atom in a cavity due to the vacuum field.

\subsection{A general correlation function}
We begin by stating the definition~\cite{Lopp2021} of the equal-time two-point Wightman tensor
\begin{align}\label{MAIN:eq:correlation-function}
    \SecondOrderTensor{\mathbfscr{W}}_{\Vect{r}\Vect{r}^\prime}^{(F)}(t) 
    \equiv
    \bra{\Vect{0}}
    \QmOp{\Vect{F}}(t,\Vect{r})
    \QmOp{\Vect{F}}^\dagger(t,\Vect{r}^\prime)
    \ket{\Vect{0}},
\end{align}
for the field $\Vect{F}$
which is the expectation value of the outer product of the field operators $\QmOp{\Vect{F}}(t,\Vect{r})\QmOp{\Vect{F}}^\dagger(t,\Vect{r}^\prime)$ at fixed time $t$ but \OrangelineChange{in different} locations $\Vect{r}$ and $\Vect{r}^\prime$. \OrangelineChange{In fact,~\cref{MAIN:eq:correlation-function}} describes the spatial correlations in the vacuum field $\Vect{F}$ at the respective positions $\Vect{r}$ and $\Vect{r}^\prime$.

For the purpose of illustrating the formalism \OrangelineChange{$\SecondOrderTensor{\mathbfscr{W}}_{\Vect{r}\Vect{r}^\prime}^{(F)}$} may be seen as a tensorial version of the correlation functions introduced by Glauber~\cite{Glauber1963,Glauber2007} in quantum optics. For example, taking the trace of the Wightman tensor yields an intensity correlation function which is a precursor of the (spatial) \OrangelineChange{first-order} coherence function $G^{(1)}(t,\Vect{r};t,\Vect{r}^\prime)$.

\subsection{Wightman tensor via mode decomposition}
We begin by expressing the Wightman tensor in terms of the $\Vect{f}_\ell$-modes,~\cref{MAIN:eq:mode-expansion-fmodes}, which yields for,~\cref{MAIN:eq:correlation-function}, the decomposition 
\begin{align}\label{MAIN:eq:wightman-tensor-mode-decomposition}
 \SecondOrderTensor{\mathbfscr{W}}_{\Vect{r}\Vect{r}^\prime}^{(F)}(t) 
    = 
    \sum_{\ell,\ell^\prime}
    \bra{\Vect{0}} 
    \QmOp{F}_\ell^{}(t)
    \QmOp{F}_{\ell^\prime}(t)\ket{\Vect{0}}
    \Vect{f}_\ell^{}(\Vect{r})
    \Vect{f}_{\ell^\prime}^\dagger(\Vect{r}^\prime)\OrangelineChange{.}
\end{align}
\OrangelineChange{Here} we have used the linearity of the mode sums and acted with the vacuum directly on the operator parts of the fields. Note, that in the process we \OrangelineChange{used} the fact that the fields are hermitian operators,~\OrangelineChange{i.e.}, $\QmOp{F}_{\ell^\prime}^\dagger=\QmOp{F}_{\ell^\prime}^{}$. 

\subsubsection{Determination of vacuum expectation value}
Proceeding from~\cref{MAIN:eq:wightman-tensor-mode-decomposition} our next task is \OrangelineChange{to calculate} the field operator expectation value with respect to the vacuum state $\mathcal{O}^{(F)}_{\ell\ell^\prime}$\OrangelineChange{, for} which we introduce the abbreviation
\begin{align}\label{MAIN:eq:operatorsandwich}
    \mathcal{O}^{(F)}_{\ell\ell^\prime}
    \equiv
    \bra{\Vect{0}}
    \QmOp{F}_\ell^{}(t)
    \QmOp{F}_{\ell^\prime}^{}(t)
    \ket{\Vect{0}}.
\end{align}

 Since the time argument is identical \OrangelineChange{for} both field operators, and is immaterial for what follows, we will suppress it going forward and simply write $\QmOp{F}_\ell^{}(t)\equiv \QmOp{F}_\ell^{}$ from now on to compactify the notation.

In order to evaluate the expectation value,~\cref{MAIN:eq:operatorsandwich}, we recall that the non-interacting vacuum ket-state $\ket{\Vect{0}}$ of the free (electromagnetic) field $\Vect{F}$ is a direct product
\begin{align}
    \ket{\Vect{0}}
    \equiv 
    \bigotimes_{k} \ket{0_k}
    = \ket{0_1}\ket{0_2}\ket{0_3}\cdots\ket{0}_\ell\cdots
\end{align}
of all ground states of all modes and that the operator $\QmOp{F}_\ell$ only acts on the $\ell$-th mode. \OrangelineChange{Other} ground states $\ket{0_k}$ with $k\neq \ell$ are not affected by $\QmOp{F}_\ell$.

Obviously, the same property holds true for the vacuum bra-vector $\bra{\Vect{0}}$, and \OrangelineChange{none of the} ground states $\bra{0_{k^\prime}}$ with $k^\prime\neq \ell^\prime$ \OrangelineChange{is} affected by $\QmOp{F}_{\ell^\prime}$, and \OrangelineChange{they} pass to the right\OrangelineChange{,} where they meet the ground states $\ket{0_k}$ from the ket-vacuum.

Since we can only take the scalar product between the same modes\OrangelineChange{,} we have to distinguish the two cases $\ell=\ell^\prime$ and $\ell\neq \ell^\prime$. 

The first case of identical modes,~\OrangelineChange{i.e.,} $\ell \equiv \ell^\prime$, leads us to the expression
\begin{align}\label{MAIN:eq:eq:Fellsquared-with-product}
    \mathcal{O}^{(F)}_{\ell\ell}
    =
    \braket{0_\ell| \QmOp{F}_\ell^2 |0_\ell} \prod_{k\neq \ell} \braket{0_k|0_k}
\end{align}
or
\begin{align}
    \mathcal{O}^{(F)}_{\ell\ell} = \braket{0_\ell| \QmOp{F}_\ell^2 |0_\ell} 
\end{align}
where we have \OrangelineChange{used} the normalization condition $\braket{0_k|0_k}=1$ of the ground state, which in the field representation reads
\begin{align}\label{MAIN:eq:Fellsquared-expectation-value}
    \int_{\mathrlap{-\infty}}^{\mathrlap{\infty}} \D F_k \braket{0_k|F_k} \braket{F_k|0_k}
    = 
    \int_{\mathrlap{-\infty}}^{\mathrlap{\infty}} \D F_k ~\qty|\psi_k(F_k)|^2
\end{align}
and is satisfied\OrangelineChange{,} since according to~\cref{APP:sec:wavefunctional-groundstate}
we find
\begin{align}\label{MAIN:eq:wave-functional-gaussian-state}
 \psi_k(F_k) = \frac{1}{\pi^{1/4}} \frac{1}{\sqrt{\mathcal{F}_k}} \exp(-\frac{1}{2}\qty(\frac{F_k}{\mathcal{F}_k})^2).
\end{align}

Moreover, the field operator of the $\ell$-th mode obeys the eigenvalue equation
\begin{align}\label{MAIN:eq:generic-F-eigenvalue-equation}
    \QmOp{F}_\ell^{}\ket{F_\ell} = F_\ell\ket{F_\ell}
\end{align}
and as a consequence\OrangelineChange{,} we have the spectral representation
\begin{align}\label{MAIN:eq:operator-eigenbasis-rep}
    g(\QmOp{F}_\ell) \equiv
    \int_{\mathrlap{-\infty}}^{\mathrlap{\infty}} \D F_\ell~
    g(F_\ell)\ket{F_\ell} \bra{F_\ell}
\end{align}
for integrable functions $g\equiv g(x)$.

When we introduce this spectral representation for the $\ell$-th \OrangelineChange{mode into} ~\cref{MAIN:eq:Fellsquared-expectation-value}, we obtain
\begin{align}
    \OrangelineChange{\mathcal{O}_{\ell\ell}^{(F)}}
    =
    \int_{\mathrlap{-\infty}}^{\mathrlap{\infty}} \D F_\ell
    F_\ell^2 ~ \qty|\psi_\ell(F_\ell)|^2
\end{align}
which with the help of the Gaussian wave function,~\cref{MAIN:eq:wave-functional-gaussian-state}
reads
\begin{align}\label{MAIN:eq:Fellsquared-expectation-value-result}
    \OrangelineChange{\mathcal{O}_{\ell\ell}^{(F)}} = \frac{1}{2}\mathcal{F}_{\ell}^2.
\end{align}

Next we consider the case $\ell\neq\ell^\prime$ which yields the expression
\begin{align}\label{MAIN:eq:Fellprimesquared-expectation-value}
    \mathcal{O}_{\ell\ell^\prime}^{(F)}
    = 
    \braket{0_{\ell^\prime}|\QmOp{F}_{\ell^\prime}|0_{\ell^\prime}}
    \braket{0_{\ell}|\QmOp{F}_\ell |0_\ell}
    \prod_{k\neq \ell,\ell^\prime} \braket{0_k|0_k}.
\end{align}

We emphasize that, in contrast to ~\cref{MAIN:eq:eq:Fellsquared-with-product}, the mode indices $\ell$ \emph{and} $\ell^\prime$ appear now. Nevertheless, the normalization condition is again $\braket{0_k|0_k}=1$ for each mode and reduces~\cref{MAIN:eq:Fellprimesquared-expectation-value} to
\begin{align}
    \mathcal{O}_{\ell\ell^\prime}^{(F)}
    = 
    \braket{0_{\ell}|\QmOp{F}_\ell |0_\ell}
    \braket{0_{\ell^\prime}|\QmOp{F}_{\ell^\prime}|0_{\ell^\prime}}.
\end{align}

When we now employ the field representation\OrangelineChange{,} again we find with the eigenvalue equation~\cref{MAIN:eq:generic-F-eigenvalue-equation} \OrangelineChange{for} $\ell\neq\ell^\prime$ the formula
\begin{align}\label{MAIN:eq:Fellprimesquared-expectation-value-result}
    \OrangelineChange{
    \braket{0_{\ell^\prime}|\QmOp{F}_{\ell^\prime}|0_{\ell^\prime}}
    =
    \int_{\mathrlap{-\infty}}^{\mathrlap{\infty}} \D F_\ell
    F_\ell ~ \qty|\psi_\ell(F_\ell)|^2 = 0
    }
\end{align}
where in the last step we have \OrangelineChange{used} the symmetric Gaussian wave function~\OrangelineChange{\cref{MAIN:eq:wave-functional-gaussian-state}} of the ground state.

When we combine the results ~\cref{MAIN:eq:Fellsquared-expectation-value-result} and ~\cref{MAIN:eq:Fellprimesquared-expectation-value-result}\OrangelineChange{,} we find
\begin{align}\label{MAIN:eq:OFell-expectation-value-result}
    \mathcal{O}_{\ell\ell^\prime}^{(F)}
    =
    \frac{1}{2} \delta_{\ell\ell^\prime} \mathcal{F}_\ell^2.
\end{align}

With the respective definitions of the vacuum fields $\mathcal{F}_\ell$ in ~\cref{MAIN:eq:vacuum-amplitude-vector-potential,MAIN:eq:scriptE-link-scriptA,MAIN:eq:link-between-script-fields} we can bring~\cref{MAIN:eq:OFell-expectation-value-result} into the final form
\begin{align}\label{MAIN:eq:final-result-vacuum-expectation-value}
    \mathcal{O}^{(F)}_{\ell\ell^\prime} 
    = 
    \frac{\delta_{\ell\ell^\prime}}{2}
    \frac{1}{\beta^{(F)}\FlabelNoIdx(k_\ell)}\frac{1}{\mathcal{V}_\ell}
\end{align}
which constitutes our result for the vacuum expectation value,~\cref{MAIN:eq:operatorsandwich}.
This expression for the field $\Vect{F}$ is determined by the physical constants contained in $\beta^{(F)}$, the wave number $k_\ell$ together with the function $F$\OrangelineChange{,} and the mode volume $\mathcal{V}_\ell$ of the $\ell$-th mode.

\subsubsection{Wightman tensor and kernels}
With the result for the vacuum expectation value\OrangelineChange{,} we are now in a position to determine the Wightman tensor of the field $\Vect{F}$. Using the result from~\cref{MAIN:eq:final-result-vacuum-expectation-value} and inserting it into~\cref{MAIN:eq:wightman-tensor-mode-decomposition}\OrangelineChange{,} we arrive at
\begin{align}
    \SecondOrderTensor{\mathbfscr{W}}_{\Vect{r}\Vect{r}^\prime}^{(F)}(t) 
    = 
    \frac{1}{2\beta^{(F)}}
    \sum_{\ell}
    \FlabelNoIdx^{-1}(k_\ell)\frac{1}{\mathcal{V}_\ell}
    \Vect{f}_\ell(\Vect{r})
    \OrangelineChange{\Vect{f}_{\ell}^\dagger}(\Vect{r}^\prime)
\end{align}
for the mode expanded version of the Wightman tensor. We observe, that this expression seems reminiscent of the expression for the transverse delta function in terms of the modes,~\cref{MAIN:eq:kernel-of-projector-in-modes}.

Actually, with the help of the \OrangelineChange{square root of the} negative Laplacian, we can move \OrangelineChange{}{$\FlabelNoIdx^{-1}(k_\ell)$} out of the sum by reversing its action on the modes via 
\begin{align}
\FlabelNoIdx^{-1}(k_\ell)\Vect{f}_\ell\OrangelineChange{(\Vect{r})}=\FlabelNoIdx^{-1}\qty(\OrangelineChange{\sqrt{-\VLaplace_{\Vect{r}}}})\Vect{f}_\ell\OrangelineChange{(\Vect{r})}
\end{align}
and using the independence of the right-hand side from the summation index $\ell$. Together with the representation of the transverse delta function,~\cref{MAIN:eq:kernel-of-projector-in-modes}, we arrive at the expression
\begin{align}\label{MAIN:eq:wightman-tensor-generic-result}
    \SecondOrderTensor{\mathbfscr{W}}_{\Vect{r}\Vect{r}^\prime}^{(F)}(t)
    =
    \frac{1}{2\beta^{(F)}}
    \FlabelNoIdx^{-1}\qty(\OrangelineChange{\sqrt{-\VLaplace_{\Vect{r}}}})
    \DeltaTransversal(\Vect{r}-\Vect{r}^\prime)
\end{align}
for the Wightman tensor\OrangelineChange{,} which is fully consistent with the results obtained in Ref.~\cite{Lopp2021} in free space for the electric or magnetic field.

When we now compare the expression~\cref{MAIN:eq:wightman-tensor-generic-result} for the Wightman tensor $\SecondOrderTensor{\mathbfscr{W}}_{\Vect{r}\Vect{r}^\prime}^{(F)}$ with the one~\cref{MAIN:eq:kernel-as-deltatransverse-derivative} of the kernel $\OrangelineChange{\SecondOrderTensor{\mathbfscr{K}}^{(F)}}$ we find that $\FlabelNoIdx(\OrangelineChange{\sqrt{-\VLaplace_{\Vect{r}}}})$ is either in the denominator or in the numerator. At the same time we obtain from the definitions ~\cref{MAIN:eq:FEB-definition} and ~\cref{MAIN:eq:FA-definition} of $\FlabelNoIdx$ for $\Vect{E}$, $\Vect{B}$ and $\Vect{A}$ the relation
\begin{align}
    \FlabelNoIdx^{(E)}=\FlabelNoIdx^{(B)} = \frac{1}{\FlabelNoIdx^{(A)}}.
\end{align}

As a result we arrive at the connection formulae
\begin{align}\label{MAIN:eq:wightman-E}
    \SecondOrderTensor{\mathbfscr{W}}_{\Vect{r}\Vect{r}^\prime}^{(E)}(t)
    =
    \frac{\hbar c}{2\varepsilon_0}
    \SecondOrderTensor{\mathbfscr{K}}^{(A)}(\Vect{r}-\Vect{r}^\prime)
    = 
    \frac{\hbar c}{2\varepsilon_0}\qty(-\VLaplace_{\Vect{r}})\SecondOrderTensor{\mathbfscr{K}}^{(E)}(\Vect{r}-\Vect{r}^\prime)
\end{align}
and
\begin{align}\label{MAIN:eq:wightman-B}
    \SecondOrderTensor{\mathbfscr{W}}_{\Vect{r}\Vect{r}^\prime}^{(B)}(t)
    =
    \frac{\hbar}{2\varepsilon_0c }
    \SecondOrderTensor{\mathbfscr{K}}^{(A)}(\Vect{r}-\Vect{r}^\prime)
    = 
    \frac{\hbar }{2\varepsilon_0c}\qty(-\VLaplace_{\Vect{r}})\SecondOrderTensor{\mathbfscr{K}}^{(B)}(\Vect{r}-\Vect{r}^\prime)
\end{align}
for the Wightman tensors $\SecondOrderTensor{\mathbfscr{W}}_{\Vect{r}\Vect{r}^\prime}^{(E)}$ and $\SecondOrderTensor{\mathbfscr{W}}_{\Vect{r}\Vect{r}^\prime}^{(B)}$.

From~\cref{MAIN:eq:wightman-E,MAIN:eq:wightman-B} we make the observation, that the Wightman tensors $\SecondOrderTensor{\mathbfscr{W}}_{\Vect{r}\Vect{r}^\prime}^{(E/B)}$ are intimately related to our kernels $\SecondOrderTensor{\mathbfscr{K}}^{(A/E/B)}$ - either via the application of a negative Laplacian or even directly identical to the Wightman tensor except for a dimensionful proportionality constant. 

While the existence of a relation like this seems initially surprising, \OrangelineChange{it} is only partially so, since the kernels can be seen as the field theoretical analogue of covariance matrices for the Gaussian vacuum state. The Wightman tensors\OrangelineChange{,} in turn collect all possible quadratic field correlation functions. Thus an intimate relationship between both quantities is to be expected.

\subsection{\OrangelineChange{Wightman tensor from functional integrals}}
While our approach to determine the explicit form of the Wightman tensor $\SecondOrderTensor{\mathbfscr{W}}_{\Vect{r}\Vect{r}^\prime}^{(F)}$ via the mode expansion and the vacuum wave functions was ultimately successful\OrangelineChange{,} it did not rely on the wave functionals themselves. Thus \OrangelineChange{the task} arises, how similar questions can be framed and answered using the wave functional. We will now give a sketch using functional methods \OrangelineChange{on} how this might be achieved.

We start by recalling the relation~\cite{Hatfield2018,Peskin2018} between the functional integration measure and the field basis
\begin{align}
    \int\!\!\FuncD[\Vect{F}]  
    \equiv 
    \prod_\ell 
    \int_{\mathrlap{-\infty}}^{\mathrlap{+\infty}} 
    \D F_\ell.
\end{align}

Moreover\OrangelineChange{, we note} that the field operators $\QmOp{\Vect{F}}=\QmOp{\Vect{F}}(t,\Vect{r})$ and $\QmOp{\Vect{F}}^\prime=\QmOp{\Vect{F}}(t,\Vect{r}^\prime)$  can be expressed as functional Schrödinger integrals via
\begin{align}\label{MAIN:eq:decompositionFfunctional}
\QmOp{\Vect{F}}(t,\Vect{r})
=
\int\!\!\FuncD[\Vect{F}] \ket{\Vect{F}} \bra{\Vect{F}} \Vect{F}(t,\Vect{r})
\end{align}
and
\begin{align}\label{MAIN:eq:decompositionFprimefunctional}
\QmOp{\Vect{F}}^{\prime\dagger} (t,\Vect{r}^\prime)
=
\int\!\!\FuncD[\Vect{F}^\prime] \ket{\Vect{F}^\prime} \bra{\Vect{F}^\prime} \Vect{F}^{\prime\dagger}(t,\Vect{r}^\prime)
\end{align}
where $\ket{\Vect{F}}\equiv\ket{\{F_\ell\}}=\ket{F_1}\ket{F_2}\cdots$ and $\ket{\Vect{F}^\prime}\equiv\ket{\{F_\ell^\prime\}}=\ket{F_1^\prime}\ket{F_2^\prime}\cdots$ correspond to the state vectors of the field.

With these preliminaries settled\OrangelineChange{,} we recall the definition of the Wightman tensor,~\cref{MAIN:eq:correlation-function}
\begin{align}
\SecondOrderTensor{\mathbfscr{W}}_{\Vect{r}\Vect{r}^\prime}^{(F)}(t)
 =
\braket{\Vect{0}|\QmOp{\Vect{F}}(t,\Vect{r})\QmOp{\Vect{F}}^\dagger(t,\Vect{r}^\prime)|\Vect{0}}
\end{align}
and obtain\OrangelineChange{,} by inserting the operator expansions from ~\cref{MAIN:eq:decompositionFfunctional} and ~\cref{MAIN:eq:decompositionFprimefunctional}\OrangelineChange{,} the double functional integral
\begin{widetext}
\begin{align}\label{MAIN:eq:double-functional-integral-wightman}
\SecondOrderTensor{\mathbfscr{W}}_{\Vect{r}\Vect{r}^\prime}^{(F)}(t)
 =
 \int\!\!\FuncD[\Vect{F}]\int\!\!\FuncD[\Vect{F}^\prime]
 \braket{\Vect{0}|\Vect{F}}\braket{\Vect{F}|\Vect{F}^\prime} \braket{\Vect{F}^\prime|\Vect{0}}\Vect{F}(t,\Vect{r})\Vect{F}^{\prime\dagger}(t,\Vect{r}^\prime)
\end{align}
\end{widetext}
representation for the Wightman tensor of the field $\Vect{F}$. 

At first\OrangelineChange{,} this result appears to be \OrangelineChange{too} cumbersome \OrangelineChange{for} actual practical use. However, with the help of the functional Dirac delta function and the relation
\begin{align}
 \delta[\Vect{F}^{}-\Vect{F}^\prime]
 \equiv
 \prod_\ell
 \delta(F_\ell^{}-F_\ell^\prime)
 =
 \prod_\ell
 \braket{\{F_\ell^{}\}|\{F_\ell^\prime \}}
 =
 \braket{\Vect{F}|\Vect{F}^\prime}
\end{align}
we can collapse one of the functional integrations in~\cref{MAIN:eq:double-functional-integral-wightman} and arrive at
\begin{align}
\SecondOrderTensor{\mathbfscr{W}}_{\Vect{r}\Vect{r}^\prime}^{(F)}(t)
 =
 \int\!\!\FuncD[\Vect{F}]
 ~\qty|\braket{\Vect{F}|\Vect{0}}|^2 \Vect{F}(t,\Vect{r})\Vect{F}^{\dagger}(t,\Vect{r}^\prime).
\end{align}

\OrangelineChange{Note that} in the process of collapsing the integration\OrangelineChange{,} only a relabeling due to the replacement $\Vect{F}^\prime\mapsto\Vect{F}$ has taken place, while the spatial dependence on $\Vect{r}^\prime$, characteristic of a two-point correlation function in the expression\OrangelineChange{,} was completely retained.

At this point in the development of the functional approach we are finally in the position to identify our wave functionals of the vacuum by
\begin{align}
    \qty|\braket{\Vect{F}|\Vect{0}}|^2
    \equiv
    \mathcal{N}
\exp(-\beta^{(F)}\Sigma^{(F)}[\Vect{F}]),
\end{align}
where we have made use of the mode sum in functional form,~\cref{MAIN:eq:double-integral-FKF}, and defined the normalization constant $\mathcal{N}\equiv \qty({\mathcal{N}^{(F)}})^2$ of the functional.

\OrangelineChange{As a consequence we are lead to a \emph{single} functional integral representation
\begin{align}
 \SecondOrderTensor{\mathbfscr{W}}_{\Vect{r}\Vect{r}^\prime}^{(F)}(t)
 =\mathcal{N}
 \int\!\!\FuncD[\Vect{F}] 
 \E^{-\beta^{(F)} \Sigma^{(F)}[\Vect{F}]}
 \Vect{F}(t,\Vect{r})
 \Vect{F}^\dagger(t,\Vect{r}^\prime)
\end{align}
for the Wightman tensor.}

Moreover, the normalization constant can be expressed~\cite{Peskin2018,Hatfield2018} as another functional integral, namely
\begin{align}\label{MAIN:eq:partition-sum}
    \mathcal{N}^{-1}=\mathcal{Z}^{(F)}[\Vect{F}] \equiv \int\!\!\FuncD[\Vect{F}] \exp(-\beta_F\Sigma[\Vect{F}])
\end{align}
which we have named $\mathcal{Z}^{(F)}$ to allude to \OrangelineChange{a} close analogy with the partition sum in statistical physics. 

In summary, \OrangelineChange{we obtain} the now purely functional expression
\begin{align}
\SecondOrderTensor{\mathbfscr{W}}_{\Vect{r}\Vect{r}^\prime}^{(F)}(t)
    \equiv
    \frac{\int\!\!\FuncD[\Vect{F}] \E^{-\beta_F\Sigma[\Vect{F}]}\Vect{F}(t,\Vect{r})\Vect{F}^\dagger(t,\Vect{r}^\prime)
    }{\mathcal{Z}^{(F)}[\Vect{F}]}
\end{align}
for the Wightman function $\SecondOrderTensor{\mathbfscr{W}}_{\Vect{r}\Vect{r}^\prime}^{(F)}$. This expression is the moment of a Gaussian functional integral~\cite{Peskin2018} and can\OrangelineChange{,} in principle\OrangelineChange{,} be computed, similar to its distant cousin -- the Gaussian integral in $\mathbb{R}^n$ \OrangelineChange{--} by completing the square and calculating a (functional) determinant. However, since we are dealing with a vector field and not the usual \OrangelineChange{case of a} scalar field~\cite{Hatfield2018}, things are a bit more complicated. Hence, we postpone this task, together with the detailed discussion of how the partition sum~\OrangelineChange{\cref{MAIN:eq:partition-sum}} may be used together with functional differentiation as a generating functional to calculate more complex correlation functions.

%%%%%%%%%%%%%%%%%%%%%%%%%%%%%%%%%%%%%%%%%%%%%%%%%%%%%%%%%%%%%%%%%%%%%%%%%%%%%%
% CONCLUSION START HERE
%%%%%%%%%%%%%%%%%%%%%%%%%%%%%%%%%%%%%%%%%%%%%%%%%%%%%%%%%%%%%%%%%%%%%%%%%%%%%%
\section{Conclusions}\label{MAIN:sec:conclusion}
Motivated by the thriving fields of cavity QED\OrangelineChange{,} and circuit QED we have analyzed the wave functional of the vacuum in a resonator. We have found expressions that are identical to those of free space discussed in the literature.

\OrangelineChange{At} first sight\OrangelineChange{,} this identity is surprising, since the two situations differ considerably in the way the frequency of the mode enters into the mode expansion. In the continuous superposition of free space, it is the integration variable governed by the wave number. In the discrete case of the resonator\OrangelineChange{,} the summation rather than the integration extends over the mode indices\OrangelineChange{,} which in turn determines the mode frequency in a nontrivial way. 

We \OrangelineChange{were} able to overcome this complication with the help of the introduction of the \OrangelineChange{square} root of the negative Laplacian. In this way we could express the mode sum by the double integral of a bilinear form of the fields and \OrangelineChange{of} a scalar kernel given by the Fourier integral of the function reflecting the difference \OrangelineChange{in} the dependence of the vacuum fields on the wave number. 

Moreover, our analysis emphasizes the important role of the choice of the modes. Although the modes have been eliminated in the wave functional, its form still depends on \OrangelineChange{them}. We have illustrated this phenomenon for the wave functional $\Psi[\Vect{A}]$ of the vector potential $\Vect{A}$ which involves either $\Vect{A}$ or $\nabla\times\Vect{A}$ resulting from the $\Vect{u}_\ell$- or $\Vect{w}_\ell$-modes. 

In hindsight of our calculation\OrangelineChange{,} one could argue, that this is not as surprising as one might have thought. Especially, since the wave functional for the \emph{quantum state} of the vacuum fields is most naturally expressed in the eigenmodes, as they correspond to the physical degrees of freedom that \emph{are quantized}. Once we retreat from employing an explicit mode expansion, all \OrangelineChange{the} information that is left to fix the \emph{quantum state} needs to be retained in the \emph{associated kernel}.

We conclude by noting that despite the beauty of the wave functional, we are not aware of any application \OrangelineChange{of evaluating}, for example, vacuum expectation values prevalent in QED. One elementary example \OrangelineChange{of} its usefulness could be the sum \OrangelineChange{of} modes appearing in the second moment of the displacement of an electron due to the vacuum electric field. This quantity determines \OrangelineChange{the Lamb shift} in the Welton picture~\cite{Welton1948} and leads to the Bethe logarithm.

Indeed, due to the integration of the \OrangelineChange{second-order} time derivative in the Lorentz equation\OrangelineChange{,} the displacement contains in the mode expansion of the electric field $\omega_\ell^{-2}$. Since we deal with the second moment the electric field appears in a bilinear way and actually $\omega_\ell^{-4}$ enters into the sum of the modes. 

\OrangelineChange{Moreover, the vacuum electric field is proportional to $\omega_\ell^{1/2}$, reducing due to the bilinearity of the second moment of the displacement in the field the power to $\omega_\ell^{-3}$.} When we replace the sum by an integration\OrangelineChange{,} the volume element contains \OrangelineChange{$\omega_\ell^2$} leaving us with $\omega_\ell^{-1}$\OrangelineChange{,} creating\OrangelineChange{,} after the integration\OrangelineChange{,} the Bethe logarithm. 

It would be interesting to see how this expression emerges from the use of the wave functional which would eliminate the need for performing the sum over the modes. For this purpose, we first note that the complication of the square of the frequencies appearing \OrangelineChange{in} the mode expansion of the free field as $\omega_\ell^{-2}$ can be removed by the use of the inverse of the negative Laplacian. Since we deal with the second moment\OrangelineChange{,} the electric field appears in a bilinear way\OrangelineChange{,} and the functional integration with respect to the wave functional should yield in a straight-forward way an expression for the displacement. 

The result we obtained for the (electric field) Wightman tensor might be a first step in such a direction\OrangelineChange{,} as its elements contain all the necessary correlation functions for such a calculation. However, it is implicitly expected that it also has a singular behavior in the coincidence limit due to it being a derivative of a transverse delta function.

Unfortunately, this topic goes beyond the scope of the present article and has to be postponed to \OrangelineChange{a} future publication.

%%%%%%%%%%%%%%%%%%%%%%%%%%%%%%%%%%%%%%%%%%%%%%%%%%%%%%%%%%%%%%%%%%%%%%%%%%%%%%
% ACKNOWLEDGMENTS START HERE
%%%%%%%%%%%%%%%%%%%%%%%%%%%%%%%%%%%%%%%%%%%%%%%%%%%%%%%%%%%%%%%%%%%%%%%%%%%%%%
\section*{Acknowledgments}
It is a great honor and pleasure for us to dedicate our article to \OrangelineChange{Professor} Iwo Bialynicki-Birula on the occasion of his 90th birthday. He has taught us to love the wave functional of the vacuum and thereby triggered our curiosity about the corresponding quantity in a resonator\OrangelineChange{,} which constitutes the topic of our paper. We are enormously grateful to him for numerous stimulating and illuminating discussions about this and other problems over the last decades. Since our first joint article~\cite{Bialynicki-Birula1993} on quantum phase uncertainties, we have learned so much from him and are proud to be his friends. Happy Birthday, Iwo, and many more healthy and happy years!

The authors \OrangelineChange{are grateful} to M.~Keck, N.~Rach, J.~Seiler, E.~Giese, A.~Wolf, R.~Lopp and Ch.~Ufrecht for many interesting and fruitful discussions. AF is grateful to R.~Lopp for pointing him to Ref.~\cite{Lopp2021}.

AF is grateful to the Carl Zeiss Foundation (Carl-Zeiss-Stiftung) and \OrangelineChange{IQST} for funding in terms of the project MuMo-RmQM.
The QUANTUS and INTENTAS projects are supported by the German Space Agency at the German Aerospace Center (Deutsche Raumfahrtagentur im Deutschen Zentrum f\"ur Luft- und Raumfahrt, DLR) with funds provided by the Federal Ministry for Economic Affairs and Climate Action (Bundesministerium f\"ur Wirtschaft und Klimaschutz, BMWK) due to an enactment of the German Bundestag under Grant Nos. 50WM2250D-2250E (QUANTUS+), as well as 50WM2177-2178 (INTENTAS).

%%%%%%%%%%%%%%%%%%%%%%%%%%%%%%%%%%%%%%%%%%%%%%%%%%%%%%%%%%%%%%%%%%%%%%%%%%%%%%
% APPENDICES START HERE
%%%%%%%%%%%%%%%%%%%%%%%%%%%%%%%%%%%%%%%%%%%%%%%%%%%%%%%%%%%%%%%%%%%%%%%%%%%%%%
\appendix
\counterwithout{equation}{section}
\renewcommand{\thesection}{\Alph{section}}
\renewcommand{\thesubsection}{\Alph{section}\arabic{subsection}}

\section{\OrangelineChange{Modes}}\label{APP:sec:mode-decomposition}
In this appendix, we briefly summarize the key ingredients of the description of the electromagnetic field in a resonator with discrete modes in the absence of charges and currents. \OrangelineChange{We} concentrate on the mode expansions and the energy of the electromagnetic field. Throughout this section and the article we employ the Coulomb gauge. Although these expressions are well-established, we present them here for the sake of completeness.

\subsection{Mode functions and amplitudes}\label{SEC:APP:Modes_and_Amplitudes}
Central to our review of the electromagnetic field in a resonator are the Maxwell equations consisting of the \OrangelineChange{two sets of equations}
\begin{align}\label{APP:eq:maxwell1}
\nabla \cdot \Vect{B} = 0
\quad \text{and} \quad
\nabla \times \Vect{E} 
=
-\frac{\partial\Vect{B}}{\partial t}
\end{align}
\OrangelineChange{and}
\begin{align}\label{APP:eq:maxwell2}
\nabla \cdot \Vect{E} = 0
\quad \text{and} \quad
\nabla \times \Vect{B} 
=
\frac{1}{c^2}\frac{\partial\Vect{E}}{\partial t}
\end{align}
in the absence of currents and charges, where $c$ denotes the speed of light.

We solve the homogenous equations by introducing the vector potential $\Vect{A}=\Vect{A}(t,\Vect{r})$ \OrangelineChange{in Coulomb} gauge
\begin{align}\label{APP:eq:coulomb-gauge}
    \nabla \cdot \Vect{A} = 0
\end{align}
and the ansatz
\begin{align}\label{APP:eq:efield-bfield-coulomb-gauge}
    \Vect{E} \equiv - \frac{\partial \Vect{A}}{\partial t}
    \quad \text{and} \quad 
    \Vect{B} \equiv \nabla \times \Vect{A}.
\end{align}
As a result,~\cref{APP:eq:maxwell2} implies the free-space wave equation 
\begin{align}\label{APP:eq:wave-equation}
    \Box \Vect{A}(t,\Vect{r})\equiv\qty[\frac{1}{c^2} \frac{\partial^2}{\partial t^2} - \VLaplace]\Vect{A}(t,\Vect{r})=0
\end{align}
for the vector potential $\Vect{A}\equiv\Vect{A}(t,\Vect{r})$, in the absence of currents and charges where $\VLaplace$ is the three-dimensional Laplacian. 

We emphasize that in the derivation of this wave equation\OrangelineChange{,} we have already used the Coulomb gauge \OrangelineChange{condition~\cref{APP:eq:coulomb-gauge} to simplify}
\begin{align}
    \OrangelineChange{
    \nabla \times \qty(\nabla \times \Vect{A}) = \nabla\qty(\nabla \cdot \Vect{A}) -\VLaplace \Vect{A}
    =-\VLaplace \Vect{A}
    .
    }
\end{align}

Next, we make the separation ansatz 
\begin{align}\label{APP:eq:separation-ansatz}
    \Vect{A}(t,\Vect{r})\equiv\mathcal{A}~q(t)\Vect{u}(\Vect{r})
\end{align}
\OrangelineChange{with a real} dimensionless spatial function $\Vect{u}=\Vect{u}(\Vect{r})$ and the real dimensionless time-dependent function $q=q(t)$. In order to ensure that $\Vect{A}$ has the appropriate units\OrangelineChange{,} we have introduced the constant $\mathcal{A}$. The vectorial nature of $\Vect{A}$ is contained in \OrangelineChange{the function} $\Vect{u}$.

When we substitute the \OrangelineChange{ansatz \cref{APP:eq:separation-ansatz} into} the wave equation, \cref{APP:eq:wave-equation}, we arrive at the Helmholtz equation 
\begin{align}\label{APP:eq:helmholtz-equation}
    \qty[\VLaplace+\qty(\frac{\omega}{c})^2]\Vect{u}(\Vect{r})=0,
\end{align}
and the harmonic oscillator equation
\begin{align}\label{APP:eq:harmonic-mode-oscillator-equation}
\ddot{q}+\omega^2 q = 0
\end{align}
with frequency $\omega$. \OrangelineChange{Here,} dots denote differentiation with respect to time.

We \OrangelineChange{emphasize that solutions} of the Helmholtz equation,~\cref{APP:eq:helmholtz-equation} become unique once we specify a proper boundary condition. \OrangelineChange{For example,} we could choose
\begin{align}\label{APP:eq:vector-potential-boundary-conditions}
    \Vect{n}(\Vect{r})\times \Vect{A}(t,\Vect{r}) \equiv \Vect{0}
\end{align}
for all points $\Vect{r}\in\partial \mathcal{V}$ making up the cavity walls, which corresponds to a perfectly conducting cavity surface $\partial \mathcal{V}$ \OrangelineChange{with normal vector $\Vect{n}(\Vect{r})$.}

When we apply the Coulomb gauge condition,~\cref{APP:eq:coulomb-gauge}, to the separation ansatz, \cref{APP:eq:separation-ansatz} we obtain the transversality constraint
\begin{align}\label{APP:eq:transversality-constraint}
    \nabla \cdot \Vect{u}(\Vect{r}) = 0.
\end{align}

While we work in the classical theory this constraint is no issue, but as \OrangelineChange{Paul Dirac} first noticed~\cite{Dirac2013}, it can come to haunt us when we quantize electromagnetism~\cite{Woolley2020,Stokes2021b,Stokes2022} or any other gauge field~\cite{Weinberg1995}.

The general solution of the harmonic oscillator equation, \cref{APP:eq:harmonic-mode-oscillator-equation} reads
\begin{align}\label{APP:eq:harmonic-mode-oscillator-solution}
    \OrangelineChange{
    q(t)=q_0 \cos{(\omega t)} + \frac{\dot{q}_0}{\omega} \sin{(\omega t)},
    }
\end{align}
where we have introduced the arbitrary initial conditions $q_0\equiv q(t=0)$ and \OrangelineChange{$\dot{q}_0\equiv \dot{q}(t=0)$}.

The time-derivative of $q$ leads us to the expression
\begin{align}\label{APP:eq:harmonic-mode-oscillator-pq-relation}
    \dot{q} = \omega p
\end{align}
with
\begin{align}
    \OrangelineChange{
     \quad p\equiv p(t)=-q_0\sin{(\omega t)}+\frac{\dot{q}_0}{\omega}\cos{(\omega t)}.
     }
\end{align}

The boundary conditions imposed by the resonator enforce a discrete set of mode functions $\Vect{u}$ of the vector potential enumerated by a set of three indices~\cite{Kakazu1995} determining an effective wave vector. Moreover, due to the Coulomb \OrangelineChange{gauge \cref{APP:eq:coulomb-gauge,APP:eq:transversality-constraint} we} find two polarization directions \OrangelineChange{for} $\Vect{u}$. 

For the sake of implementing a concise notation we abbreviate these indices consisting of wave vector \emph{and} polarization indices by a single quantity $\ell$, and use the set $\qty{\Vect{u}_\ell}$ for the eigenmodes of the vector potential.

\subsection{Vector potential}
As a result of the linearity of the wave \OrangelineChange{equation \cref{APP:eq:wave-equation} the}  vector potential $\Vect{A}$ in the resonator is the superposition
\begin{align}\label{APP:eq:vector-potential-separated-mode-expansion}
    \Vect{A}(t,\Vect{r})=\sum_{\ell} A_\ell(t) \Vect{u}_\ell(\Vect{r})
\end{align}
of all modes \OrangelineChange{$\qty{\Vect{u}_\ell}$ which are the (eigen)-mode functions of $\Vect{A}$. Here we} have introduced the abbreviation 
\begin{align}\label{APP:eq:vector-potential-mode-coefficient}
    A_\ell(t)\equiv \mathcal{A}_\ell q_\ell(t)
\end{align} 
for the vector potential contribution originating from the mode $\Vect{u}_\ell$.
 
The mode functions $\qty{\Vect{u}_\ell}$ of the vector potential form an orthonormal basis of transverse vector fields inside the resonator with the orthogonality relation 
\begin{align}\label{APP:eq:orthonormality-condition-of-modes}
    \frac{1}{\mathcal{V}_{\ell}}
    \int \!\! \D^3 r \; {\Vect{u}_{\ell}}^\dagger (\Vect{r}) \Vect{u}_{m}(\Vect{r}) = \delta_{\ell m}
\end{align}
\OrangelineChange{where $\mathcal{V}_\ell$ denotes the mode volume.}

\OrangelineChange{A more general definition for the mode volume is for example given by}
\begin{align}
    \OrangelineChange{
    \overline{\mathcal{V}}_\ell 
    \equiv
    \frac{\int \D^3 r ~\Abs{\Vect{u}_\ell (\Vect{r})}^2}{ \Abs{\Vect{u}_\ell(\Vect{r}_c)}^2}   }
\end{align}
where $\Vect{r}_c$ is a point of special interest of a given resonator. 

For example, in a box resonator with perfectly reflecting and conducting surfaces exhibiting sinusoidal modes one typically~\cite{Schleich2001} picks $\Vect{r}_c$ as the point of maximal mode amplitude. Alternatively, in the presence of an atomic dipole at a fixed location inside the cavity one can also use its position. Such choices can be directly linked to single-atom cavity QED analogues of the Purcell effect~\cite{Purcell1946}, that is the enhancement (or suppression) of the spontaneous emission rate of the dipole in a resonant cavity environment. For recent generalizations to more complicated systems and open cavities we refer to~\cite{Muljarov2016,Ren2021}. 

\subsection{Electric field}
Since there are no charges and currents \OrangelineChange{present, the electric field~\cref{APP:eq:efield-bfield-coulomb-gauge} in Coulomb gauge takes the explicit form}
 \begin{align}\label{APP:eq:electric-field-decomposition-modes}
    \OrangelineChange{
    \Vect{E}(t,\Vect{r}) = -\sum_\ell \mathcal{A}_\ell \dot{q}_\ell(t) \Vect{u}_\ell(\Vect{r}),
    }
 \end{align}
 \OrangelineChange{
 where we made use of the mode expansion of the vector potential~\cref{APP:eq:vector-potential-separated-mode-expansion}.
 }
 
\OrangelineChange{With the general solution~\cref{APP:eq:harmonic-mode-oscillator-solution} of the harmonic oscillator equation \cref{APP:eq:harmonic-mode-oscillator-equation}, and the connection \cref{APP:eq:harmonic-mode-oscillator-pq-relation} between $\dot{q}_\ell$ and $p_\ell$ we find}
\begin{align}
    \Vect{E}(t,\Vect{r})= \sum_\ell \mathcal{E}_\ell p_\ell(t) \Vect{u}_\ell(\Vect{r})
\end{align}
where we have introduced the relation
\begin{align}\label{APP:eq:electric-field-mode-coefficient}
    \mathcal{E}_\ell\equiv\mathcal{A}_\ell \omega_\ell.
\end{align}

Hence, the contribution \OrangelineChange{of each mode to the total electric field is} determined by the amplitude 
\begin{align}\label{APP:eq:electric-field-amplitude}
    E_\ell(t) \equiv \mathcal{E}_\ell p_\ell(t)
\end{align}
in the mode expansion
\begin{align}\label{APP:eq:electric-field-mode-decomposition}
    \Vect{E}(t,\Vect{r}) = \sum_\ell E_\ell(t) \Vect{u}_\ell(\Vect{r}).
\end{align}
A comparison of this expression to the expansion of the electric field
\begin{align}
    \Vect{E}(t,\Vect{r})=\sum_{\ell} E_\ell(t) \Vect{v}_\ell(\Vect{r})
\end{align}
in its eigenmodes $\qty{\Vect{v}_\ell}$, reveals that $\Vect{A}$ and the $\Vect{E}$ share the same set of eigenmodes. Consequently, the set $\qty{\Vect{u}_\ell}$ of modes of the vector potential can be mapped \emph{one--to--one} to the set $\qty{\Vect{v}_\ell}$ of eigenmodes of the electric \OrangelineChange{field.} We emphasize, that this property is only true in the absence of currents and charges, within and on the resonator boundary, because otherwise the wave equations for both fields $\Vect{A}$ and $\Vect{E}$ might differ in their boundary conditions \OrangelineChange{and thus} lead to different eigenmode expansions.

\subsection{Magnetic induction}
We conclude this discussion of the fields by presenting a similar representation for the magnetic induction $\Vect{B}$ in terms of the mode functions of the vector potential $\Vect{A}$. However, in contrast to the electric field $\Vect{E}$, \OrangelineChange{linked} to $\Vect{A}$ by differentiation in time, the field $\Vect{B}$ is linked to the vector potential by taking the curl, that is a coordinate derivative.

Indeed, we find from the definition $\Vect{B}\equiv\nabla\times\Vect{A}$ of $\Vect{B}$ in terms of $\Vect{A}$ given by ~\cref{APP:eq:vector-potential-separated-mode-expansion} the expression
\begin{align}
    \Vect{B}(t,\Vect{r}) = \sum_\ell \mathcal{A}_\ell q_\ell(t) \qty[\nabla\times\Vect{u}_\ell(\Vect{r})].
\end{align}

In order to bring out the analogy to $\Vect{E}$, we multiply and divide in the expansion the mode function by \OrangelineChange{$\omega_\ell/c$,} which yields
\begin{align}\label{APP:eq:magnetic-field-decomposition-modes}
    \Vect{B}(t,\Vect{r}) = \sum_\ell B_\ell(t) \frac{c}{\omega_\ell} \qty[\nabla\times \Vect{u}_\ell(\Vect{r})]   
\end{align}
where we have introduced the magnetic induction in the mode 
\begin{align}\label{APP:eq:magnetic-field-amplitude}
    B_\ell \equiv \mathcal{B}_\ell q_\ell(t)
\end{align}
 with the vacuum magnetic induction 
\begin{align}\label{APP:eq:magnetic-field-mode-coefficient}
    \mathcal{B}_\ell \equiv \frac{\mathcal{A}_\ell \omega_\ell}{c}  = \frac{\mathcal{E}_\ell}{c}.
\end{align}
 In the last step, we have recalled from~\cref{APP:eq:electric-field-mode-coefficient} the definition of the vacuum electric field. 

 \OrangelineChange{When we compare~\cref{APP:eq:magnetic-field-decomposition-modes} to the eigenmode expansion,}
 \begin{align}
     \Vect{B}(t,\Vect{r})
     =
     \sum_\ell
    B_\ell(t)\Vect{w_\ell}(\Vect{r})
 \end{align}
 of $\Vect{B}$\OrangelineChange{,} we can again find a \emph{one--to--one} mapping between eigenmodes. However, now we have to make the matching by comparing the expressions
 \begin{align}\label{APP:eq:eigenmode-link-Amodes-to-B-modes}
     \sum_\ell B_\ell(t) \frac{c}{\omega_\ell} \qty[\nabla\times\Vect{u}_\ell(\Vect{r})] \stackrel{!}{=}
     \sum_\ell
     B_\ell(t)\Vect{w_\ell}(\Vect{r}).
 \end{align}
 
When we \OrangelineChange{note that} there can be \OrangelineChange{\emph{no reshuffling}} of the sequence of mode indices since only the coefficient $B_\ell(t)$ contributes to the field energy, the eigenmodes of $\Vect{B}$ must be related to the eigenmodes of $\Vect{A}$ by making the identification
\begin{align}\label{APP:eq-link-ul-wl}
    \Vect{w}_\ell(\Vect{r}) \equiv \frac{c}{\omega_\ell} \qty[\nabla\times\Vect{u}_\ell(\Vect{r})].
\end{align}

However, when we recall that (eigen)-modes are determined by the boundary conditions resulting from~\cref{APP:eq-link-ul-wl}, this is not surprising. The magnetic induction has to fulfill different boundary conditions to be consistent with Maxwell's equations on the resonator surface. We \OrangelineChange{emphasize again} that our elementary treatment is valid only in the absence of currents and charges within and on the resonator surface. \OrangelineChange{Otherwise,} significant changes can arise. For more \OrangelineChange{details,} we \OrangelineChange{refer,} for \OrangelineChange{example,} to the classic text~\cite{Joannopoulos2008} on nano-photonics, or more recent work referenced therein.

\subsection{Determination of the vacuum field amplitude} 
In order to define the quantity $\mathcal{A}_\ell$\OrangelineChange{,} we recall from \OrangelineChange{Appendix}~\ref{APP:sec:field-energy-em-field} that the energy
\begin{align}\label{APP:eq:hamiltonian-density-electric-magnetic}
H(t) =\frac{1}{2}\int \!\!\! \D^3 r ~ \varepsilon_0\Big[\Vect{E}(t,\Vect{r})^2+ \big(c\Vect{B}(t,\Vect{r})\big)^2\Big]
\end{align}
of the electromagnetic field in the resonator takes the form
\begin{align}\label{APP:eq:hamiltonian-density-electric-magnetic-mode-decomposed}
    H =   \sum_\ell \varepsilon_0\mathcal{A}_\ell^2  \omega_\ell^2 \frac{\mathcal{V}_\ell}{2} \Big[p_\ell^2(t)+q_\ell^2(t)\Big]
\end{align}
where we \OrangelineChange{used} the \OrangelineChange{expansions \cref{APP:eq:electric-field-decomposition-modes,APP:eq:magnetic-field-decomposition-modes} for} $\Vect{E}$ and $\Vect{B}$. 

When we compare~\cref{APP:eq:hamiltonian-density-electric-magnetic-mode-decomposed} to the representation
\begin{align}
    H = \sum_\ell \frac{\hbar \omega_\ell}{2} \Big[p_\ell^2(t)+q_\ell^2(t)\Big]
\end{align}
of the total energy as a sum of all modes, where each mode contains the energy $\hbar \omega_\ell$ we obtain the explicit expression
\begin{align}\label{APP:eq:vacuum-vector-potential-amplitude}
    \mathcal{A}_\ell \equiv \sqrt{\frac{\hbar}{\varepsilon_0 \omega_\ell \mathcal{V}_\ell}}
\end{align}
for the amplitude $\mathcal{A}_\ell$ of the vector potential due to a single mode.

Due to the connection, \cref{APP:eq:electric-field-mode-coefficient}, between $\mathcal{E}_\ell$ and $\mathcal{A}_\ell$ we find the corresponding relation
\begin{align}\label{eq:vacuum-electic-field-amplitude}
    \mathcal{E}_\ell \equiv \sqrt{\frac{\hbar\omega_\ell}{\varepsilon_0 \mathcal{V}_\ell}}
\end{align}
for the electric field. In the quantized theory, discussed in~\cref{APP:sec:wavefunctional-groundstate}, $\mathcal{E}_\ell$ will become the amplitude of the vacuum field.

In \cref{tab:table1}, we summarize key features of the mode expansions based on the eigenmodes or the $\Vect{u}_\ell$-modes, such as the strength of the fields and the vacuum field amplitude in each mode. Here, we emphasize the different power laws of the mode frequency $\omega_\ell$ in $\mathcal{A}_\ell$,~$\mathcal{E}_\ell$ and~$\mathcal{B}_\ell$.

\subsection{Natural modes}
In this appendix we have expanded the three fields $\Vect{A}$, $\Vect{E}$ and $\Vect{B}$ into the modes $\Vect{u}_\ell$ of $\Vect{A}$. However, since we focus on a situation with no charges and \OrangelineChange{currents,} we can also express $\Vect{E}$ and $\Vect{B}$ in their natural modes $\Vect{v}_\ell$ and $\Vect{w}_\ell$. Indeed, $\Vect{E}$ and $\Vect{B}$ also satisfy the homogeneous wave equations, that is
\begin{align}
    \Box \Vect{E}(t,\Vect{r}) = \qty[\frac{1}{c^2}\frac{\partial^2}{\partial t^2}-\VLaplace]\Vect{E}(t,\Vect{r}) = 0
\end{align}
and 
\begin{align}
    \Box \Vect{B}(t,\Vect{r}) = \qty[\frac{1}{c^2}\frac{\partial^2}{\partial t^2}-\VLaplace]\Vect{B}(t,\Vect{r}) =0
\end{align}
following from the Maxwell equations,~\cref{APP:eq:maxwell1,APP:eq:maxwell2}, in the absence of currents and charges.

Needless to say, $\Vect{E}$ and $\Vect{B}$ have to obey boundary conditions imposed by the resonator, leading us to the natural modes $\Vect{f}_\ell=\Vect{f}_\ell(\Vect{r})$ defined by the Helmholtz equation
\begin{align}
    \qty(\VLaplace+k_\ell^2)\Vect{f}_\ell = 0,
\end{align}
and the boundary conditions \OrangelineChange{with $k_\ell=\omega_\ell/c$.}

For the sake of simplicity\OrangelineChange{,} we have not included in the modes $\Vect{f}_\ell$ a superscript $A$, $E$ or $B$ as~\OrangelineChange{to} express the fact that they depend on the choice of the field. Indeed, for $\Vect{A}$ and $\Vect{E}$ the natural modes are obviously $\Vect{u}_\ell$, \OrangelineChange{i.e.,}
\begin{align}
    \Vect{u}_\ell \equiv \Vect{f}_\ell^{(A)}
    =
    \Vect{f}_\ell^{(E)}
    \equiv \Vect{v}_\ell,
\end{align}
but for $\Vect{B}$\OrangelineChange{, we} find
\begin{align}
    \Vect{f}_\ell^{(B)} \equiv k_\ell^{-1}\qty(\nabla\times \Vect{u}_\ell)\equiv \Vect{w}_\ell.
\end{align}

The introduction of natural modes allows us to represent the mode expansions of all three fields by the single expression
\begin{align}
    \Vect{F}=\sum_\ell F_\ell \Vect{f}_\ell
\end{align}
where $\Vect{F}$ denotes either $\Vect{A}$, $\Vect{E}$ or $\Vect{B}$, and the modes $\Vect{f}_\ell$ depend on the choice of $\Vect{F}$.
\section{Field energy in a resonator}\label{APP:sec:field-energy-em-field}
In order to bring out most clearly the similarities and differences between the total energy $H$ of the radiation field and the mode sum $\Sigma^{(F)}$ defining the wave functional of the vacuum, and in particular, the difference in the powers of the frequency of the mode in $H$ and $\Sigma^{(F)}$, we re-derive in this appendix the energy
\begin{align}
    H= \frac{1}{2}\varepsilon_0 \int\D^3 r~\qty(\Vect{E}^2+\qty(c\Vect{B})^2)
\end{align}
of the electromagnetic field in a resonator in two slightly different ways: \OrangelineChange{First we calculate, in typical textbook fashion, the electric and magnetic contribution to the field energy and then use the previously defined eigenmodes of the field $\Vect{B}$ to find the magnetic contribution to the field energy.}

\subsection{Textbook quantum optics approach}
We begin with the textbook treatment, following along the lines of Ref.~\cite{Schleich2001}. The contribution
\begin{align}
    H^{(E)}\equiv \frac{1}{2}\varepsilon_0 \int ~\D^3 r~\Vect{E}^2
\end{align}
to $H$ due to the electric field
\begin{align}
    \Vect{E}=\sum_\ell E_\ell \Vect{u}_\ell
\end{align}
leads us immediately to the expression
\begin{align}
    H^{(E)}= \frac{1}{2}\varepsilon_0 \sum_{\ell,\ell^\prime}E_\ell E_{\ell^\prime}\int \D^3 r~\Vect{u}_\ell(\Vect{r})\cdot \Vect{u}_{\ell^\prime}(\Vect{r}),
\end{align}
which reduces with the orthonormality relation,~\cref{APP:eq:orthonormality-condition-of-modes} of the modes to
\begin{align}\label{APP:eq:electric-field-energy}
    H^{(E)}= \frac{1}{2}\varepsilon_0 \sum_\ell E_\ell^2 \mathcal{V}_\ell.
\end{align}

It is slightly more complicated to calculate the term 
\begin{align}\label{APP:eq:magnetic-field-energy}
\OrangelineChange{
H^{(B)}\equiv \frac{\varepsilon_0}{2}  \int~\D^3 r~\qty(c\Vect{B})^2
}
\end{align}
associated with the magnetic induction
\begin{align}\label{APP:eq:magnetic-field-mode-decomposition}
    \Vect{B}=\sum_\ell B_\ell\frac{c}{\omega_\ell} \qty[\nabla\times\Vect{u}_\ell].
\end{align}

\OrangelineChange{Indeed, when we substitute the mode representation \cref{APP:eq:magnetic-field-mode-decomposition} into $H^{(B)}$, given by \cref{APP:eq:magnetic-field-energy}, we find the expression}
\begin{align}\label{APP:eq:magnetic-field-mode-expansion-uncalculated}
    H^{(B)}=\frac{1}{2}\varepsilon_0 c^2 \sum_{\ell,\ell^\prime}
    B_\ell B_{\ell^\prime} \frac{c^2}{\omega_\ell \omega_{\ell^\prime}} \mathcal{J}_{\ell\ell^\prime},
\end{align}
where we have introduced the abbreviation
\begin{align}\label{APP:eq:integral-bfield}
    \OrangelineChange{
    \mathcal{J}_{\ell\ell^\prime} \equiv\int~\D^3 r~ \qty[\nabla\times\Vect{u}_\ell(\Vect{r})] \cdot \qty[\nabla\times\Vect{u}_{\ell^\prime}(\Vect{r})].
    }
\end{align}

\OrangelineChange{With the help of the identity proven in appendix~\ref{APP:sec:mathematical-complement}. The integrand in~\cref{APP:eq:integral-bfield} can be rewritten as}
\OrangelineChange{%
\begin{align}\label{APP:eq:mode-identity-for-transverse-modes}
    \qty[\nabla\times\Vect{u}_\ell]\cdot \qty[\nabla\times\Vect{u}_{\ell^\prime}] &=
    \nabla\cdot\qty[\Vect{u}_{\ell^\prime}\times(\nabla\times\Vect{u}_{\ell}] \notag \\  &\quad\quad\quad +\Vect{u}_{\ell^\prime}\cdot\qty[\nabla\times(\nabla\times\Vect{u}_\ell)]
\end{align}
}%
where the first term on the right hand side is a complete divergence. Hence, the application of the Gauss theorem converts the volume integral $\mathcal{J}_{\ell\ell^\prime}$ into a surface integral which vanishes due to the mode functions respecting the boundary conditions of the resonator.

The remaining term
\begin{align}\label{APP:eq:mode-identity-vlaplace-and-rotrot-of-Amodes}
    \OrangelineChange{
    \nabla\times(\nabla\times\Vect{u}_\ell)=\nabla(\nabla\cdot\Vect{u}_\ell)-\VLaplace \Vect{u}_\ell
    }
\end{align}
in~\cref{APP:eq:mode-identity-for-transverse-modes} reduces with the Coulomb gauge condition~\cref{APP:eq:transversality-constraint} and the Helmholtz wave \OrangelineChange{equation \cref{APP:eq:helmholtz-equation} to}
\begin{align}\label{APP:eq:mode-identity-rot-rot-of-Amodes}
    \qty[\nabla\times(\nabla\times\Vect{u}_\ell)]=\qty(\frac{\omega_\ell}{c})^2 \Vect{u}_\ell
\end{align}

Hence, the integral $\mathcal{J}_{\ell\ell^\prime}$, \OrangelineChange{given by~\cref{APP:eq:integral-bfield} yields}
\begin{align}
    \mathcal{J}_{\ell\ell^\prime}=\qty(\frac{\omega_\ell}{c})^2 \int~\D^3 r~ \Vect{u}_\ell(\Vect{r}) \cdot \Vect{u}_{\ell^\prime}(\Vect{r})=\qty(\frac{\omega_\ell}{c})^2 \mathcal{V}_\ell\delta_{\ell\ell^\prime},
\end{align}
where in the last step we have used the orthonormality relation,~\cref{APP:eq:orthonormality-condition-of-modes} of the mode functions.

Consequently, we arrive at the expression
\begin{align}\label{eq:app:magnetic-field-energy}
    H^{(B)}=\frac{1}{2}\varepsilon_0 \sum_\ell c^2 B_\ell^2 \mathcal{V}_\ell.
\end{align}
for the magnetic field energy, \cref{APP:eq:magnetic-field-mode-expansion-uncalculated}.

We conclude by combining the formulae for the electric $H^{(E)}$ and magnetic part $H^{(B)}$ given by \cref{APP:eq:electric-field-energy} and \cref{eq:app:magnetic-field-energy}, and arrive at the representation
\begin{align}\label{APP:eq:full:hamiltonian-mode-decomposed}
    H =\frac{1}{2}\sum_\ell \mathcal{A}_\ell^2 \varepsilon_0 \omega_\ell^2 \mathcal{V}_\ell\qty(p_\ell^2+q_\ell^2)
\end{align}
of the energy in terms of modes. Here, we have recalled the definitions \OrangelineChange{\cref{APP:eq:electric-field-amplitude} and \cref{APP:eq:magnetic-field-amplitude} of $E_\ell$
and $B_\ell$ respectively, together with the connections \cref{APP:eq:electric-field-mode-coefficient} and \cref{APP:eq:magnetic-field-mode-coefficient}.}

\subsection{Magnetic field energy via eigenmodes}
When we recall our discussion of the respective eigenmodes of $\Vect{E}$ and $\Vect{B}$, and their relation to the eigenmodes of $\Vect{A}$, one might think that we could have avoided the cumbersome calculation of the scalar product of the curls of the modes entirely. However, this suspicion is not quite true, and to show why, we perform the relevant calculation in this section. 

When we expand the magnetic induction in its eigenmodes \OrangelineChange{$\qty{\Vect{w}_\ell}$,} we directly obtain for the magnetic field energy, \cref{APP:eq:magnetic-field-energy}, the expression
\begin{align}
    H^{(B)}
    =
    \frac{\varepsilon_0 }{2}  
    \sum_{\ell,\ell^\prime} c^2 B_\ell B_{\ell^\prime}
    \int~\D^3 r~ 
    {\Vect{w}_\ell}^\dagger(\Vect{r}) \Vect{w}_{\ell^\prime}(\Vect{r})
\end{align}

\OrangelineChange{Next,} we make use of the orthonormality of the eigenmodes $\Vect{w}_\ell$, \OrangelineChange{i.e.,}
\begin{align}
    \frac{1}{\tilde{\mathcal{V}}_\ell}
    \int~\D^3 r~ 
    {\Vect{w}_\ell}^\dagger(\Vect{r}){\Vect{w}_{\ell^\prime}}(\Vect{r}) 
    = 
    \delta_{\ell \ell^\prime}
\end{align}
which leads us to the preliminary result
\begin{align}
    H^{(B)}
    =
    \frac{\varepsilon_0 }{2}  
    \sum_{\ell} c^2 B_{\ell}^2 \tilde{\mathcal{V}}_\ell .
\end{align}

We emphasize \OrangelineChange{that,} instead of the mode volume $\mathcal{V}_\ell$ of the vector potential modes $\Vect{u}_\ell$, the mode volume $\tilde{\mathcal{V}}_\ell$ corresponding to the eigenmodes $\Vect{w}_\ell$ of $\Vect{B}$ has appeared. Hence, if one wants to express the total field energy $H$ solely in terms of one mode \OrangelineChange{volume,} a connection between $\mathcal{V}_\ell$ and $\tilde{\mathcal{V}}_\ell$ is needed. 

However, the only link available between  the eigenmodes $\Vect{w}_\ell$ and $\Vect{u}_\ell$ \OrangelineChange{is \cref{APP:eq-link-ul-wl}, i.e.,}
\begin{align}
    \Vect{w}_\ell(\Vect{r}) = \frac{c}{\omega_\ell} \qty[\nabla\times\Vect{u}_\ell(\Vect{r})].
\end{align}
When we take the scalar product of this equation with itself and integrate over the resonator volume we obtain the relation
\begin{align}
     \tilde{\mathcal{V}}_\ell 
     = 
     \int ~\D^3 r~ 
     \Abs{\Vect{w}_\ell(\Vect{r})}^2 
     = \frac{c^2}{\omega_\ell^2} 
     \int ~\D^3 r~ \qty[\nabla\times\Vect{u}_\ell(\Vect{r})]\cdot\qty[\nabla\times\Vect{u}_\ell(\Vect{r})].
\end{align}
The integrand on the right-hand side of this equation is an old acquaintance of \OrangelineChange{ours --~\cref{APP:eq:integral-bfield} evaluated} at $\ell=\ell^\prime$. 

Hence, even in the approach with the eigenmodes ultimatly no true simplification is gained, but it is just a slightly different detour. As a consequence, we again need to \OrangelineChange{apply \cref{APP:eq:mode-identity-for-transverse-modes,APP:eq:mode-identity-vlaplace-and-rotrot-of-Amodes,APP:eq:mode-identity-rot-rot-of-Amodes} to} simplify the scalar product of the two curls, and we obtain
\begin{align}
    \OrangelineChange{
     \tilde{\mathcal{V}}_\ell 
     =
     \int \D^3 r~ 
     \Abs{\Vect{w}_\ell(\Vect{r})}^2 
     = 
     \qty(\frac{\omega_\ell}{c})^2
     \qty(\frac{c}{\omega_\ell})^2 \!\!
     \int~\D^3 r~ 
     \Abs{\Vect{u}_\ell(\Vect{r})}^2  
     = 
     \mathcal{V}_\ell,
     }
\end{align}
where we have made use of \OrangelineChange{\cref{APP:eq:orthonormality-condition-of-modes} definining the mode volume of the vector potential modes $\Vect{u}_\ell$.}

As a consequence of the identity $\mathcal{V}_\ell=\tilde{\mathcal{V}}_\ell$, we also arrive at the expression
\begin{align}
    H^{(B)}
    =
    \frac{1}{2}\varepsilon_0  
    \sum_{\ell}  c^2 B_{\ell}^2 \tilde{\mathcal{V}}_\ell
    =
    \frac{\varepsilon_0 }{2}  
    \sum_{\ell}  c^2 B_{\ell}^2 \mathcal{V}_\ell
\end{align}
for the field energy $H^{(B)}$ due to the magnetic induction. 

As an afterthought we note that
naively one could have imagined that the mode volumes might be defined independently such that they differ by a numeric factor -- maybe via choosing different reference points in their respective definition of the mode volume. However, then the expression for the \OrangelineChange{Hamiltonian \cref{APP:eq:full:hamiltonian-mode-decomposed} would} be rescaled in the mode oscillator coordinate $q_\ell$ corresponding to the magnetic field \OrangelineChange{by} the \OrangelineChange{factor $\tilde{\mathcal{V}}_\ell/\mathcal{V}_\ell$.} In turn, this feature would lead to problems in the Hamilton equations of motion since the symmetry between $q_\ell$ and $p_\ell$ would be broken leading to a rescaled Poisson \OrangelineChange{bracket. This would} directly impact quantization by also rescaling the commutator $[\QmOp{q}_\ell,\QmOp{p}_\ell]=\I$ by the factor $\tilde{\mathcal{V}}_\ell/\mathcal{V}_\ell$ which is undesirable. Nevertheless, we note that the simple argument we have formulated here might not be as \OrangelineChange{clear-cut} when complicated boundary conditions \OrangelineChange{enter, or open resonators in the presence of currents and charges are considered.}
\section{Wave function representations of the ground state}\label{APP:sec:wavefunctional-groundstate}
In this appendix we derive the wave function $\psi_\ell$ of the ground state of the electromagnetic field in the modes $\Vect{u}_\ell$, $\Vect{v}_\ell$ or $\Vect{w}_\ell$ specified by the mode index $\ell$ and the field. \OrangelineChange{Indeed, for the vector potential $\Vect{A}$ and electric field $\Vect{E}$ the eigenmodes are $\Vect{u}_\ell$.} However, for the magnetic induction $\Vect{B}$ they are $\Vect{w}_\ell \equiv k_\ell^{-1} \nabla \times\Vect{u}_\ell$.

Although the material in this appendix is partially contained in standard textbooks on quantum optics~\cite{Schleich2001}\OrangelineChange{, we} find it useful to include it in our article to gain a complete understanding of the origin and the form of the dimensionless arguments of the Gaussian ground state wave function in the different representations. We first address in detail the case of $\Vect{E}$, and then turn briefly to the analogous calculations for $\Vect{B}$ and $\Vect{A}$.

\subsection{Electric field representation}\label{APP:sec:electricfield-amplitude}
We start from the mode decomposition
\begin{align}
    \Vect{E}(t,\Vect{r}) = \sum_\ell \mathcal{E}_\ell p_\ell(t) \Vect{u}_\ell(\Vect{r})
\end{align}
\OrangelineChange{of the electric field and make the transition to quantum mechanics, namely to the electric field operator $\QmOp{\Vect{E}}$, by promoting the dimensionless amplitude functions $q_\ell$ and $p_\ell$ of the harmonic field oscillator of the $\ell$-th mode defined by the mode function $\Vect{u}_\ell$ to operators
$q_\ell\mapsto\QmOp{q}_\ell$  and $p_\ell\mapsto\QmOp{p}_\ell$,} and demanding the canonical commutation relations
\begin{align}\label{APP:eq:canonical-commutation-relation}
    \qty[\QmOp{p}_\ell,\QmOp{q}_{\ell^\prime}]=\frac{1}{\I}\delta_{\ell\ell^\prime}.
\end{align}

Hence,~$\QmOp{\Vect{E}}$ takes the form
\begin{align}
    \QmOp{\Vect{E}}(t,\Vect{r}) =\sum_\ell \QmOp{E}_\ell(t)\Vect{u}_\ell(\Vect{r})
\end{align}
with
\begin{align}
    \QmOp{E}_\ell(t)\equiv\mathcal{E}_\ell \QmOp{p}_\ell(t)
\end{align}
which forces us to introduce a quantum state space for each mode.

A representative state could be, for example, the eigenstate $\Ket{E_\ell}$ defined by the eigenvalue equation
\begin{align}\label{APP:eq:electric-field-eigenvalue-equation}
    \QmOp{E}_\ell\Ket{E_\ell} \equiv E_\ell\Ket{E_\ell}
\end{align}
for the electric field operator, where $E_\ell\equiv \mathcal{E}_\ell q_\ell$ corresponds to the eigenvalue. Thus, $\Ket{E_\ell}$ describes a state where the electric field in the $\ell$-th mode assumes the well-defined value $E_\ell$.

The ground state $\Ket{0_\ell}$ of the $\ell$-th field oscillator is determined by the condition
\begin{align}\label{APP:eq:electric-field-groundstate-condition}
    \QmOp{a}_\ell \Ket{0_\ell} = 0,
\end{align}
where the linear combination
\begin{align}\label{APP:eq:electric-field-annihiliation-operator}
    \QmOp{a}_\ell \equiv \frac{1}{\sqrt{2}}\qty(\QmOp{q}_\ell+\I \QmOp{p}_\ell)
\end{align}
of $\QmOp{q}_\ell$ and $\QmOp{p}_\ell$ represents the annihilation operator $\QmOp{a}_\ell$.

When we now substitute the expression for $\QmOp{a}_\ell$ given by \cref{APP:eq:electric-field-annihiliation-operator} into the definition, \cref{APP:eq:electric-field-groundstate-condition}, of the ground state, and multiply by the bra-vector $\Bra{E_\ell}$ we arrive at the \OrangelineChange{equation}
\begin{align}
    \Bra{E_\ell} \QmOp{q}_\ell+\I \QmOp{p}_\ell \Ket{0_\ell}=0
\end{align}
\OrangelineChange{determining the ground state wave function
\begin{align}
    \psi_\ell(E_\ell)\equiv \Braket{E_\ell|0_\ell},
\end{align}
in the electric field representation, which corresponds to the first order differential equation
\begin{align}\label{APP:eq:electric-field-groundstate-diffeq}
    \qty[-\frac{1}{\I} \frac{\D}{\D (E_\ell/\mathcal{E}_\ell)}  + \I (E_\ell/\mathcal{E}_\ell) ]\psi_\ell(E_\ell) = 0.
\end{align}
}

Here we have used the fact that, according to ~\cref{APP:eq:electric-field-eigenvalue-equation}, $\Ket{E_\ell}$ is an eigenstate of $\QmOp{E}_\ell$\OrangelineChange{, and therefore} of $\QmOp{p}_\ell$, \OrangelineChange{leading us to} the identifications
\begin{align}
    \QmOp{p}_\ell \mapsto p_\ell
    \quad \text{and} \quad
    \QmOp{q}_\ell \mapsto -\frac{1}{\I}\frac{\D}{\D p_\ell}
\end{align}
to satisfy the canonical commutation relation, \cref{APP:eq:canonical-commutation-relation}. \OrangelineChange{Moreover, in \cref{APP:eq:electric-field-groundstate-diffeq} we have expressed} the derivative with respect to $p_\ell$ by $\mathcal{E}_\ell p_\ell \equiv E_\ell$.

Hence, we arrive at the Gaussian wave function
 \begin{align}\label{APP:eq:electric-field-groundstate-distribution}
     \OrangelineChange{
     \psi_\ell (E_\ell) = \mathcal{N}_\ell^{(E)} \exp[-\frac{1}{2}\qty(\frac{E_\ell}{\mathcal{E}_\ell})^2],
     }
 \end{align}
where the normalization constant 
\begin{align}
    \OrangelineChange{
    \mathcal{N}_\ell^{(E)}\equiv \frac{1}{\sqrt[4]{\pi}\sqrt{\mathcal{E}_\ell}}
    }
\end{align}
follows from the condition
\begin{align}
    \int_{\mathrlap{-\infty}}^{\mathrlap{\infty}} ~\D{E_\ell} ~\Abs{\psi_\ell(E_\ell)}^2 =1,
\end{align}
imposed by the Born interpretation.

\subsection{Magnetic induction representation}
\label{APP:sec:magneticfield-amplitude}
Next we turn to the magnetic induction $\Vect{B}$ where the corresponding operator reads
\begin{align}
    \QmOp{\Vect{B}}(t,\Vect{r})= \sum_\ell \QmOp{B}_\ell(t) \Vect{w}_\ell(\Vect{r})
\end{align}
with 
\begin{align}
\OrangelineChange{\QmOp{B}_\ell(t)\equiv \mathcal{B}_\ell \QmOp{q}_\ell(t).}
\end{align}
\OrangelineChange{This decomposition} leads us to the eigenvalue equation
\begin{align}
    \QmOp{B}_\ell \ket{B_\ell} = B_\ell \ket{B_\ell}
\end{align}
for the state $\ket{B_\ell}$ of a well-defined value $B_\ell$ of the magnetic induction 
$\Vect{B}$ in the $\ell$-th mode $\Vect{w}_\ell(\Vect{r})\equiv k_\ell^{-1} (\nabla \times \Vect{u}_\ell)$. 
Here, similarly to the electric field case, the expression
\begin{align}
    B_\ell\equiv \mathcal{B}_\ell q_\ell
\end{align}
denotes the eigenvalue. 

Indeed, in this representation we have to make the identification
\begin{align}
    \QmOp{p}_\ell \mapsto \frac{1}{\I} \frac{\D}{\D q_\ell}
    \quad \text{and} \quad
    \QmOp{q}\mapsto q_\ell
\end{align}
leading us directly to the differential equation
\begin{align}\label{APP:eq:magnetic-field-groundstate-diffeq}
    \frac{\D}{\D (B_\ell/\mathcal{B}_\ell)} \psi_\ell(B_\ell) = - (B_\ell/\mathcal{B}_\ell) \psi_\ell(B_\ell)
\end{align}
for the wave function
\begin{align}
    \psi_\ell(B_\ell) \equiv \Braket{B_\ell|0_\ell}
\end{align}
of the ground state of the $\ell$-th mode in the magnetic induction representation. 

The differential equation,~\cref{APP:eq:magnetic-field-groundstate-diffeq} also admits a solution in form of a Gaussian
\begin{align}\label{APP:eq:magnetic-field-groundstate-distribution}
    \psi_\ell(B_\ell) \equiv \mathcal{N}_{\ell}^{(B)} \exp[-\frac{1}{2}\qty(\frac{B_\ell}{\mathcal{B}_\ell})^2]
\end{align}
with the normalization constant
\begin{align}
    \OrangelineChange{
    \mathcal{N}_\ell^{(B)}\equiv\frac{1}{\sqrt[4]{\pi}\sqrt{\mathcal{B}_\ell}}.
    }
\end{align}

The only difference \OrangelineChange{from} the electric field representation discussed in the preceding section is the fact that the eigenstates $\Ket{B_\ell}$ are now, apart from the vacuum fields $\mathcal{B}_\ell$, eigenstates of $\QmOp{q}_\ell$ rather than of $\QmOp{p}_\ell$.

\subsection{Vector potential representation}
\label{APP:sec:vectorpotential-amplitude}
We conclude by briefly \OrangelineChange{discussing} the vector potential representation
\begin{align}
    \psi_\ell(A_\ell) \equiv \Braket{A_\ell|0_\ell}
\end{align}
of the ground state wave function in the $\ell$-th mode resulting from the operator
\begin{align}
    \QmOp{A}(t,\Vect{r})\equiv \sum_\ell \QmOp{A}_\ell(t) \QmOp{u}_\ell(\Vect{r})
\end{align}
of the vector potential with
\begin{align}
    \QmOp{A}_\ell(t) \equiv \mathcal{A}_\ell \QmOp{q}_\ell(t).
\end{align}

Since the operator $\QmOp{A}_\ell$ like $\QmOp{B}_\ell$ is also proportional to $\QmOp{q}_\ell$ we find immediately
\begin{align}\label{eq:app:vector-potential-groundstate-distribution}
    \psi_\ell(A_\ell) = \mathcal{N}_\ell^{(A)} \exp[-\frac{1}{2}\qty(\frac{A_\ell}{\mathcal{A}_\ell})^2]
\end{align}
with the normalization constant
\begin{align}
    \OrangelineChange{
    \mathcal{N}_\ell^{(A)}\equiv \frac{1}{\sqrt[4]{\pi}\sqrt{\mathcal{A}_\ell}},
    }
\end{align}
in complete analogy to the \OrangelineChange{distributions \cref{APP:eq:electric-field-groundstate-distribution} and \cref{APP:eq:magnetic-field-groundstate-distribution} in} the electric field and magnetic induction \OrangelineChange{variables $E_\ell$,} and $B_\ell$, respectively.
\section{Reduction scheme for the kernel}\label{APP:sec:reductionscheme}

In the main body of this article we have derived an exact expression for the kernel $\SecondOrderTensor{\mathbfscr{K}}$ of the wave functional of the vacuum in a resonator represented by the field $\Vect{F}$ in terms of the natural modes $\Vect{f}_\ell$. \OrangelineChange{According to
~\cref{MAIN:eq:kernel-as-deltatransverse-derivative} this kernel is a matrix, defined by the action of the function $F$ containing the root of the negative Laplacian on the transverse delta function.}
Since the fields in the double integral are transverse, we can replace it \OrangelineChange{with} the familiar Dirac delta function of free space. As a result, the kernel reduces to a scalar \OrangelineChange{$\mathbfscr{K}^{(F)}$}.

In this appendix we rederive the expression for the scalar kernel from a slightly different perspective. From the \OrangelineChange{outset,} we assume the kernel to be a scalar in the form of a Fourier representation of a root of the negative Laplacian. We first obtain an exact expression for the double integral containing the bilinear form of a field $\Vect{F}$ and the scalar kernel $\mathbfscr{K}^{(F)}$ \OrangelineChange{expressed as} a single integral of the square of $F$ containing the \OrangelineChange{fourth} root of the negative Laplacian acting on $\Vect{F}$. Then we evaluate this integral for a given mode representation and match the result with the formula for the mode sum.

This procedure yields the individual scalar kernels. We conclude by comparing and contrasting this approach to the diagonal and non-diagonal representation of the density operator in terms of coherent states, and given by the $P$- and $R$-distribution ~\cite{Glauber1963}, respectively.

\subsection{A general identity for Fourier transformable kernels}
We now verify the identity
\begin{align}\label{APP:eq:G.GprimeK}
    \OrangelineChange{
    \tilde{\mathcal{I}}^{(F)}\equiv\int\!\!\!\D^3 r \!\! \int\!\!\!\D^3 r^\prime 
    \Vect{F}\cdot\Vect{F}^\prime
    \mathbfscr{K}^{(F)}(\Vect{r}-\Vect{r}^\prime)
    =
    \int \D^3 r \qty|\FlabelNoIdx\qty(\sqrt[4]{-\VLaplace})\Vect{F}|^2
    }
\end{align}
for a vector field $\Vect{F}=\Vect{F}(t,\Vect{r})$, \OrangelineChange{where} the kernel
\begin{align}
    \mathbfscr{K}^{(F)}(\Vect{r})
    \equiv
    \frac{1}{(2\pi)^3}
    \int \D^3 k
    ~\FlabelNoIdx(k)
    \E^{\I \Vect{k}\cdot\Vect{r}}
\end{align}
appears in the double integral with the difference $\Vect{r}-\Vect{r}^\prime$ of the integration variables $\Vect{r}$ and $\Vect{r}^\prime$. \OrangelineChange{Here $F$ is not a generic scalar function but the function $\FlabelNoIdx(k)=1/k$ or $\FlabelNoIdx(k)=k$} appearing in the mode sum $\Sigma^{(F)}$, defined by~\cref{MAIN:eq:sigmaF-exponential-sum}, and given for $\Vect{E}$ and $\Vect{B}$ by~\cref{MAIN:eq:FEB-definition}, and for $\Vect{A}$ by~\cref{MAIN:eq:FA-definition}.

Central to the \OrangelineChange{relation~\cref{APP:eq:G.GprimeK}  is} the eigenvalue~\OrangelineChange{equation ~\cref{MAIN:eq:vlaplace-of-plane-wave} of} \OrangelineChange{$\E^{\I \Vect{k}\Vect{r}}$} leading us immediately to the representation
\begin{align}\label{APP:eq:KFkernelondelta}
    \OrangelineChange{
    \mathbfscr{K}^{(F)}(\Vect{r})=\FlabelNoIdx\qty(\sqrt{-\VLaplace_{\Vect{r}}})\delta(\Vect{r})
    }
\end{align}
where we have recalled the Fourier representation,~\cref{APP:eq:dirac-delta-definition} of the Dirac delta function.

When we \OrangelineChange{substitute~\cref{APP:eq:KFkernelondelta} into} the left--hand side of ~\cref{APP:eq:G.GprimeK} we arrive at the expression
\OrangelineChange{%
\begin{align}
    \tilde{\mathcal{I}}^{(F)}
    &\equiv 
    \int_{\mathrlap{}} \D^3 r
    \int_{\mathrlap{}} \D^3 r^\prime~
    \Vect{F} \cdot 
    \Big[\FlabelNoIdx\qty(\sqrt[4]{-\VLaplace_{\Vect{r}}}) \notag \\
    &\quad \quad \quad     
    \FlabelNoIdx\qty(\sqrt[4]{-\VLaplace_{\Vect{r}^\prime}})\delta(\Vect{r}-\Vect{r}^\prime)
    \Big]
    \Vect{F}^\prime.
\end{align}
}%
Here we have used the relation
\begin{align}\label{APP:eq:delta-factoring-laplacian}
    \OrangelineChange{
    \FlabelNoIdx\qty(\sqrt{-\VLaplace_{\Vect{r}}})\delta(\Vect{r}-\Vect{r}^\prime)
    =
    \FlabelNoIdx\qty(\sqrt[4]{-\VLaplace_{\Vect{r}}})\FlabelNoIdx\qty(\sqrt[4]{-\VLaplace_{\Vect{r}^\prime}})
    \delta(\Vect{r}-\Vect{r}^\prime),
    }
\end{align}
which is \OrangelineChange{only true for $\FlabelNoIdx(k)=1/k$ and $\FlabelNoIdx(k)=k$} and follows from the fact that the delta function is in the difference of the integration variables, \OrangelineChange{i.e.,} $\Vect{r}-\Vect{r}^\prime$.

\OrangelineChange{When we recall that the field $\Vect{F}$ vanishes outside of the resonator we can integrate both integrals by part. As a result, we arrive at the representation}
\OrangelineChange{%
\begin{align}\label{APP:eq:kernel-integral-indentity}
    \tilde{\mathcal{I}}^{(F)}
    &=
    \int \D^3 r
    \int \D^3 r^\prime ~ \delta(\Vect{r}-\Vect{r}^\prime)
     \notag \\
    &\quad \quad \times
    \qty[\FlabelNoIdx\qty(\sqrt[4]{-\VLaplace_{\Vect{r}}})\Vect{F}]
    \cdot
    \qty[\FlabelNoIdx\qty(\sqrt[4]{-\VLaplace_{\Vect{r}^\prime}})\Vect{F}^\prime]
\end{align}
}%
\OrangelineChange{of the integral $\tilde{\mathcal{I}}^{(F)}$. The Dirac delta function} allows us now to reduce the double integral into a single one leading us to the \OrangelineChange{identity~\cref{APP:eq:G.GprimeK}}.

\subsection{Evaluation of the integral}
Next we evaluate the integral on the right-hand side of the identity,~\cref{APP:eq:G.GprimeK}, using the expansion
\begin{align}
    \Vect{F}=\sum_\ell F_\ell \Vect{f}_\ell
\end{align}
of $\Vect{F}$ into the natural modes $\Vect{f}_\ell$, and find
\OrangelineChange{%
\begin{align}\label{APP:eq:equality-to-mode-sum-of-frac-laplace-kernel}
    \int \D^3 r \qty|
    \FlabelNoIdx\qty(\sqrt[4]{-\VLaplace})\Vect{F}
    |^2
    =
    \sum_\ell
    F_\ell^2 \FlabelNoIdx(k_\ell) \mathcal{V}_\ell = \Sigma^{(F)},
\end{align}
}%
where we have used the \OrangelineChange{identity~\cref{MAIN:eq:Laplacian-on-mode-function} for} the action of the fourth root of the negative Laplacian on $\Vect{f}_\ell$, and the orthonormality relation, ~\cref{MAIN:eq:mode-expansion-fmodes-orthonormality}. \OrangelineChange{In the last two steps in~\cref{APP:eq:equality-to-mode-sum-of-frac-laplace-kernel}, we used the identities $\FlabelNoIdx(k)=1/k$ and $\FlabelNoIdx(k)=k$ and have recalled the definition ~\cref{MAIN:eq:sigmaF-exponential-sum} of the mode sum $\Sigma^{(F)}$.}

Together with the identity, ~\cref{APP:eq:G.GprimeK}, we finally arrive at the relation
\begin{align}
    \Sigma^{(F)}
    =
    \int \D^3 r
    \int \D^3 r^\prime ~
    \Vect{F} \cdot \Vect{F}^\prime ~\mathbfscr{K}^{(F)}(\Vect{r}-\Vect{r}^\prime)
\end{align}
with the kernels
\begin{align}
    \mathbfscr{K}^{(A)}(\Vect{r}) 
    \equiv    
    \frac{1}{(2\pi)^3} \int \D^3 k ~k ~\E^{\I \Vect{k}\Vect{r}}
\end{align}
and
\begin{align}
    \mathbfscr{K}^{(E/B)}(\Vect{r})
    \equiv    
    \frac{1}{(2\pi)^3} \int \D^3 k ~\frac{1}{k}~\E^{\I \Vect{k}\Vect{r}},
\end{align}
in complete agreeement with the derivation in~\cref{MAIN:sec:bilinear-forms-and-kernels}.

\subsection{A curious analogy}
This approach is reminiscent of the representation~\cite{Schleich2001} of the density operator $\QmOp{\varrho}$ in terms of coherent states. By multiplying the completeness relation of the coherent states from the left and from the right onto the density operator $\QmOp{\varrho}$, we obtain the non-diagonal representation
\begin{align}
    \QmOp{\varrho} = \frac{1}{\pi^2} \int\D^2 \alpha \int\D^2 \beta ~\ket{\alpha}\braket{\alpha|\QmOp{\varrho}|\beta} \bra{\beta}
\end{align}

When we compare this expression to the corresponding one of the double integral \OrangelineChange{$\tilde{\mathcal{I}}^{(F)}$}, given by \cref{APP:eq:G.GprimeK} we note three similarities: (\textit{i}) \OrangelineChange{the two} different coherent states $\ket{\alpha}$ and $\ket{\beta}$ play the role of the fields $\Vect{F}$ and $\Vect{F}^\prime$, (\textit{ii}) \OrangelineChange{the} matrix element $\braket{\alpha|\QmOp{\varrho}|\beta}$ corresponds to the kernel, and (\textit{iii}) \OrangelineChange{the two integrations} over the coherent states translate into the double integral over the coordinates. 

Needless to say, there are also fundamental differences between the two expressions. For example, the coherent states live in state space and describe the \emph{quantum mechanics} of a \emph{single mode}. In contrast, the bilinear form involves the \emph{classical total fields}. Nevertheless, in both cases, the states and the fields are associated with vector spaces and therefore take advantage of similar mathematical tools.

Roy Glauber and George Sudarshan, independently, introduced the diagonal representation
\begin{align}
    \OrangelineChange{
    \QmOp{\varrho} = \frac{1}{\pi} \int\D^2 \alpha ~P(\alpha)\ket{\alpha}\bra{\alpha}.
    }
\end{align}    
of the density operator $\QmOp{\varrho}$ involving the $P$-distribution.

In our problem, this concept corresponds to the right-hand side of ~\cref{APP:eq:G.GprimeK} \OrangelineChange{which, according to ~\cref{APP:eq:equality-to-mode-sum-of-frac-laplace-kernel},} is identical to the mode sum $\Sigma^{(F)}$ which only contains the squares of the field strength and is therefore diagonal. This transition from a non-diagonal to a diagonal representation is made possible by derivatives acting on delta functions. Indeed, the $P$-distribution of a coherent state is already a Dirac delta function, and non-classical states are more singular~\cite{Schleich2001}.
\section{Explicit expressions for kernels}\label{APP:sec:explicit-j-kernels}
In this \OrangelineChange{appendix,} we derive an explicit expression for the kernel
\begin{align}\label{APP:eq:definition-jkernel}
    \mathbfscr{K}^{(j)}(\Vect{r})
    \equiv
    \frac{1}{(2\pi)^3}
    \int \D^3 k k^j \E^{\I \Vect{k}\Vect{r}},
\end{align}
and consider especially the two cases $j=1$ and $j=-1$ corresponding to
$\mathbfscr{K}^{(A)}$ and $\mathbfscr{K}^{(E/B)}$. 

We note, that while we formally calculate the integral for all integer values of $j$ in this section, the resulting expressions and integrals are obviously problematic from the simple viewpoint \OrangelineChange{of Riemann or Lebesgue integration} of functions since they either are singular at the origin or at infinity depending on the value of $j$. Methods to deal with such singular integrals have been developed in the theory of generalized functions~\cite{Zemanian1987,Estrada1989,Galapon2016} in terms of Hadamard finite part regularization. This is the framework in which the following calculation should be understood.

In case of an integral with a singularity at the \OrangelineChange{origin,} standard Hadamard regularization~\cite{Zemanian1987} can be directly applied. In case of a singularity at infinity tools with similar scope were developed in Ref.~\cite{Jones1996}. For an example of the necessary \OrangelineChange{procedures,} we refer to Ref.~\cite{Estrada1989} \OrangelineChange{where the regularization of $1/r^j$ is discussed in detail.} In our calculation we implicitly \OrangelineChange{assume that} such a regularization is performed and the kernel expressions are understood in this way. After the dust settles, the resulting kernel may be made sense of as a pseudo-function/generalized function induced by the meromorphic continuation of the remaining finite part, with the singular parts removed.   

We begin the the formal integration by choosing spherical coordinates $k\equiv \Abs{\Vect{k}}$, $\vartheta$ and \OrangelineChange{$\varphi$,} noting that the integrand does not depend on $\varphi$.
Thus we arrive immediately at the two-dimensional integral
\begin{align}
    \mathbfscr{K}^{(j)}
    =
   \frac{1}{(2\pi)^2}
   \int\limits_{0}^\infty \!\!\!\! ~\D k k^{j+2}
   \int\limits_{0}^\pi \!\!\!\! ~\D \vartheta \sin{\vartheta}
   \E^{\I kr \cos\vartheta},
\end{align}
which after integration over $\vartheta$ yields the expression
\begin{align}
    \mathbfscr{K}^{(j)}
    =
    \frac{1}{(2\pi)^2}
   \frac{1}{r}
   \frac{1}{\I}
   \int\limits_{0}^\infty \!\!\! ~\D k k^{j+1}
   \qty(\E^{\I kr}-\E^{-\I kr}).
\end{align}

\OrangelineChange{Next we eliminate the power $k^{j+1}$ by differentiating the radial wave $\exp(\pm\I kr)$ with respect to $r$ in total of $j+1$-times and find}
\begin{align}
    \mathbfscr{K}^{(j)}
    &=
    \frac{(-1)}{(2\pi)^2}
    \frac{1}{r}
   \frac{1}{\I^j}
   \frac{\partial^{j+1}}{\partial r^{j+1}}
   \qty[
   \int\limits_{0}^\infty \!\!\!\!~\D k
   \E^{\I kr}+(-1)^j \int\limits_{0}^\infty \!\!\!\!~\D k \E^{-\I kr}
   ].
\end{align}

In order to evaluate the two remaining integrals we introduce the convergence factor $\exp(-\epsilon k)$, to calculate the resulting \OrangelineChange{integral,} and then let $\epsilon>0$ approach zero afterwards. With the help of the relation
\begin{align}
    \OrangelineChange{
    \int_{0}^\infty \!\!\!\!~\D k  ~ \E^{-(\epsilon\mp \I r)k} = \frac{1}{\epsilon\mp \I r},
    }
\end{align}
we finally obtain
\begin{align}
    \mathbfscr{K}^{(j)}(\Vect{r})
    =
    \text{Pf}~
    \frac{1}{2\pi}
    \frac{1}{r}
    \frac{(-1)}{\I^j}
    \frac{\partial^{j+1}}{\partial r^{j+1}}
    d^{(j)}_\epsilon(r)
\end{align}
\OrangelineChange{Here we} have introduced the abbreviation
\begin{align}
    d_\epsilon^{(j)}(r)
    \equiv
    \frac{1}{\pi} \frac{\epsilon}{\epsilon^2+r^2}
    \frac{1+(-1)^j}{2}
    + \frac{\I}{\pi}
    \frac{r}{\epsilon^2+r^2}
    \frac{1-(-1)^j}{2}
\end{align}
and added the pseudo-function~\cite{Zemanian1987,Estrada1989} operator $\text{Pf}$ to remind us that the kernel is a pseudo-functions/generalized function resulting from implicitly performing Hadamard finite part regularization on the integral leading \OrangelineChange{to it,} if necessary.

With the representation
\begin{align}
    \lim_{\epsilon\to 0} \frac{1}{\pi}\frac{\epsilon}{\epsilon^2+r^2} = \delta(r)
\end{align}
of the Dirac delta function and the identity
\begin{align}
    \lim_{\epsilon\to 0} \frac{r}{\epsilon^2+r^2} =\mathcal{P}\qty(\frac{1}{r}),
\end{align}
where $\mathcal{P}$ denotes the Cauchy principal part, we obtain the expression
\begin{align}
    d^{(j)}(r)= \frac{1+(-1)^j}{2}\delta(r)+\frac{\I}{\pi}\frac{1-(-1)^j}{2} \mathcal{P}\qty(\frac{1}{r}).
\end{align}

Hence, for even values of $j$ only the delta function contributes to 
\begin{align}
    d^{(j)}\equiv \lim_{\epsilon\to 0} d_\epsilon^{(j)},
\end{align} whereas for odd ones only the contribution due to the derivatives of the Cauchy principal part appears, leading us to the expressions
\begin{align}
    \mathbfscr{K}^{(2n)}
    = \text{Pf}~
    \frac{1}{2\pi}
    \frac{(-1)^{n+1}}{r}
    \frac{\partial^{2n+1}}{\partial r^{2n+1}}
    \delta{(r)}
\end{align}
and
\begin{align}
    \mathbfscr{K}^{(2n+1)}
    = \text{Pf}~
    \frac{1}{2\pi^2}
    \frac{(-1)^{n+1}}{r}
    \frac{\partial^{2(n+1)}}{\partial r^{2(n+1)}} \mathcal{P}\qty(\frac{1}{r}).
\end{align}
Both kernel expressions should be understood as pseudo-functions including an implicit regularization lending the needed context~\cite{Zemanian1987} in which e.g. derivatives of the Cauchy principal part are to be interpreted. As \OrangelineChange{is} often done in physics we will from now on suppress the pseudo-function operator again for brevity in notation, assuming the resulting kernels and objects involving them are understood implicitly in that sense from now on. 

With these \OrangelineChange{considerations, after performing} the derivatives for $j=-1$, that is \OrangelineChange{$n=-1$, we find} the kernel
\begin{align}\label{APP:eq:kernel-result-1-over-r}
    \OrangelineChange{
    \mathbfscr{K}^{(-1)}=\mathbfscr{K}^{(E/B)} = \frac{1}{2\pi^2} \frac{1}{r^2},
    }
\end{align}
whereas for $j=+1$, that is \OrangelineChange{$n=0$, we} arrive at the kernel
\begin{align}\label{APP:eq:kernel-result-1-over-r4}
    \mathbfscr{K}^{(1)}=\mathbfscr{K}^{(A)} = \frac{(-1)}{\pi^2}\frac{1}{r^4}.
\end{align}

This expression for $\mathbfscr{K}^{(1)}$ also follows in a straight-forward way when we note from the definition~\cref{APP:eq:definition-jkernel} of $\mathbfscr{K}^{(j)}$ the connection
\begin{align}
    (-\VLaplace)~\mathbfscr{K}^{(-1)}=\mathbfscr{K}^{(1)}
\end{align}
between $\mathbfscr{K}^{(-1)}$ and $\mathbfscr{K}^{(1)}$, that is between $\mathbfscr{K}^{(E/B)}$ and $\mathbfscr{K}^{(A)}$.

Indeed, by direct differentiation of \cref{APP:eq:kernel-result-1-over-r} we obtain
\begin{align}\label{APP:eq:differential-eq-between-kernels}
    \mathbfscr{K}^{(A)}=(-\VLaplace)\mathbfscr{K}^{(E/B)}
    =
    -\frac{1}{2\pi^2}\left(\frac{\partial^2}{\partial r^2}+\frac{2}{r}\frac{\partial}{\partial r}\right)\frac{1}{r^2} 
    = 
    -\frac{1}{\pi^2}\frac{1}{r^4},
\end{align}
in complete agreement with \cref{APP:eq:kernel-result-1-over-r4}.

\section{Scalar product of two mode functions}\label{APP:sec:mathematical-complement}
\OrangelineChange{The scalar product of the curls of two mode functions is crucial for calculating the contribution} $H^{(B)}$ of \OrangelineChange{the} magnetic induction to the total energy $H$ of the electromagnetic field in a resonator performed in~\cref{APP:sec:field-energy-em-field}.

In~\cref{APP:eq:mode-identity-for-transverse-modes} we applied an identity for the scalar product of the \OrangelineChange{curls} of two vector fields which we derive here. We start with a more general identity for the three vector fields $\Vect{f}=\Vect{f}(\Vect{r},\Vect{r}^\prime)$ and $\Vect{g}=\Vect{g}(\Vect{r},\Vect{r}^\prime)$ and $\Vect{h}=\Vect{h}(\Vect{r},\Vect{r}^\prime)$.

When we take the divergence of the cross product between $\Vect{f}$ and $\Vect{h}$ we obtain
\OrangelineChange{
\begin{align}
    \nabla_{\Vect{r}}\cdot\qty[\Vect{f}(\Vect{r},\Vect{r}^\prime)\times\Vect{h}(\Vect{r},\Vect{r}^\prime)]
    &=\Vect{h}\cdot(\nabla_{\Vect{r}}\times\Vect{f}) \notag \\
    &\quad\quad
    -\Vect{f}\cdot(\nabla_{\Vect{r}}\times \Vect{h})
\end{align}
}
where from now on we suppress the functional dependencies of the fields for brevity. 

Replacing $\Vect{h}\mapsto \nabla_{\Vect{r}^\prime} \times\Vect{g}$ yields
\OrangelineChange{
\begin{align}\label{APP:eq:double-curl-different-initial}
    (\nabla_{\Vect{r}}\times \Vect{f})
    \cdot
    (\nabla_{\Vect{r}^\prime}\times \Vect{g})
    &=
    \nabla_{\Vect{r}}
    \cdot 
    \big[\Vect{f}\times(\nabla_{\Vect{r}^\prime}\times\Vect{g})\big] \notag \\
    &\quad \quad +
    \Vect{f}\cdot(\nabla_{\Vect{r}}\times(\nabla_{\Vect{r}^\prime}
    \times\Vect{g}).
\end{align}
}%

Using the definition of the cross product in terms of the Levi-Civita symbol\OrangelineChange{, i.e.} $\Vect{a}\times\Vect{b}=\Vect{e}_j \epsilon_{jk\ell} a_k b_\ell$ with summmation over double-indices implied, the terms on the right-hand side of the previous equation can be transformed into
\begin{align}
    \Vect{f}\times(\nabla_{\Vect{r}^\prime}\times\Vect{g})
    &= \nabla_{\Vect{r}^\prime}(\Vect{f}\cdot \Vect{g})-(\Vect{f}\cdot \nabla_{\Vect{r}^\prime})\Vect{g}
\end{align}
and
\begin{align}
    \nabla_{\Vect{r}}\times(\nabla_{\Vect{r}^\prime}\times\Vect{g})
    &=
    \nabla_{\Vect{r}^\prime}(\nabla_{\Vect{r}}\cdot \Vect{g})
    -(\nabla_{\Vect{r}}\cdot \nabla_{\Vect{r}^\prime})\Vect{g}.
\end{align}

Reinsertion of these identities into \cref{APP:eq:double-curl-different-initial} leads to the desired identity
\begin{align}\label{APP:eq:double-curl-different-arguments-identity}
    (\nabla_{\Vect{r}}\times\Vect{f})\cdot(\nabla_{\Vect{r}^\prime}\times\Vect{g})
    &=
    \nabla_{\Vect{r}}\cdot\qty[\nabla_{\Vect{r}^\prime}(\Vect{f}\cdot\Vect{g})] \notag \\
    &\quad -\nabla_{\Vect{r}}\cdot\qty[\qty(\Vect{f}\cdot\nabla_{\Vect{r}^\prime})\Vect{g}] \notag\\
    &\quad \quad  +
     \qty(\Vect{f}\cdot \nabla_{\Vect{r}^\prime})
     \qty(\nabla_{\Vect{r}}\cdot\Vect{g})\notag \\
     &\quad \quad\quad  -\Vect{f}\cdot\qty[\qty(\nabla_{\Vect{r}}\cdot\nabla_{\Vect{r}^\prime})\Vect{g}]
\end{align}
for the scalar product of two curls with differentiation with respect to different arguments $\Vect{r}$ and $\Vect{r}^\prime$.

Alternatively, starting from \cref{APP:eq:double-curl-different-initial} and using the case of $\Vect{r}\equiv \Vect{r}^\prime$ and the definition of the vector Laplacian we obtain the identity
\OrangelineChange{
\begin{align}\label{APP:eq:double-curl-same-arguments-identity}
    (\nabla_{\Vect{r}}\times\Vect{f})\cdot(\nabla_{\Vect{r}}\times\Vect{g})
    &=
     \nabla_{\Vect{r}}
    \cdot 
    \big[\Vect{f}\times(\nabla_{\Vect{r}}\times\Vect{g})\big] \notag\\
    &\quad \quad  +
     \Vect{f}\cdot\qty[\nabla_{\Vect{r}}(\nabla_{\Vect{r}}\cdot \Vect{g})-\VLaplace_{\Vect{r}} \Vect{g}],
\end{align}
used in~\cref{APP:eq:mode-identity-for-transverse-modes}.}

%\nocite{*}
\bibliography{00_main_doc.bib}

\begin{thebibliography}{66}%
\makeatletter
\providecommand \@ifxundefined [1]{%
 \@ifx{#1\undefined}
}%
\providecommand \@ifnum [1]{%
 \ifnum #1\expandafter \@firstoftwo
 \else \expandafter \@secondoftwo
 \fi
}%
\providecommand \@ifx [1]{%
 \ifx #1\expandafter \@firstoftwo
 \else \expandafter \@secondoftwo
 \fi
}%
\providecommand \natexlab [1]{#1}%
\providecommand \enquote  [1]{``#1''}%
\providecommand \bibnamefont  [1]{#1}%
\providecommand \bibfnamefont [1]{#1}%
\providecommand \citenamefont [1]{#1}%
\providecommand \href@noop [0]{\@secondoftwo}%
\providecommand \href [0]{\begingroup \@sanitize@url \@href}%
\providecommand \@href[1]{\@@startlink{#1}\@@href}%
\providecommand \@@href[1]{\endgroup#1\@@endlink}%
\providecommand \@sanitize@url [0]{\catcode `\\12\catcode `\$12\catcode
  `\&12\catcode `\#12\catcode `\^12\catcode `\_12\catcode `\%12\relax}%
\providecommand \@@startlink[1]{}%
\providecommand \@@endlink[0]{}%
\providecommand \url  [0]{\begingroup\@sanitize@url \@url }%
\providecommand \@url [1]{\endgroup\@href {#1}{\urlprefix }}%
\providecommand \urlprefix  [0]{URL }%
\providecommand \Eprint [0]{\href }%
\providecommand \doibase [0]{https://doi.org/}%
\providecommand \selectlanguage [0]{\@gobble}%
\providecommand \bibinfo  [0]{\@secondoftwo}%
\providecommand \bibfield  [0]{\@secondoftwo}%
\providecommand \translation [1]{[#1]}%
\providecommand \BibitemOpen [0]{}%
\providecommand \bibitemStop [0]{}%
\providecommand \bibitemNoStop [0]{.\EOS\space}%
\providecommand \EOS [0]{\spacefactor3000\relax}%
\providecommand \BibitemShut  [1]{\csname bibitem#1\endcsname}%
\let\auto@bib@innerbib\@empty
%</preamble>
\bibitem [{\citenamefont {Fermi}(1932)}]{Fermi1932}%
  \BibitemOpen
  \bibfield  {author} {\bibinfo {author} {\bibfnamefont {E.}~\bibnamefont
  {Fermi}},\ }\bibfield  {title} {\bibinfo {title} {{Quantum Theory of
  Radiation}},\ }\href {https://doi.org/10.1103/RevModPhys.4.87} {\bibfield
  {journal} {\bibinfo  {journal} {Rev. Mod. Phys.}\ }\textbf {\bibinfo {volume}
  {4}},\ \bibinfo {pages} {87} (\bibinfo {year} {1932})}\BibitemShut {NoStop}%
\bibitem [{\citenamefont {Lamb}(1995)}]{Lamb1995}%
  \BibitemOpen
  \bibfield  {author} {\bibinfo {author} {\bibfnamefont {W.~E.}\ \bibnamefont
  {Lamb}},\ }\bibfield  {title} {\bibinfo {title} {{Anti-photon}},\ }\href
  {https://doi.org/10.1007/BF01135846} {\bibfield  {journal} {\bibinfo
  {journal} {Appl. Phys. B}\ }\textbf {\bibinfo {volume} {60}},\ \bibinfo
  {pages} {77} (\bibinfo {year} {1995})}\BibitemShut {NoStop}%
\bibitem [{\citenamefont {Wheeler}(1957)}]{Wheeler1957}%
  \BibitemOpen
  \bibfield  {author} {\bibinfo {author} {\bibfnamefont {J.~A.}\ \bibnamefont
  {Wheeler}},\ }\bibfield  {title} {\bibinfo {title} {{On the nature of quantum
  geometrodynamics}},\ }\href {https://doi.org/10.1016/0003-4916(57)90050-7}
  {\bibfield  {journal} {\bibinfo  {journal} {Ann. Phys.}\ }\textbf {\bibinfo
  {volume} {2}},\ \bibinfo {pages} {604} (\bibinfo {year} {1957})}\BibitemShut
  {NoStop}%
\bibitem [{\citenamefont {Wheeler}(1962)}]{Wheeler1962}%
  \BibitemOpen
  \bibfield  {author} {\bibinfo {author} {\bibfnamefont {J.~A.}\ \bibnamefont
  {Wheeler}},\ }\href@noop {} {\emph {\bibinfo {title} {{Geometrodynamics}}}}\
  (\bibinfo  {publisher} {Academic Press},\ \bibinfo {address} {London},\
  \bibinfo {year} {1962})\BibitemShut {NoStop}%
\bibitem [{\citenamefont {Misner}\ \emph {et~al.}(1973)\citenamefont {Misner},
  \citenamefont {Thorne},\ and\ \citenamefont {Wheeler}}]{Misner2017}%
  \BibitemOpen
  \bibfield  {author} {\bibinfo {author} {\bibfnamefont {C.~W.}\ \bibnamefont
  {Misner}}, \bibinfo {author} {\bibfnamefont {K.~S.}\ \bibnamefont {Thorne}},\
  and\ \bibinfo {author} {\bibfnamefont {J.~A.}\ \bibnamefont {Wheeler}},\
  }\href@noop {} {\emph {\bibinfo {title} {{Gravitation}}}}\ (\bibinfo
  {publisher} {W. H. Freeman},\ \bibinfo {address} {San Fransisco},\ \bibinfo
  {year} {1973})\BibitemShut {NoStop}%
\bibitem [{\citenamefont {Bia{\l}ynicka-Birula}\ and\ \citenamefont
  {Bia{\l}ynicki-Birula}(1987)}]{Bialynicka-Birula1987}%
  \BibitemOpen
  \bibfield  {author} {\bibinfo {author} {\bibfnamefont {Z.}~\bibnamefont
  {Bia{\l}ynicka-Birula}}\ and\ \bibinfo {author} {\bibfnamefont
  {I.}~\bibnamefont {Bia{\l}ynicki-Birula}},\ }\bibfield  {title} {\bibinfo
  {title} {{Space{\textendash}time description of squeezing}},\ }\href
  {https://doi.org/10.1364/JOSAB.4.001621} {\bibfield  {journal} {\bibinfo
  {journal} {J. Opt. Soc. Am. B.}\ }\textbf {\bibinfo {volume} {4}},\ \bibinfo
  {pages} {1621} (\bibinfo {year} {1987})}\BibitemShut {NoStop}%
\bibitem [{\citenamefont {Bia{\l}ynicki-Birula}(2000)}]{Bialynicki-Birula2000}%
  \BibitemOpen
  \bibfield  {author} {\bibinfo {author} {\bibfnamefont {I.}~\bibnamefont
  {Bia{\l}ynicki-Birula}},\ }\bibfield  {title} {\bibinfo {title} {{The Wigner
  functional of the electromagnetic field}},\ }\href
  {https://doi.org/10.1016/S0030-4018(99)00563-5} {\bibfield  {journal}
  {\bibinfo  {journal} {Opt. Commun.}\ }\textbf {\bibinfo {volume} {179}},\
  \bibinfo {pages} {237} (\bibinfo {year} {2000})}\BibitemShut {NoStop}%
\bibitem [{\citenamefont {Bia{\l}ynicki-Birula}(1996)}]{Bialynicki-Birula1996}%
  \BibitemOpen
  \bibfield  {author} {\bibinfo {author} {\bibfnamefont {I.}~\bibnamefont
  {Bia{\l}ynicki-Birula}},\ }\bibfield  {title} {\bibinfo {title} {{Photon Wave
  Function}},\ }in\ \href {https://doi.org/10.1016/S0079-6638(08)70316-0}
  {\emph {\bibinfo {booktitle} {{Progress in Optics}}}},\ Vol.~\bibinfo
  {volume} {36}\ (\bibinfo  {publisher} {Elsevier},\ \bibinfo {address}
  {Amsterdam},\ \bibinfo {year} {1996})\ pp.\ \bibinfo {pages}
  {245--294}\BibitemShut {NoStop}%
\bibitem [{\citenamefont {Bia{\l}ynicki-Birula}(2003)}]{Bialynicki-Birula2003}%
  \BibitemOpen
  \bibfield  {author} {\bibinfo {author} {\bibfnamefont {I.}~\bibnamefont
  {Bia{\l}ynicki-Birula}},\ }\bibfield  {title} {\bibinfo {title} {{The
  Structure of the Vacuum and the Photon Number}},\ }in\ \href
  {https://doi.org/10.1007/978-3-540-40968-7_20} {\emph {\bibinfo {booktitle}
  {{Decoherence and Entropy in Complex Systems}}}}\ (\bibinfo  {publisher}
  {Springer},\ \bibinfo {address} {Berlin},\ \bibinfo {year} {2003})\ pp.\
  \bibinfo {pages} {287--295}\BibitemShut {NoStop}%
\bibitem [{\citenamefont {Bia{\l}ynicki-Birula}\ and\ \citenamefont
  {Bia{\l}ynicki-Birula}(2023)}]{Bialynicki-Birula2023}%
  \BibitemOpen
  \bibfield  {author} {\bibinfo {author} {\bibfnamefont {I.}~\bibnamefont
  {Bia{\l}ynicki-Birula}}\ and\ \bibinfo {author} {\bibfnamefont
  {Z.}~\bibnamefont {Bia{\l}ynicki-Birula}},\ }\bibfield  {title} {\bibinfo
  {title} {{The Zeldovich number: A universal dimensionless measure for the
  electromagnetic field}},\ }\bibfield  {journal} {\bibinfo  {journal} {arXiv}\
  }\href {https://doi.org/10.48550/arXiv.2303.12183}
  {10.48550/arXiv.2303.12183} (\bibinfo {year} {2023})\BibitemShut {NoStop}%
\bibitem [{\citenamefont {Born}\ \emph {et~al.}(1926)\citenamefont {Born},
  \citenamefont {Heisenberg},\ and\ \citenamefont {Jordan}}]{Born1926}%
  \BibitemOpen
  \bibfield  {author} {\bibinfo {author} {\bibfnamefont {M.}~\bibnamefont
  {Born}}, \bibinfo {author} {\bibfnamefont {W.}~\bibnamefont {Heisenberg}},\
  and\ \bibinfo {author} {\bibfnamefont {P.}~\bibnamefont {Jordan}},\
  }\bibfield  {title} {\bibinfo {title} {{Zur Quantenmechanik. II.}},\ }\href
  {https://doi.org/10.1007/BF01379806} {\bibfield  {journal} {\bibinfo
  {journal} {Z. Phys.}\ }\textbf {\bibinfo {volume} {35}},\ \bibinfo {pages}
  {557} (\bibinfo {year} {1926})}\BibitemShut {NoStop}%
\bibitem [{\citenamefont {Dirac}(1927)}]{Dirac1927}%
  \BibitemOpen
  \bibfield  {author} {\bibinfo {author} {\bibfnamefont {P.~A.~M.}\
  \bibnamefont {Dirac}},\ }\bibfield  {title} {\bibinfo {title} {{The quantum
  theory of the emission and absorption of radiation}},\ }\href
  {https://doi.org/10.1098/rspa.1927.0039} {\bibfield  {journal} {\bibinfo
  {journal} {Proc. R. Soc. London A}\ }\textbf {\bibinfo {volume} {114}},\
  \bibinfo {pages} {243} (\bibinfo {year} {1927})}\BibitemShut {NoStop}%
\bibitem [{\citenamefont {Lamb}\ and\ \citenamefont
  {Retherford}(1947)}]{Lamb1947}%
  \BibitemOpen
  \bibfield  {author} {\bibinfo {author} {\bibfnamefont {W.~E.}\ \bibnamefont
  {Lamb}}\ and\ \bibinfo {author} {\bibfnamefont {R.~C.}\ \bibnamefont
  {Retherford}},\ }\bibfield  {title} {\bibinfo {title} {{Fine Structure of the
  Hydrogen Atom by a Microwave Method}},\ }\href
  {https://doi.org/10.1103/PhysRev.72.241} {\bibfield  {journal} {\bibinfo
  {journal} {Phys. Rev.}\ }\textbf {\bibinfo {volume} {72}},\ \bibinfo {pages}
  {241} (\bibinfo {year} {1947})}\BibitemShut {NoStop}%
\bibitem [{\citenamefont {Foley}\ and\ \citenamefont
  {Kusch}(1948)}]{Foley1948}%
  \BibitemOpen
  \bibfield  {author} {\bibinfo {author} {\bibfnamefont {H.~M.}\ \bibnamefont
  {Foley}}\ and\ \bibinfo {author} {\bibfnamefont {P.}~\bibnamefont {Kusch}},\
  }\bibfield  {title} {\bibinfo {title} {{On the Intrinsic Moment of the
  Electron}},\ }\href {https://doi.org/10.1103/PhysRev.73.412} {\bibfield
  {journal} {\bibinfo  {journal} {Phys. Rev.}\ }\textbf {\bibinfo {volume}
  {73}},\ \bibinfo {pages} {412} (\bibinfo {year} {1948})}\BibitemShut
  {NoStop}%
\bibitem [{\citenamefont {Schwinger}(1958)}]{Schwinger1958}%
  \BibitemOpen
  \bibfield  {author} {\bibinfo {author} {\bibfnamefont {J.}~\bibnamefont
  {Schwinger}},\ }\href@noop {} {\emph {\bibinfo {title} {{Selected Papers on
  Quantum Electrodynamics}}}}\ (\bibinfo  {publisher} {Dover Publications},\
  \bibinfo {address} {New York},\ \bibinfo {year} {1958})\BibitemShut {NoStop}%
\bibitem [{\citenamefont {Schweber}(1994)}]{Schweber1994}%
  \BibitemOpen
  \bibfield  {author} {\bibinfo {author} {\bibfnamefont {S.~S.}\ \bibnamefont
  {Schweber}},\ }\href
  {https://press.princeton.edu/books/paperback/9780691033273/qed-and-the-men-who-made-it}
  {\emph {\bibinfo {title} {{QED and the Men Who Made It}}}}\ (\bibinfo
  {publisher} {Princeton University Press},\ \bibinfo {address} {Princeton},\
  \bibinfo {year} {1994})\BibitemShut {NoStop}%
\bibitem [{\citenamefont {Bia{\l}ynicki-Birula}\ and\ \citenamefont
  {Bia{\l}ynicka-Birula}(1975)}]{Bialynicki-Birula1975}%
  \BibitemOpen
  \bibfield  {author} {\bibinfo {author} {\bibfnamefont {I.}~\bibnamefont
  {Bia{\l}ynicki-Birula}}\ and\ \bibinfo {author} {\bibfnamefont
  {Z.}~\bibnamefont {Bia{\l}ynicka-Birula}},\ }\href@noop {} {\emph {\bibinfo
  {title} {{Quantum Electrodynamics}}}}\ (\bibinfo  {publisher} {Pergamon
  Press},\ \bibinfo {address} {New York},\ \bibinfo {year} {1975})\BibitemShut
  {NoStop}%
\bibitem [{\citenamefont {Haroche}(2013)}]{Haroche2013}%
  \BibitemOpen
  \bibfield  {author} {\bibinfo {author} {\bibfnamefont {S.}~\bibnamefont
  {Haroche}},\ }\bibfield  {title} {\bibinfo {title} {{Nobel Lecture:
  Controlling photons in a box and exploring the quantum to classical
  boundary}},\ }\href {https://doi.org/10.1103/RevModPhys.85.1083} {\bibfield
  {journal} {\bibinfo  {journal} {Rev. Mod. Phys.}\ }\textbf {\bibinfo {volume}
  {85}},\ \bibinfo {pages} {1083} (\bibinfo {year} {2013})}\BibitemShut
  {NoStop}%
\bibitem [{\citenamefont {Walther}\ \emph {et~al.}(2006)\citenamefont
  {Walther}, \citenamefont {Varcoe}, \citenamefont {Englert},\ and\
  \citenamefont {Becker}}]{Walther2006}%
  \BibitemOpen
  \bibfield  {author} {\bibinfo {author} {\bibfnamefont {H.}~\bibnamefont
  {Walther}}, \bibinfo {author} {\bibfnamefont {B.~T.~H.}\ \bibnamefont
  {Varcoe}}, \bibinfo {author} {\bibfnamefont {B.-G.}\ \bibnamefont
  {Englert}},\ and\ \bibinfo {author} {\bibfnamefont {T.}~\bibnamefont
  {Becker}},\ }\bibfield  {title} {\bibinfo {title} {{Cavity quantum
  electrodynamics}},\ }\href {https://doi.org/10.1088/0034-4885/69/5/R02}
  {\bibfield  {journal} {\bibinfo  {journal} {Rep. Prog. Phys.}\ }\textbf
  {\bibinfo {volume} {69}},\ \bibinfo {pages} {1325} (\bibinfo {year}
  {2006})}\BibitemShut {NoStop}%
\bibitem [{\citenamefont {Blais}\ \emph {et~al.}(2021)\citenamefont {Blais},
  \citenamefont {Grimsmo}, \citenamefont {Girvin},\ and\ \citenamefont
  {Wallraff}}]{Blais2021}%
  \BibitemOpen
  \bibfield  {author} {\bibinfo {author} {\bibfnamefont {A.}~\bibnamefont
  {Blais}}, \bibinfo {author} {\bibfnamefont {A.~L.}\ \bibnamefont {Grimsmo}},
  \bibinfo {author} {\bibfnamefont {S.~M.}\ \bibnamefont {Girvin}},\ and\
  \bibinfo {author} {\bibfnamefont {A.}~\bibnamefont {Wallraff}},\ }\bibfield
  {title} {\bibinfo {title} {{Circuit quantum electrodynamics}},\ }\href
  {https://doi.org/10.1103/RevModPhys.93.025005} {\bibfield  {journal}
  {\bibinfo  {journal} {Rev. Mod. Phys.}\ }\textbf {\bibinfo {volume} {93}},\
  \bibinfo {pages} {025005} (\bibinfo {year} {2021})}\BibitemShut {NoStop}%
\bibitem [{\citenamefont {Sheremet}\ \emph {et~al.}(2023)\citenamefont
  {Sheremet}, \citenamefont {Petrov}, \citenamefont {Iorsh}, \citenamefont
  {Poshakinskiy},\ and\ \citenamefont {Poddubny}}]{Sheremet2023}%
  \BibitemOpen
  \bibfield  {author} {\bibinfo {author} {\bibfnamefont {A.~S.}\ \bibnamefont
  {Sheremet}}, \bibinfo {author} {\bibfnamefont {M.~I.}\ \bibnamefont
  {Petrov}}, \bibinfo {author} {\bibfnamefont {I.~V.}\ \bibnamefont {Iorsh}},
  \bibinfo {author} {\bibfnamefont {A.~V.}\ \bibnamefont {Poshakinskiy}},\ and\
  \bibinfo {author} {\bibfnamefont {A.~N.}\ \bibnamefont {Poddubny}},\
  }\bibfield  {title} {\bibinfo {title} {{Waveguide quantum electrodynamics:
  Collective radiance and photon-photon correlations}},\ }\href
  {https://doi.org/10.1103/RevModPhys.95.015002} {\bibfield  {journal}
  {\bibinfo  {journal} {Rev. Mod. Phys.}\ }\textbf {\bibinfo {volume} {95}},\
  \bibinfo {pages} {015002} (\bibinfo {year} {2023})}\BibitemShut {NoStop}%
\bibitem [{\citenamefont {Landau}\ and\ \citenamefont
  {Peierls}(1931)}]{Landau1931}%
  \BibitemOpen
  \bibfield  {author} {\bibinfo {author} {\bibfnamefont {L.}~\bibnamefont
  {Landau}}\ and\ \bibinfo {author} {\bibfnamefont {R.}~\bibnamefont
  {Peierls}},\ }\bibfield  {title} {\bibinfo {title} {{Erweiterung des
  Unbestimmtheitsprinzips f{\"{u}}r die relativistische Quantentheorie}},\
  }\href {https://doi.org/10.1007/BF01391513} {\bibfield  {journal} {\bibinfo
  {journal} {Z. Phys.}\ }\textbf {\bibinfo {volume} {69}},\ \bibinfo {pages}
  {56} (\bibinfo {year} {1931})}\BibitemShut {NoStop}%
\bibitem [{\citenamefont {Bohr}\ and\ \citenamefont
  {Rosenfeld}(1933)}]{Bohr1933}%
  \BibitemOpen
  \bibfield  {author} {\bibinfo {author} {\bibfnamefont {N.}~\bibnamefont
  {Bohr}}\ and\ \bibinfo {author} {\bibfnamefont {L.}~\bibnamefont
  {Rosenfeld}},\ }\bibfield  {title} {\bibinfo {title} {{Zur Frage der
  Messbarkeit der elektromagnetischen Feldgrößen}},\ }\href
  {https://ui.adsabs.harvard.edu/abs/1933KDVS...12....3B/abstract} {\bibfield
  {journal} {\bibinfo  {journal} {Kgl. Danske Vidensk. Selskab. Math.-Fys.
  Medd.}\ }\textbf {\bibinfo {volume} {12}},\ \bibinfo {pages} {3} (\bibinfo
  {year} {1933})}\BibitemShut {NoStop}%
\bibitem [{\citenamefont {Bohr}\ and\ \citenamefont
  {Rosenfeld}(1950)}]{Bohr1950}%
  \BibitemOpen
  \bibfield  {author} {\bibinfo {author} {\bibfnamefont {N.}~\bibnamefont
  {Bohr}}\ and\ \bibinfo {author} {\bibfnamefont {L.}~\bibnamefont
  {Rosenfeld}},\ }\bibfield  {title} {\bibinfo {title} {{Field and Charge
  Measurements in Quantum Electrodynamics}},\ }\href
  {https://doi.org/10.1103/PhysRev.78.794} {\bibfield  {journal} {\bibinfo
  {journal} {Phys. Rev.}\ }\textbf {\bibinfo {volume} {78}},\ \bibinfo {pages}
  {794} (\bibinfo {year} {1950})}\BibitemShut {NoStop}%
\bibitem [{\citenamefont {Wheeler}\ and\ \citenamefont
  {Zurek}(2016)}]{Wheeler2016}%
  \BibitemOpen
  \bibfield  {author} {\bibinfo {author} {\bibfnamefont {J.~A.}\ \bibnamefont
  {Wheeler}}\ and\ \bibinfo {author} {\bibfnamefont {W.~H.}\ \bibnamefont
  {Zurek}},\ }\href
  {https://press.princeton.edu/books/hardcover/9780691641027/quantum-theory-and-measurement}
  {\emph {\bibinfo {title} {{Quantum Theory and Measurement}}}}\ (\bibinfo
  {publisher} {Princeton University Press},\ \bibinfo {year}
  {2016})\BibitemShut {NoStop}%
\bibitem [{\citenamefont {Salecker}\ and\ \citenamefont
  {Wigner}(1958)}]{Salecker1958}%
  \BibitemOpen
  \bibfield  {author} {\bibinfo {author} {\bibfnamefont {H.}~\bibnamefont
  {Salecker}}\ and\ \bibinfo {author} {\bibfnamefont {E.~P.}\ \bibnamefont
  {Wigner}},\ }\bibfield  {title} {\bibinfo {title} {{Quantum Limitations of
  the Measurement of Space-Time Distances}},\ }\href
  {https://doi.org/10.1103/PhysRev.109.571} {\bibfield  {journal} {\bibinfo
  {journal} {Phys. Rev.}\ }\textbf {\bibinfo {volume} {109}},\ \bibinfo {pages}
  {571} (\bibinfo {year} {1958})}\BibitemShut {NoStop}%
\bibitem [{\citenamefont {Kucha{\v{r}}}(1970)}]{Kuchar1970}%
  \BibitemOpen
  \bibfield  {author} {\bibinfo {author} {\bibfnamefont {K.}~\bibnamefont
  {Kucha{\v{r}}}},\ }\bibfield  {title} {\bibinfo {title} {{Ground State
  Functional of the Linearized Gravitational Field}},\ }\href
  {https://doi.org/10.1063/1.1665133} {\bibfield  {journal} {\bibinfo
  {journal} {J. Math. Phys.}\ }\textbf {\bibinfo {volume} {11}},\ \bibinfo
  {pages} {3322} (\bibinfo {year} {1970})}\BibitemShut {NoStop}%
\bibitem [{\citenamefont {Schwinger}(1967)}]{Schwinger1967}%
  \BibitemOpen
  \bibfield  {author} {\bibinfo {author} {\bibfnamefont {J.}~\bibnamefont
  {Schwinger}},\ }\bibfield  {title} {\bibinfo {title} {{Sources and
  Electrodynamics}},\ }\href {https://doi.org/10.1103/PhysRev.158.1391}
  {\bibfield  {journal} {\bibinfo  {journal} {Phys. Rev.}\ }\textbf {\bibinfo
  {volume} {158}},\ \bibinfo {pages} {1391} (\bibinfo {year}
  {1967})}\BibitemShut {NoStop}%
\bibitem [{\citenamefont {Weinberg}(1995)}]{Weinberg1995}%
  \BibitemOpen
  \bibfield  {author} {\bibinfo {author} {\bibfnamefont {S.}~\bibnamefont
  {Weinberg}},\ }\href {https://doi.org/10.1017/cbo9781139644167} {\emph
  {\bibinfo {title} {{The Quantum Theory of Fields - Volume 1: Foundations}}}}\
  (\bibinfo  {publisher} {Cambridge University Press},\ \bibinfo {address}
  {Cambridge},\ \bibinfo {year} {1995})\BibitemShut {NoStop}%
\bibitem [{\citenamefont {Padmanabhan}(2016)}]{Padmanabhan2016}%
  \BibitemOpen
  \bibfield  {author} {\bibinfo {author} {\bibfnamefont {T.}~\bibnamefont
  {Padmanabhan}},\ }\href
  {https://link.springer.com/book/10.1007/978-3-319-28173-5} {\emph {\bibinfo
  {title} {{Quantum Field Theory}}}}\ (\bibinfo  {publisher} {Springer},\
  \bibinfo {address} {Cham},\ \bibinfo {year} {2016})\BibitemShut {NoStop}%
\bibitem [{\citenamefont {Jackiw}(1987)}]{Jackiw1987}%
  \BibitemOpen
  \bibfield  {author} {\bibinfo {author} {\bibfnamefont {R.~W.}\ \bibnamefont
  {Jackiw}},\ }\href {https://cds.cern.ch/record/181728} {\emph {\bibinfo
  {title} {{Schrödinger picture analysis of boson and fermion quantum field
  theories}}}},\ \bibinfo {type} {Tech. Rep.}\ (\bibinfo  {institution} {MIT.
  Cent. Theor. Phys.},\ \bibinfo {address} {Cambridge, MA},\ \bibinfo {year}
  {1987})\BibitemShut {NoStop}%
\bibitem [{\citenamefont {Hatfield}(2018)}]{Hatfield2018}%
  \BibitemOpen
  \bibfield  {author} {\bibinfo {author} {\bibfnamefont {B.}~\bibnamefont
  {Hatfield}},\ }\href {https://doi.org/10.1201/9780429493232} {\emph {\bibinfo
  {title} {{Quantum Field Theory Of Point Particles And Strings}}}}\ (\bibinfo
  {publisher} {CRC Press},\ \bibinfo {address} {Boca Raton},\ \bibinfo {year}
  {2018})\BibitemShut {NoStop}%
\bibitem [{\citenamefont {Bose}\ \emph {et~al.}(2017)\citenamefont {Bose},
  \citenamefont {Mazumdar}, \citenamefont {Morley}, \citenamefont {Ulbricht},
  \citenamefont {Toro{\ifmmode\check{s}\else\v{s}\fi}}, \citenamefont
  {Paternostro}, \citenamefont {Geraci}, \citenamefont {Barker}, \citenamefont
  {Kim},\ and\ \citenamefont {Milburn}}]{Bose2017}%
  \BibitemOpen
  \bibfield  {author} {\bibinfo {author} {\bibfnamefont {S.}~\bibnamefont
  {Bose}}, \bibinfo {author} {\bibfnamefont {A.}~\bibnamefont {Mazumdar}},
  \bibinfo {author} {\bibfnamefont {G.~W.}\ \bibnamefont {Morley}}, \bibinfo
  {author} {\bibfnamefont {H.}~\bibnamefont {Ulbricht}}, \bibinfo {author}
  {\bibfnamefont {M.}~\bibnamefont {Toro{\ifmmode\check{s}\else\v{s}\fi}}},
  \bibinfo {author} {\bibfnamefont {M.}~\bibnamefont {Paternostro}}, \bibinfo
  {author} {\bibfnamefont {A.~A.}\ \bibnamefont {Geraci}}, \bibinfo {author}
  {\bibfnamefont {P.~F.}\ \bibnamefont {Barker}}, \bibinfo {author}
  {\bibfnamefont {M.~S.}\ \bibnamefont {Kim}},\ and\ \bibinfo {author}
  {\bibfnamefont {G.}~\bibnamefont {Milburn}},\ }\bibfield  {title} {\bibinfo
  {title} {{Spin Entanglement Witness for Quantum Gravity}},\ }\href
  {https://doi.org/10.1103/PhysRevLett.119.240401} {\bibfield  {journal}
  {\bibinfo  {journal} {Phys. Rev. Lett.}\ }\textbf {\bibinfo {volume} {119}},\
  \bibinfo {pages} {240401} (\bibinfo {year} {2017})}\BibitemShut {NoStop}%
\bibitem [{\citenamefont {Marletto}\ and\ \citenamefont
  {Vedral}(2017)}]{Marletto2017}%
  \BibitemOpen
  \bibfield  {author} {\bibinfo {author} {\bibfnamefont {C.}~\bibnamefont
  {Marletto}}\ and\ \bibinfo {author} {\bibfnamefont {V.}~\bibnamefont
  {Vedral}},\ }\bibfield  {title} {\bibinfo {title} {{Gravitationally Induced
  Entanglement between Two Massive Particles is Sufficient Evidence of Quantum
  Effects in Gravity}},\ }\href
  {https://doi.org/10.1103/PhysRevLett.119.240402} {\bibfield  {journal}
  {\bibinfo  {journal} {Phys. Rev. Lett.}\ }\textbf {\bibinfo {volume} {119}},\
  \bibinfo {pages} {240402} (\bibinfo {year} {2017})}\BibitemShut {NoStop}%
\bibitem [{\citenamefont {Rickles}\ and\ \citenamefont
  {DeWitt}(2011)}]{Rickles2011}%
  \BibitemOpen
  \bibfield  {author} {\bibinfo {author} {\bibfnamefont {D.}~\bibnamefont
  {Rickles}}\ and\ \bibinfo {author} {\bibfnamefont {C.~M.}\ \bibnamefont
  {DeWitt}},\ }\href@noop {} {\emph {\bibinfo {title} {{The Role of Gravitation
  in Physics: Report from the 1957 Chapel Hill Conference}}}}\ (\bibinfo
  {publisher} {Max-Planck-Gesellschaft zur Förderung der Wissenschaften},\
  \bibinfo {address} {Berlin},\ \bibinfo {year} {2011})\BibitemShut {NoStop}%
\bibitem [{\citenamefont {Chen}\ \emph {et~al.}(2023)\citenamefont {Chen},
  \citenamefont {Giacomini},\ and\ \citenamefont {Rovelli}}]{Chen2023}%
  \BibitemOpen
  \bibfield  {author} {\bibinfo {author} {\bibfnamefont {L.-Q.}\ \bibnamefont
  {Chen}}, \bibinfo {author} {\bibfnamefont {F.}~\bibnamefont {Giacomini}},\
  and\ \bibinfo {author} {\bibfnamefont {C.}~\bibnamefont {Rovelli}},\
  }\bibfield  {title} {\bibinfo {title} {{Quantum States of Fields for Quantum
  Split Sources}},\ }\href {https://doi.org/10.22331/q-2023-03-20-958}
  {\bibfield  {journal} {\bibinfo  {journal} {Quantum}\ }\textbf {\bibinfo
  {volume} {7}},\ \bibinfo {pages} {958} (\bibinfo {year} {2023})}\BibitemShut
  {NoStop}%
\bibitem [{\citenamefont {Dyson}(2013)}]{Dyson2013}%
  \BibitemOpen
  \bibfield  {author} {\bibinfo {author} {\bibfnamefont {F.}~\bibnamefont
  {Dyson}},\ }\bibfield  {title} {\bibinfo {title} {{Is a graviton
  detectable?}},\ }\href {https://doi.org/10.1142/S0217751X1330041X} {\bibfield
   {journal} {\bibinfo  {journal} {Int. J. Mod. Phys. A}\ }\textbf {\bibinfo
  {volume} {28}},\ \bibinfo {pages} {1330041} (\bibinfo {year}
  {2013})}\BibitemShut {NoStop}%
\bibitem [{\citenamefont {Glauber}(1963)}]{Glauber1963}%
  \BibitemOpen
  \bibfield  {author} {\bibinfo {author} {\bibfnamefont {R.~J.}\ \bibnamefont
  {Glauber}},\ }\bibfield  {title} {\bibinfo {title} {{Coherent and Incoherent
  States of the Radiation Field}},\ }\href
  {https://doi.org/10.1103/PhysRev.131.2766} {\bibfield  {journal} {\bibinfo
  {journal} {Phys. Rev.}\ }\textbf {\bibinfo {volume} {131}},\ \bibinfo {pages}
  {2766} (\bibinfo {year} {1963})}\BibitemShut {NoStop}%
\bibitem [{\citenamefont {Rasch}(1934)}]{Rasch1934}%
  \BibitemOpen
  \bibfield  {author} {\bibinfo {author} {\bibfnamefont {G.}~\bibnamefont
  {Rasch}},\ }\bibfield  {title} {\bibinfo {title} {{Zur Theorie und Anwendung
  des Produktintegrals.}},\ }\href {https://doi.org/10.1515/crll.1934.171.65}
  {\bibfield  {journal} {\bibinfo  {journal} {J. rein. u. angew. Math.}\
  }\textbf {\bibinfo {volume} {171}},\ \bibinfo {pages} {65} (\bibinfo {year}
  {1934})}\BibitemShut {NoStop}%
\bibitem [{\citenamefont {Salecker}(1950)}]{Salecker1950}%
  \BibitemOpen
  \bibfield  {author} {\bibinfo {author} {\bibfnamefont {H.}~\bibnamefont
  {Salecker}},\ }\bibfield  {title} {\bibinfo {title}
  {{Quantenelektrodynamische Selbstenergie und exakte
  L{\ifmmode\ddot{o}\else\"{o}\fi}sungen der
  Schr{\ifmmode\ddot{o}\else\"{o}\fi}dinger-Gleichung II}},\ }\href
  {https://doi.org/10.1515/zna-1950-0902} {\bibfield  {journal} {\bibinfo
  {journal} {Zeitschrift f{\ifmmode\ddot{u}\else\"{u}\fi}r Naturforschung A}\
  }\textbf {\bibinfo {volume} {5}},\ \bibinfo {pages} {480} (\bibinfo {year}
  {1950})}\BibitemShut {NoStop}%
\bibitem [{\citenamefont {Bia{\l}ynicki-Birula}(1998)}]{Bialynicki-Birula1998}%
  \BibitemOpen
  \bibfield  {author} {\bibinfo {author} {\bibfnamefont {I.}~\bibnamefont
  {Bia{\l}ynicki-Birula}},\ }\bibfield  {title} {\bibinfo {title} {Nonstandard
  introduction to squeezing of the electromagnetic field},\ }\href@noop {}
  {\bibfield  {journal} {\bibinfo  {journal} {Acta Phys. Polon. B}\ }\textbf
  {\bibinfo {volume} {29}},\ \bibinfo {pages} {3569} (\bibinfo {year}
  {1998})},\ \Eprint {https://arxiv.org/abs/quant-ph/9809069}
  {arXiv:quant-ph/9809069} \BibitemShut {NoStop}%
\bibitem [{\citenamefont {Anastopoulos}\ and\ \citenamefont
  {Hu}(2022)}]{Anastopoulos2022}%
  \BibitemOpen
  \bibfield  {author} {\bibinfo {author} {\bibfnamefont {C.}~\bibnamefont
  {Anastopoulos}}\ and\ \bibinfo {author} {\bibfnamefont {B.-L.}\ \bibnamefont
  {Hu}},\ }\bibfield  {title} {\bibinfo {title} {{Gravity, Quantum Fields and
  Quantum Information: Problems with Classical Channel and Stochastic
  Theories}},\ }\href {https://doi.org/10.3390/e24040490} {\bibfield  {journal}
  {\bibinfo  {journal} {Entropy}\ }\textbf {\bibinfo {volume} {24}},\ \bibinfo
  {pages} {490} (\bibinfo {year} {2022})}\BibitemShut {NoStop}%
\bibitem [{\citenamefont {Weinberg}(1996)}]{Weinberg1996}%
  \BibitemOpen
  \bibfield  {author} {\bibinfo {author} {\bibfnamefont {S.}~\bibnamefont
  {Weinberg}},\ }\href {https://doi.org/10.1017/cbo9781139644174} {\emph
  {\bibinfo {title} {{The Quantum Theory of Fields - Volume 2: Modern
  Applications}}}}\ (\bibinfo  {publisher} {Cambridge University Press},\
  \bibinfo {address} {Cambridge},\ \bibinfo {year} {1996})\BibitemShut
  {NoStop}%
\bibitem [{\citenamefont {Gupta}(1950)}]{Gupta1950}%
  \BibitemOpen
  \bibfield  {author} {\bibinfo {author} {\bibfnamefont {S.~N.}\ \bibnamefont
  {Gupta}},\ }\bibfield  {title} {\bibinfo {title} {{Theory of Longitudinal
  Photons in Quantum Electrodynamics}},\ }\href
  {https://doi.org/10.1088/0370-1298/63/7/301} {\bibfield  {journal} {\bibinfo
  {journal} {Proc. Phys. Soc. A}\ }\textbf {\bibinfo {volume} {63}},\ \bibinfo
  {pages} {681} (\bibinfo {year} {1950})}\BibitemShut {NoStop}%
\bibitem [{\citenamefont {Bleuler}(1950)}]{Bleuler1950}%
  \BibitemOpen
  \bibfield  {author} {\bibinfo {author} {\bibfnamefont {K.}~\bibnamefont
  {Bleuler}},\ }\bibfield  {title} {\bibinfo {title} {{Eine neue Methode zur
  Behandlung der longitudinalen und skalaren Photonen}},\ }\href
  {https://doi.org/10.5169/seals-112124} {\bibfield  {journal} {\bibinfo
  {journal} {Helv. Phys. Acta}\ }\textbf {\bibinfo {volume} {23}},\ \bibinfo
  {pages} {567} (\bibinfo {year} {1950})}\BibitemShut {NoStop}%
\bibitem [{\citenamefont {Fuster}\ \emph {et~al.}(2005)\citenamefont {Fuster},
  \citenamefont {Henneaux},\ and\ \citenamefont {Maas}}]{Fuster2005}%
  \BibitemOpen
  \bibfield  {author} {\bibinfo {author} {\bibfnamefont {A.}~\bibnamefont
  {Fuster}}, \bibinfo {author} {\bibfnamefont {M.}~\bibnamefont {Henneaux}},\
  and\ \bibinfo {author} {\bibfnamefont {A.}~\bibnamefont {Maas}},\ }\bibfield
  {title} {\bibinfo {title} {{BRST quantization: A short review}},\ }\href
  {https://doi.org/10.1142/S0219887805000892} {\bibfield  {journal} {\bibinfo
  {journal} {Int. J. Geom. Meth. Mod. Phys.}\ }\textbf {\bibinfo {volume}
  {2}},\ \bibinfo {pages} {939} (\bibinfo {year} {2005})}\BibitemShut {NoStop}%
\bibitem [{\citenamefont {Falceto}(2022)}]{Falceto2022}%
  \BibitemOpen
  \bibfield  {author} {\bibinfo {author} {\bibfnamefont {F.}~\bibnamefont
  {Falceto}},\ }\bibfield  {title} {\bibinfo {title} {{Canonical Quantization
  of the Electromagnetic Field in Arbitrary $\xi$-Gauge}},\ }\href
  {https://doi.org/10.1007/s00023-022-01259-w} {\bibfield  {journal} {\bibinfo
  {journal} {Ann. Henri Poincar{\'{e}}}\ ,\ \bibinfo {pages} {1}} (\bibinfo
  {year} {2022})}\BibitemShut {NoStop}%
\bibitem [{\citenamefont {Schleich}(2001)}]{Schleich2001}%
  \BibitemOpen
  \bibfield  {author} {\bibinfo {author} {\bibfnamefont {W.~P.}\ \bibnamefont
  {Schleich}},\ }\href {https://doi.org/10.1002/3527602976} {\emph {\bibinfo
  {title} {{Quantum Optics in Phase Space}}}}\ (\bibinfo  {publisher}
  {Wiley‐VCH},\ \bibinfo {address} {Weinheim},\ \bibinfo {year}
  {2001})\BibitemShut {NoStop}%
\bibitem [{\citenamefont {Lopp}\ and\ \citenamefont
  {Mart{\ifmmode\acute{\imath}\else\'{\i}\fi}n-Mart{\ifmmode\acute{\imath}\else\'{\i}\fi}nez}(2021)}]{Lopp2021}%
  \BibitemOpen
  \bibfield  {author} {\bibinfo {author} {\bibfnamefont {R.}~\bibnamefont
  {Lopp}}\ and\ \bibinfo {author} {\bibfnamefont {E.}~\bibnamefont
  {Mart{\ifmmode\acute{\imath}\else\'{\i}\fi}n-Mart{\ifmmode\acute{\imath}\else\'{\i}\fi}nez}},\
  }\bibfield  {title} {\bibinfo {title} {{Quantum delocalization, gauge, and
  quantum optics: Light-matter interaction in relativistic quantum
  information}},\ }\href {https://doi.org/10.1103/PhysRevA.103.013703}
  {\bibfield  {journal} {\bibinfo  {journal} {Phys. Rev. A}\ }\textbf {\bibinfo
  {volume} {103}},\ \bibinfo {pages} {013703} (\bibinfo {year}
  {2021})}\BibitemShut {NoStop}%
\bibitem [{\citenamefont {Glauber}(2007)}]{Glauber2007}%
  \BibitemOpen
  \bibfield  {author} {\bibinfo {author} {\bibfnamefont {R.~J.}\ \bibnamefont
  {Glauber}},\ }\href {https://doi.org/10.1002/9783527610075} {\emph {\bibinfo
  {title} {{Quantum Theory of Optical Coherence: Selected Papers and
  Lectures}}}}\ (\bibinfo  {publisher} {Wiley},\ \bibinfo {address}
  {Weinheim},\ \bibinfo {year} {2007})\BibitemShut {NoStop}%
\bibitem [{\citenamefont {Peskin}\ and\ \citenamefont
  {Schroeder}(2018)}]{Peskin2018}%
  \BibitemOpen
  \bibfield  {author} {\bibinfo {author} {\bibfnamefont {M.~E.}\ \bibnamefont
  {Peskin}}\ and\ \bibinfo {author} {\bibfnamefont {D.~V.}\ \bibnamefont
  {Schroeder}},\ }\href {https://doi.org/10.1201/9780429503559} {\emph
  {\bibinfo {title} {{An Introduction To Quantum Field Theory}}}}\ (\bibinfo
  {publisher} {Taylor {\&} Francis},\ \bibinfo {address} {Andover},\ \bibinfo
  {year} {2018})\BibitemShut {NoStop}%
\bibitem [{\citenamefont {Welton}(1948)}]{Welton1948}%
  \BibitemOpen
  \bibfield  {author} {\bibinfo {author} {\bibfnamefont {T.~A.}\ \bibnamefont
  {Welton}},\ }\bibfield  {title} {\bibinfo {title} {{Some Observable Effects
  of the Quantum-Mechanical Fluctuations of the Electromagnetic Field}},\
  }\href {https://doi.org/10.1103/PhysRev.74.1157} {\bibfield  {journal}
  {\bibinfo  {journal} {Phys. Rev.}\ }\textbf {\bibinfo {volume} {74}},\
  \bibinfo {pages} {1157} (\bibinfo {year} {1948})}\BibitemShut {NoStop}%
\bibitem [{\citenamefont {Bia{\l}ynicki-Birula}\ \emph
  {et~al.}(1993)\citenamefont {Bia{\l}ynicki-Birula}, \citenamefont
  {Freyberger},\ and\ \citenamefont {Schleich}}]{Bialynicki-Birula1993}%
  \BibitemOpen
  \bibfield  {author} {\bibinfo {author} {\bibfnamefont {I.}~\bibnamefont
  {Bia{\l}ynicki-Birula}}, \bibinfo {author} {\bibfnamefont {M.}~\bibnamefont
  {Freyberger}},\ and\ \bibinfo {author} {\bibfnamefont {W.}~\bibnamefont
  {Schleich}},\ }\bibfield  {title} {\bibinfo {title} {{Various measures of
  quantum phase uncertainty: a comparative study}},\ }\href
  {https://doi.org/10.1088/0031-8949/1993/T48/017} {\bibfield  {journal}
  {\bibinfo  {journal} {Phys. Scr.}\ }\textbf {\bibinfo {volume} {T48}},\
  \bibinfo {pages} {113} (\bibinfo {year} {1993})}\BibitemShut {NoStop}%
\bibitem [{\citenamefont {Dirac}(2013)}]{Dirac2013}%
  \BibitemOpen
  \bibfield  {author} {\bibinfo {author} {\bibfnamefont {P.~A.~M.}\
  \bibnamefont {Dirac}},\ }\href
  {https://books.google.de/books?id=Z3XCAgAAQBAJ} {\emph {\bibinfo {title}
  {{Lectures on Quantum Mechanics: Quantization with Constraints}}}}\ (\bibinfo
   {publisher} {Dover Publications},\ \bibinfo {address} {New York},\ \bibinfo
  {year} {2013})\BibitemShut {NoStop}%
\bibitem [{\citenamefont {Woolley}(2020)}]{Woolley2020}%
  \BibitemOpen
  \bibfield  {author} {\bibinfo {author} {\bibfnamefont {R.~G.}\ \bibnamefont
  {Woolley}},\ }\bibfield  {title} {\bibinfo {title} {{Power-Zienau-Woolley
  representations of nonrelativistic QED for atoms and molecules}},\ }\href
  {https://doi.org/10.1103/PhysRevResearch.2.013206} {\bibfield  {journal}
  {\bibinfo  {journal} {Phys. Rev. Res.}\ }\textbf {\bibinfo {volume} {2}},\
  \bibinfo {pages} {013206} (\bibinfo {year} {2020})}\BibitemShut {NoStop}%
\bibitem [{\citenamefont {Stokes}\ and\ \citenamefont
  {Nazir}(2021)}]{Stokes2021b}%
  \BibitemOpen
  \bibfield  {author} {\bibinfo {author} {\bibfnamefont {A.}~\bibnamefont
  {Stokes}}\ and\ \bibinfo {author} {\bibfnamefont {A.}~\bibnamefont {Nazir}},\
  }\bibfield  {title} {\bibinfo {title} {{Identification of
  Poincar{\'{e}}-gauge and multipolar nonrelativistic theories of QED}},\
  }\href {https://doi.org/10.1103/PhysRevA.104.032227} {\bibfield  {journal}
  {\bibinfo  {journal} {Phys. Rev. A}\ }\textbf {\bibinfo {volume} {104}},\
  \bibinfo {pages} {032227} (\bibinfo {year} {2021})}\BibitemShut {NoStop}%
\bibitem [{\citenamefont {Stokes}\ and\ \citenamefont
  {Nazir}(2022)}]{Stokes2022}%
  \BibitemOpen
  \bibfield  {author} {\bibinfo {author} {\bibfnamefont {A.}~\bibnamefont
  {Stokes}}\ and\ \bibinfo {author} {\bibfnamefont {A.}~\bibnamefont {Nazir}},\
  }\bibfield  {title} {\bibinfo {title} {{Implications of gauge freedom for
  nonrelativistic quantum electrodynamics}},\ }\href
  {https://doi.org/10.1103/RevModPhys.94.045003} {\bibfield  {journal}
  {\bibinfo  {journal} {Rev. Mod. Phys.}\ }\textbf {\bibinfo {volume} {94}},\
  \bibinfo {pages} {045003} (\bibinfo {year} {2022})}\BibitemShut {NoStop}%
\bibitem [{\citenamefont {Kakazu}\ and\ \citenamefont
  {Kim}(1995)}]{Kakazu1995}%
  \BibitemOpen
  \bibfield  {author} {\bibinfo {author} {\bibfnamefont {K.}~\bibnamefont
  {Kakazu}}\ and\ \bibinfo {author} {\bibfnamefont {Y.~S.}\ \bibnamefont
  {Kim}},\ }\bibfield  {title} {\bibinfo {title} {{Quantization of
  electromagnetic fields in a circular cylindrical cavity}},\ }\bibfield
  {journal} {\bibinfo  {journal} {arXiv}\ }\href
  {https://doi.org/10.48550/arXiv.quant-ph/9511012}
  {10.48550/arXiv.quant-ph/9511012} (\bibinfo {year} {1995})\BibitemShut
  {NoStop}%
\bibitem [{\citenamefont {Purcell}\ \emph {et~al.}(1946)\citenamefont
  {Purcell}, \citenamefont {Torrey},\ and\ \citenamefont
  {Pound}}]{Purcell1946}%
  \BibitemOpen
  \bibfield  {author} {\bibinfo {author} {\bibfnamefont {E.~M.}\ \bibnamefont
  {Purcell}}, \bibinfo {author} {\bibfnamefont {H.~C.}\ \bibnamefont
  {Torrey}},\ and\ \bibinfo {author} {\bibfnamefont {R.~V.}\ \bibnamefont
  {Pound}},\ }\bibfield  {title} {\bibinfo {title} {{Resonance Absorption by
  Nuclear Magnetic Moments in a Solid}},\ }\href
  {https://doi.org/10.1103/PhysRev.69.37} {\bibfield  {journal} {\bibinfo
  {journal} {Phys. Rev.}\ }\textbf {\bibinfo {volume} {69}},\ \bibinfo {pages}
  {37} (\bibinfo {year} {1946})}\BibitemShut {NoStop}%
\bibitem [{\citenamefont {Muljarov}\ and\ \citenamefont
  {Langbein}(2016)}]{Muljarov2016}%
  \BibitemOpen
  \bibfield  {author} {\bibinfo {author} {\bibfnamefont {E.~A.}\ \bibnamefont
  {Muljarov}}\ and\ \bibinfo {author} {\bibfnamefont {W.}~\bibnamefont
  {Langbein}},\ }\bibfield  {title} {\bibinfo {title} {{Exact mode volume and
  Purcell factor of open optical systems}},\ }\href
  {https://doi.org/10.1103/PhysRevB.94.235438} {\bibfield  {journal} {\bibinfo
  {journal} {Phys. Rev. B}\ }\textbf {\bibinfo {volume} {94}},\ \bibinfo
  {pages} {235438} (\bibinfo {year} {2016})}\BibitemShut {NoStop}%
\bibitem [{\citenamefont {Ren}\ \emph {et~al.}(2021)\citenamefont {Ren},
  \citenamefont {Franke},\ and\ \citenamefont {Hughes}}]{Ren2021}%
  \BibitemOpen
  \bibfield  {author} {\bibinfo {author} {\bibfnamefont {J.}~\bibnamefont
  {Ren}}, \bibinfo {author} {\bibfnamefont {S.}~\bibnamefont {Franke}},\ and\
  \bibinfo {author} {\bibfnamefont {S.}~\bibnamefont {Hughes}},\ }\bibfield
  {title} {\bibinfo {title} {{Quasinormal Modes, Local Density of States, and
  Classical Purcell Factors for Coupled Loss-Gain Resonators}},\ }\href
  {https://doi.org/10.1103/PhysRevX.11.041020} {\bibfield  {journal} {\bibinfo
  {journal} {Phys. Rev. X}\ }\textbf {\bibinfo {volume} {11}},\ \bibinfo
  {pages} {041020} (\bibinfo {year} {2021})}\BibitemShut {NoStop}%
\bibitem [{\citenamefont {Joannopoulos}\ \emph {et~al.}(2008)\citenamefont
  {Joannopoulos}, \citenamefont {Johnson}, \citenamefont {Winn},\ and\
  \citenamefont {Meade}}]{Joannopoulos2008}%
  \BibitemOpen
  \bibfield  {author} {\bibinfo {author} {\bibfnamefont {J.~D.}\ \bibnamefont
  {Joannopoulos}}, \bibinfo {author} {\bibfnamefont {S.~G.}\ \bibnamefont
  {Johnson}}, \bibinfo {author} {\bibfnamefont {J.~N.}\ \bibnamefont {Winn}},\
  and\ \bibinfo {author} {\bibfnamefont {R.~D.}\ \bibnamefont {Meade}},\ }\href
  {https://press.princeton.edu/books/hardcover/9780691124568/photonic-crystals}
  {\emph {\bibinfo {title} {{Photonic Crystals}}}}\ (\bibinfo  {publisher}
  {Princeton University Press},\ \bibinfo {address} {Princeton},\ \bibinfo
  {year} {2008})\BibitemShut {NoStop}%
\bibitem [{\citenamefont {Zemanian}(1987)}]{Zemanian1987}%
  \BibitemOpen
  \bibfield  {author} {\bibinfo {author} {\bibfnamefont {A.~H.}\ \bibnamefont
  {Zemanian}},\ }\href {https://doi.org/10.5555/40028} {\emph {\bibinfo {title}
  {{Distribution theory and transform analysis: an introduction to generalized
  functions, with applications}}}}\ (\bibinfo  {publisher} {Dover
  Publications},\ \bibinfo {address} {New York},\ \bibinfo {year}
  {1987})\BibitemShut {NoStop}%
\bibitem [{\citenamefont {Estrada}\ and\ \citenamefont
  {Kanwal}(1989)}]{Estrada1989}%
  \BibitemOpen
  \bibfield  {author} {\bibinfo {author} {\bibfnamefont {R.}~\bibnamefont
  {Estrada}}\ and\ \bibinfo {author} {\bibfnamefont {R.~P.}\ \bibnamefont
  {Kanwal}},\ }\bibfield  {title} {\bibinfo {title} {{Regularization,
  pseudofunction, and hadamard finite part}},\ }\href
  {https://doi.org/10.1016/0022-247X(89)90216-3} {\bibfield  {journal}
  {\bibinfo  {journal} {J. Math. Anal. Appl.}\ }\textbf {\bibinfo {volume}
  {141}},\ \bibinfo {pages} {195} (\bibinfo {year} {1989})}\BibitemShut
  {NoStop}%
\bibitem [{\citenamefont {Galapon}(2016)}]{Galapon2016}%
  \BibitemOpen
  \bibfield  {author} {\bibinfo {author} {\bibfnamefont {E.~A.}\ \bibnamefont
  {Galapon}},\ }\bibfield  {title} {\bibinfo {title} {{The Cauchy principal
  value and the Hadamard finite part integral as values of absolutely
  convergent integrals}},\ }\bibfield  {journal} {\bibinfo  {journal} {J. Math.
  Phys.}\ }\textbf {\bibinfo {volume} {57}},\ \href
  {https://doi.org/10.1063/1.4943300} {10.1063/1.4943300} (\bibinfo {year}
  {2016})\BibitemShut {NoStop}%
\bibitem [{\citenamefont {Jones}(1996)}]{Jones1996}%
  \BibitemOpen
  \bibfield  {author} {\bibinfo {author} {\bibfnamefont {D.~S.}\ \bibnamefont
  {Jones}},\ }\bibfield  {title} {\bibinfo {title} {{Hadamard's Finite Part}},\
  }\href
  {https://doi.org/10.1002/(SICI)1099-1476(19960910)19:13<1017::AID-MMA723>3.0.CO;2-2}
  {\bibfield  {journal} {\bibinfo  {journal} {Math. Methods Appl. Sci.}\
  }\textbf {\bibinfo {volume} {19}},\ \bibinfo {pages} {1017} (\bibinfo {year}
  {1996})}\BibitemShut {NoStop}%
\end{thebibliography}%
\end{document}